\documentclass[12pt,a4paper,twoside,openany]{book}
\usepackage[margin=3cm]{geometry}
\usepackage[utf8]{inputenc}

\usepackage[english]{babel}
\usepackage{mathtools}
\usepackage{amsmath}
\usepackage{amsfonts}
\usepackage{amssymb}
\usepackage{graphicx}
\usepackage{subcaption}
\usepackage[font=small,labelfont=bf]{caption}
\usepackage{float}
\usepackage[colorlinks=true,urlcolor=blue,citecolor=blue,linkcolor=black]{hyperref}
\usepackage[font=footnotesize,labelfont=bf]{caption}
\usepackage{emptypage}
\usepackage{fancyhdr}
\usepackage[toc]{appendix}
\numberwithin{equation}{chapter}

\usepackage{tensor}
\usepackage{accents}

\usepackage{verbatim}

\usepackage{amsthm}
\theoremstyle{definition}
 
\theoremstyle{remark}

\usepackage{titlesec}
\titleformat{\chapter}[display]   
{\normalfont\huge\bfseries}{\chaptertitlename\ \thechapter}{20pt}{\Huge}   
\titlespacing*{\chapter}{0pt}{-50pt}{40pt}

\usepackage[en-US]{datetime2}
\DTMlangsetup{showdayofmonth=false}
\usepackage{lastpage}
\usepackage[english]{HYtitle}
\title{Higgs inflation and higher-order gravity in Palatini formulation}
\date{April 2020}
\author{Jaakko Annala}
\level{Master's Thesis}
\subject{Theoretical Physics}
\faculty{Faculty of Science}
\programme{}
\department{Department of Physics}
\address{P.O. Box 64 (Gustaf Hällströmin katu 2a)\\FI-00014 University of Helsinki}
\prof{Syksy Räsänen}
\censors{Syksy Räsänen}{Mark Hindmarsh}{}
\keywords{cosmology, inflation, Higgs, modifications of gravity, Palatini formulation}
\depositeplace{}
\additionalinformation{}
\classification{}
\numberofpages{\pageref{LastPage}}

\newcommand{\pd}{\partial}
\newcommand{\cd}{\nabla}
\newcommand{\half}{\frac{1}{2}}
\newcommand{\Lagr}{\mathcal{L}}

\newcommand{\lcd}{\accentset{\circ}{\cd}}
\newcommand{\lcc}{\accentset{\circ}{\Gamma}}
\newcommand{\lBox}{\accentset{\circ}{\Box}}

\newcommand{\sg }{\sqrt{-g}}
\newcommand{\sR}{R}
\newcommand{\sRc}{\hat{\sR}}

\newcommand{\lcR}{\accentset{\circ}{R}}

\newcommand{\TI}{\indices}

\newcommand{\sq}{\sqrt{-q}}
\newcommand{\Ut}{\accentset{\sim}{U}} 

\begin{document}
\frontmatter

\hymaketitle
\begin{abstract}
We study how higher-order gravity affects Higgs inflation in the Palatini formulation. We first review the metric and Palatini formulations in comparative manner and discuss their differences. Next cosmic inflation driven by a scalar field and inflationary observables are discussed. After this we review the Higgs inflation and compute the inflationary observables both in the metric and Palatini formulations. We then consider adding higher-order terms of the curvature to the action. We derive the equations of motion for the most general action quadratic in the curvature that does not violate parity in both the metric and Palatini formulations. Finally we present a new result. We analyse Higgs inflation in the Palatini formulation with higher-order curvature terms. We consider a simplified scenario where only terms constructed from the symmetric part of the Ricci tensor are added to the action. This implies that there are no new gravitational degrees of freedom, which makes the analysis easier. As a new result we found out that the scalar perturbation spectrum is unchanged, but the tensor perturbation spectrum is suppressed by the higher-order curvature couplings.
\end{abstract}
\tableofcontents

\chapter{Notation and conventions} \label{sec:notation} 

    We follow the conventions of \cite{Carroll04}. The metric tensor with the signature $(-+++)$ is denoted by $g_{\mu\nu}$ and its determinant by
    \begin{equation}
        g = \mathrm{det}(g_{\mu\nu}).
    \end{equation}
    Covariant derivative with respect to a general connection is denoted by $\cd_\mu$ and its connection coefficients by $\Gamma^{\sigma}_{\mu\nu}$. The connection coefficients are defined such that the covariant derivative is
    \begin{subequations} \label{eq:defCD}
        \begin{equation} 
            \cd_\mu T^\nu = \pd_\mu T^\nu + \Gamma^{\nu}_{\mu\lambda}T^{\lambda},
        \end{equation}
        \begin{equation}
            \cd_\mu T_\nu = \pd_\mu T_\nu - \Gamma^{\lambda}_{\mu\nu}T_{\lambda}.
        \end{equation}    
    \end{subequations}
    When it is more convenient we use the comma notation for partial derivatives, $\frac{\pd \phi}{\pd x^\mu} = \pd_\mu \phi = \phi_{,\mu}$, and a semicolon for covariant derivatives, $\cd_\mu T^{...}_{...} = T^{...}_{...;\mu}$. The Levi--Civita connection, which is metric compatible and symmetric, is denoted by a circle over the symbol and the coefficients are
    \begin{equation}\label{eq:lcc0}
        \lcc^{\sigma}_{\mu\nu} = \half g^{\sigma\rho}\left(\pd_\mu g_{\nu\rho} + \pd_\nu g_{\rho\mu} - \pd_\rho g_{\mu\nu}\right).
    \end{equation}
    Furthermore, quantities constructed from the Levi--Civita connection are denoted by a circle over the symbols, e.g. $\lcd_\mu$. The Riemann tensor is defined as
    \begin{equation} \label{eq:defRie}
        \sR\TI{^\rho_{\sigma\mu\nu}}= \pd_\mu\Gamma^{\rho}_{\nu\sigma}-\pd_\nu\Gamma^{\rho}_{\mu\sigma} + \Gamma^{\rho}_{\mu\lambda}\Gamma^{\lambda}_{\nu\sigma} - \Gamma^{\rho}_{\nu\lambda}\Gamma^{\lambda}_{\mu\sigma}.
    \end{equation}
    The symmetrization of indices is defined as
    \begin{equation} \label{eq:defSym}
        T^{(\mu_1...\mu_n)} = \frac{1}{n!}\sum_{\sigma \in S_n}T^{\mu_{\sigma(1)}...\mu_{\sigma(n)}},
    \end{equation}
    where $S_n$ is the symmetric group of $n$ elements. The anti-symmetrization of indices is defined as
    \begin{equation} \label{eq:defaSym}
        T^{[\mu_1...\mu_n]} = \frac{1}{n!}\sum_{\sigma \in S_n} \mathrm{sgn}(\sigma)T^{\mu_{\sigma(1)}...\mu_{\sigma(n)}},
    \end{equation}
    where $\mathrm{sgn}(\sigma)$ gives the sign of the permutation $\sigma$: $+1$ for even and $-1$ for odd permutation. The torsion tensor is defined as
    \begin{equation}\label{eq:torsion}
        T\TI{^\sigma_{\mu\nu}} = 2\Gamma^\sigma_{[\mu\nu]}.
    \end{equation}
    The Weyl tensor is defined in dimension $d>2$ as
    \begin{equation}\label{eq:weyl}
        \begin{aligned}
            W_{\rho\sigma\mu\nu} = R_{\rho\sigma\mu\nu}&-\frac{2}{(d-2)}\left(g_{\rho[\mu}R_{\nu]\sigma}-g_{\sigma[\mu}R_{\nu]\rho}\right)\\ &+ \frac{2}{(d-1)(d-2)}g_{\rho[\mu}g_{\nu]\sigma}R.    
        \end{aligned}
    \end{equation}
    The Levi--Civita symbol in right handed coordinate systems is defined
    \begin{equation} \label{eq:eps}
        \epsilon_{\mu_1\mu_2...\mu_n} =
        \begin{cases}
            +1 & \quad \text{if} \quad \mu_1...\mu_n \quad \text{is an even permutation of} \quad 0,1,...,(n-1) \\
            -1 & \quad \text{if} \quad \mu_1...\mu_n \quad \text{is an odd permutation of} \quad 0,1,...,(n-1) \\
            0 & \quad \text{otherwise}.
        \end{cases}
    \end{equation}
    The generalized Kronecker delta is defined
    \begin{equation} \label{eq:genKronecker}
        \delta^{\mu_1\mu_2...\mu_n}_{\nu_1\nu_2...\nu_n} \equiv n!\delta^{[\mu_1}_{\nu_1}\delta^{\mu_2}_{\nu_2}...\delta^{\mu_n]}_{\nu_n}.
    \end{equation}
    We use units where 
    \begin{equation}
        \hbar = c = 1.
    \end{equation}
    Additionally, unless otherwise stated, we set the Planck mass to unity
    \begin{equation}
        M_p\equiv\frac{1}{\sqrt{8\pi G_N}}=1,
    \end{equation}
    where $G_N$ is the Newton's gravitational constant.
\mainmatter

\chapter{Introduction} \label{sec:Intro}

Inflation is a hypothetical period of exponential expansion in the very early universe, which is able to explain problems that the standard Hot Big Bang model cannot answer. Most notably, it can explain the origin of the primordial perturbations that ultimately give rise to the structure formation of our universe and predict the form of the fluctuations we observe in the cosmic microwave background (CMB). Inflation is a paradigm, and there is a plethora of different models that have various ways of producing the exponential expansion of the universe, see e.g. \cite{Martin13}. 
In Higgs inflation the Standard Model (SM) Higgs is non-minimally coupled to gravity and drives inflation. The advantage of this scenario is that it is very minimalistic: no new fields are added to the SM and only one new parameter is added, namely the non-minimal coupling to gravity. Higgs inflation is reviewed in chapter \ref{sec:HI}.

In higher-order gravity, higher-order terms in the curvature are present in the gravity sector. The appearance of these terms can be motivated in several ways, more on this in chapter \ref{sec:higherG}. Modified gravity sector (even just the addition of non-minimal coupling in Higgs inflation) makes the question: which are the fundamental degrees of freedom in General Relativity (GR), even more relevant. GR can be formulated in different ways by starting with different independent variables and applying the variational principle to the Einstein--Hilbert action. In the metric formulation of GR the spacetime manifold is described by a metric and the Levi--Civita connection, which is completely described by the metric. The metric is assumed to be the only degree of freedom of gravity and varying the Einstein--Hilbert action with respect to the metric gives the Einstein field equations. However, a priori there is no reason for the connection to be described purely by the metric. In the Palatini formulation\footnote{In fact Einstein was the first to consider the Palatini formulation \cite{Ferraris82}.} the metric and a general connection are treated as independent degrees of freedom. In the case of the Einstein--Hilbert action, variation with respect to the connection makes the connection become the Levi--Civita connection, and thus in the case of the Einstein--Hilbert action these formulations are equivalent. However, when the gravity sector is modified the two formulations differ. These two formulations lead to different inflationary predictions. The aim of this thesis is to investigate how higher-order curvature terms in the Palatini formulation change the predictions of Higgs inflation.

The structure of this thesis is as follows. First, in chapter \ref{sec:varGR} we introduce the metric and Palatini formulations of gravity. We show that the formulations are equivalent with the Einstein--Hilbert action, however with subtle differences. When introducing the metric formulation we discuss non-trivial boundary terms. In Palatini formulation we consider the subtleties arising with the most general connection.

In chapter \ref{sec:Inflation}, we introduce the topics of cosmology relevant for this thesis, namely the FLRW model and inflation. In the case of inflation we focus on the generation of primordial perturbations and inflationary observables.

In chapter \ref{sec:HI}, we review Higgs inflation in the metric and Palatini formulations. We also discuss conformal transformations that make the analysis easier.

In chapter \ref{sec:higherG} we review the basic properties of higher-order gravity in the metric and Palatini formulations. By considering $f(R)$ gravity we illustrate the differences between the two formulations. We also present the general parity preserving action that is quadratic in the curvature tensor in the metric and Palatini formulation and derive the equations of motion for these terms.

Finally, in chapter \ref{sec:QGHI} we present a new result. We tackle the question how the higher-order curvature terms change Higgs inflation. We consider a simple case where only the symmetric part of the Ricci tensor appears in the action. We also discuss disformal transformations, which turn out to be relevant for our analysis. We then derive the slow-roll equations and finally compute how the inflationary predictions are modified by the added higher-order curvature terms.

\chapter[Variational principle in general relativity]{Variational principle in \\general relativity} \label{sec:varGR}

The variational principle has an important role in physics. Most field theories are formulated by writing down an action in terms of the fields of the theory. GR is no exception and it can be formulated using an action and applying the variational principle.

The formulation of gravity that we are focusing in this thesis is the Palatini formulation (sometimes also called the first order formalism) in which the metric tensor and the connection are independent degrees of freedom. In this chapter we review the main features of the metric and the Palatini formulations and see how the procedure works with the Einstein--Hilbert action. We also consider the subtle differences of these formulations.

\section{Metric formulation} \label{sec:metricForm}
The assumption in metric formulation is that the only degree of freedom describing the spacetime is the metric $g_{\mu\nu}$ and hence the connection on the manifold is metric compatible $\lcd_\sigma g^{\mu\nu}=0$ (parallel transport preserves the metric) and torsion free $\lcc^{\sigma}_{\mu\nu} = \lcc^\sigma_{\nu\mu}$. Thus, the connection coefficients are
\begin{equation}\label{eq:lcc}
    \lcc^{\sigma}_{\mu\nu} = \half g^{\sigma\rho}\left(\pd_\mu g_{\nu\rho} + \pd_\nu g_{\rho\mu} - \pd_\rho g_{\mu\nu}\right).
\end{equation}
The formulation proceeds by defining a Lagrangian density $\Lagr$, which is a functional depending purely on the metric and its derivatives: $\Lagr = \Lagr\left(g_{\mu\nu},\pd g_{\mu\nu}, \pd \pd g_{\mu\nu},...\right)$. The Lagrangian has to be Lorentz invariant, also other mathematical or physical constraints may be imposed e.g. simplicity and invariance under other transformations. The action integral is the Lagrangian integrated over a compact region of the spacetime $\Sigma$ with respect to the invariant spacetime volume element $\sg d^nx$. The action reads
\begin{equation} \label{eq:met:act}
    S[g_{\mu\nu}] = \int_{\Sigma}d^nx \sg \Lagr.
\end{equation}
Then the classical variational principle is applied: requiring that an arbitrary variation of the action with respect to the metric should vanish identically, while the variation on the boundary $\pd\Sigma$ vanishes. In other words, the functional derivative of the Lagrangian $\Lagr$ with respect to the metric vanishes,

\begin{equation}
    \frac{\delta\Lagr}{\delta g^{\mu\nu}} = 0.
\end{equation}

The simplest Lagrangian constructed only from the metric and its derivatives, that leads to the Einstein field equations, is the Ricci scalar. This gives the Einstein--Hilbert action
\begin{equation}\label{eq:EH}
    S_{EH}=\frac{1}{2}\int d^4x \sg \lcR,
\end{equation}
where $\lcR = g^{\mu\nu}\lcR\TI{^\lambda_{\mu\lambda\nu}}$ is the Ricci scalar. Variation of this action produces Einstein field equations in vacuum. One may add in to the action a cosmological constant $\Lambda$ and a matter part $\Lagr_{M}$. The variation of the matter part will yield the stress-energy tensor
\begin{equation}
    T_{\mu\nu} = \frac{-2}{\sg}\frac{\delta S_M}{\delta g^{\mu\nu}},
\end{equation}
where $S_M$ is the matter part of the action containing $\Lagr_M$. This leads to the Einstein field equations with a cosmological constant and a source term
\begin{equation}\label{eq:EFE}
    \lcR_{\mu\nu} - \half g_{\mu\nu}\lcR + \Lambda g_{\mu\nu} = T_{\mu\nu}.
\end{equation}

There is, however, a slight caveat in the derivation that concerns the boundary terms. When comparing different formulations it is useful to take the boundaries into account, since the resulting boundary conditions might be different.

\subsection{Einstein--Hilbert action in metric formulation}\label{sec:EH-metric-var}

We will shortly review the process of variation with the Einstein--Hilbert action. We need to take the variation of $\sg g^{\mu\nu}\lcR_{\mu\nu}$ with respect to the inverse metric $g^{\mu\nu}$. To accomplish this we make use of a couple of useful identities, namely:
\begin{equation}\label{eq:var_sg}
    \delta \sg  = -\half \sg  g_{\mu\nu} \delta g^{\mu\nu}
\end{equation} 
and the Palatini identity, which in this case reads
\begin{equation} \label{eq:pal_idMetric}
    \delta \lcR\TI{^\rho_{\mu\lambda\nu}} = \lcd_\lambda\delta \lcc^{\rho}_{\nu\mu} - \lcd_\nu\delta \lcc^{\rho}_{\lambda\mu},
\end{equation}
see appendix \ref{app:identities}. First of all we have:
\begin{equation}
    \delta(\sg g^{\mu\nu}\lcR_{\mu\nu}) = \sg \lcR_{\mu\nu}\delta g^{\mu\nu} + \lcR\delta \sg  + \sg g^{\mu\nu}\delta \lcR_{\mu\nu}.
\end{equation}
Taking the variation of the connection \eqref{eq:lcc} gives
\begin{equation}\label{eq:levicivitaVariation}
    \delta\lcc^\sigma­_{\mu\nu} = \half\left[ g_{\mu\alpha}g_{\nu\beta}\lcd^\sigma\delta g^{\alpha\beta}-g_{\alpha\mu}\lcd_\nu\delta g^{\alpha\sigma}-g_{\alpha\nu}\lcd_\mu\delta g^{\alpha\sigma} \right].
\end{equation}
Plugging \eqref{eq:levicivitaVariation} into the Palatini identity \eqref{eq:pal_idMetric} and taking the contraction between the first and the third index we end up with the variation of the Ricci tensor, finally contracting this we have

\begin{equation} \label{eq:RicciScalarVar}
    g^{\mu\nu}\delta \lcR_{\mu\nu} = \lcd_\lambda\left[g_{\mu\nu}\lcd^\lambda\delta g^{\mu\nu} - \lcd_\alpha\delta g^{\alpha\lambda}\right].
\end{equation}
The variation of the action then reads

\begin{equation} \label{eq:EHvar}
\begin{aligned}
    \delta S_{EH} &= \int d^4x \sg \left[ \lcR_{\mu\nu} - \half g_{\mu\nu}\lcR \right]\delta g^{\mu\nu}\\
    &+ \int d^4x \sg \lcd_\lambda\left[g_{\mu\nu}\lcd^\lambda\delta g^{\mu\nu} - \lcd_\alpha\delta g^{\alpha\lambda}\right].
\end{aligned}
\end{equation}
From this we immediately see that the first term leads to the Einstein field equations in vacuum and the second term leads to a boundary term which we will consider next.

\subsection{York-Gibbons-Hawking boundary term}\label{sec:YGH-term}
The second term in \eqref{eq:EHvar} is a total divergence and leads to a boundary term by using the Stokes' theorem

\begin{equation}\label{eq:stokes}
    \int_{\Sigma}d^nx\sg \lcd_\mu V^\mu = \int_{\pd\Sigma}d^{n-1}x\sqrt{|\gamma|}n_\mu V^{\mu},
\end{equation}
where $V^\mu$ is a vector field defined in the region $\Sigma$ with a boundary $\pd\Sigma$, $\gamma_{ab}$ is the induced metric on the boundary and $n^\mu$ is the unit normal to the surface. Note that this form of the Stokes' theorem only applies with the Levi--Civita connection; this turns out to be important later when we consider the Palatini formulation. Using this, the second term in \eqref{eq:EHvar} equals

\begin{equation}\label{eq:EHboundary}
    \int_{\pd\Sigma}d^3x\sqrt{|\gamma|}n_\lambda\left[g_{\mu\nu}\lcd^\lambda\delta g^{\mu\nu} - \lcd_\alpha\delta g^{\alpha\lambda}\right],
\end{equation}
from which we see that the boundary term does not vanish if we only assume that the variation vanishes on the boundary. By imposing $\delta g^{\mu\nu}|_{\pd\Sigma}=0$ we get
\begin{equation}\label{eq:EHboundary2}
    \int_{\pd\Sigma}d^3x\sqrt{|\gamma|}n_\lambda\left[g_{\mu\nu}\pd^\lambda\delta g^{\mu\nu} - \pd_\alpha\delta g^{\alpha\lambda}\right].
\end{equation}
We would also have to require $\pd_\lambda\delta g^{\mu\nu}|_{\pd\Sigma}=0$ for this term to vanish. An other way is to modify the action by a term that will cancel this boundary contribution. This additional term is called the York-Gibbons-Hawking (YGH) boundary term \cite{HawGib77,York72}. The expression \eqref{eq:EHboundary2} can be further simplified by using the following identity for the induced metric, assuming that the surface $\pd\Sigma$ is not null (see e.g. \cite[ch. 2.7]{Hawking73})

\begin{equation} \label{eq:inducedMet}
    g^{\mu\nu} = \epsilon n^\mu n^\nu + \gamma^{ab}\frac{\pd x^\mu}{\pd y^a}\frac{\pd x^\nu}{\pd y^b},
\end{equation}
where $\epsilon=n^\mu n_\mu$ equals $+1$ if $\pd\Sigma$ is time-like and $-1$ if spacelike and $y^a$ are the coordinates on the surface. We also denote 

\begin{equation}
    \gamma^{\mu\nu} = \gamma^{ab}\frac{\pd x^\mu}{\pd y^a}\frac{\pd x^\nu}{\pd y^b}.
\end{equation}
The requirement that the variation of the metric vanishes on the boundary implies that the tangential derivative $n_\lambda\gamma_{\mu\nu}\pd^\nu \delta g^{\mu\lambda}=0$ vanishes. The surface term \eqref{eq:EHboundary2} then becomes
\begin{equation} \label{eq:YGH1}
    \int_{\pd\Sigma} d^{3}x\sqrt{|\gamma|}\gamma_{\mu\nu}n_\lambda\pd^\lambda \delta g^{\mu\nu},
\end{equation}
which can be be canceled by introducing a YGH--boundary term

\begin{equation}\label{eq:YGH}
    S_{YGH}=\int_{\pd\Sigma}d^{3}x\epsilon\sqrt{|\gamma|}K,
\end{equation}
where $K = \lcd_\mu n^\mu$ is the trace of the extrinsic curvature of the boundary. It is straightforward to verify this by taking the variation of this boundary term \eqref{eq:YGH} and see that it yields \eqref{eq:YGH1} up to a constant. For more details see e.g. \cite{GCT10} and \cite[Appendix E]{WaldGR}. Thus when taking into account the boundaries, the action that produces the Einstein field equations in the metric formulation is
\begin{equation} \label{eq:EHwYGH}
    S_{GR}=\int_{\Sigma} d^4x \sg \lcR + 2\int_{\pd\Sigma}d^{3}x\epsilon\sqrt{|\gamma|}K.
\end{equation}
This is one method of dealing with the boundary term. In this thesis we are not concerned about the details of these terms; for the physical significance of these terms and further consideration see e.g. \cite{Dyer08}. In the next section we will see that in the Palatini formulation no YGH boundary term need to be added.

\section{Palatini formulation} \label{sec:palatini}

In the Palatini formulation the metric and the connection are considered as independent degrees of freedom. Thus we have a spacetime manifold with a symmetric Lorentzian metric $g_{\mu\nu}$ and a general connection, for which the metric compatibility does not hold, $\cd_\sigma g_{\mu\nu}\neq 0$, and the torsion does not necessarily vanish $T\TI{^\sigma_{\mu\nu}}\neq 0$. 

The formulation is proceeded by constructing a Lagrangian density, which now is a functional of both the metric and the connection and their derivatives\\$\Lagr = \Lagr(g, \pd g, \pd \pd g,...,\Gamma, \pd\Gamma, \pd\pd\Gamma,...)$. The action is now,

\begin{equation}\label{eq:palatiniAction}
    S[g_{\mu\nu},\Gamma^\sigma_{\mu\nu}] = \int_\Sigma d^nx\sg \Lagr.
\end{equation}
Next we apply the classical variational principle: first we vary both the metric and the connection separately with the variations vanishing on the boundary $\pd\Sigma$, and then require that the variation of the action vanishes. This yields the equations of motion for both the metric and the connection

\begin{equation}
    \frac{\delta \Lagr}{\delta g^{\mu\nu}} = 0, \quad\mathrm{and}\quad \frac{\delta \Lagr}{\delta\Gamma^\sigma_{\mu\nu}}=0.
\end{equation}

The appeal of the Palatini formulation is that in the case of Einstein--Hilbert action the dynamical equation for the connection imposes the connection to be Levi--Civita and thus we recover the pseudo-Riemannian space with Einstein equations, with no prior assumptions. Furthermore there is no need to introduce YGH--boundary terms to the action. There are, however, some unappealing properties in this formulation when considering the most general connection and when introducing matter into the Lagrangian. We will discuss these problems after we review the variation of the Einstein--Hilbert action and compare the procedure to the metric case.

\subsection{Einstein--Hilbert action in the Palatini formulation}\label{sec:EH-Palatini}
We consider the case of a completely general connection. We denote the Riemann tensor constructed purely from the connection (and tensors derived from it) with $\sR\TI{^\rho_{\mu\sigma\nu}}$ to differentiate from the Levi--Civita counterpart denoted with $\lcR\TI{^\rho_{\mu\sigma\nu}}$. With a completely general connection the Riemann tensor does not have all the symmetry properties as with the Levi--Civita connection. The only property that remains is the antisymmetry of the last two indices, which is apparent from the definition \eqref{eq:defRie}. This then implies that there is more than one unique contraction that we can take of the Riemann tensor in contrast to the one unique choice in the Levi--Civita case: the Ricci tensor. Thus it is interesting to see whether the Ricci scalar is unique or not. There are in fact three different non vanishing contractions of the Riemann tensor. The Ricci tensor is defined as
\begin{equation}\label{eq:Ricci}
    \sR_{\mu\nu} = \sR\TI{^\lambda_{\mu\lambda\nu}},
\end{equation}
which in general is not symmetric. We define the co-Ricci tensor as
\begin{equation}\label{eq:coRicci}
    \hat{\sR}\TI{^\mu_\nu} = g^{\lambda\sigma}\sR\TI{^\mu_{\sigma\nu\lambda}},
\end{equation}
which in general does not have any symmetry properties. Finally we define the antisymmetric Ricci tensor
\begin{equation}\label{eq:antisymRicci}
    \sR^\prime_{\mu\nu}=\sR\TI{^\lambda_{\lambda\mu\nu}},
\end{equation}
which is sometimes also called the homothetic curvature and it is antisymmetric from the definition of the Riemann tensor.
Since the metric tensor is symmetric the trace of $g^{\mu\nu}\sR^\prime_{\mu\nu}=0$ vanishes, furthermore 
\begin{equation}
    \hat{\sR}\TI{^\mu_\mu} = g^{\lambda\sigma}\sR\TI{^\mu_{\sigma\mu\lambda}} = g^{\lambda\sigma}\sR_{\lambda\sigma} = \sR. 
\end{equation}
Thus the Ricci scalar still remains unique. These different Ricci tensors come into play in the higher-order Gravity in section \ref{sec:higherG}, but now we can say that the Einstein--Hilbert action is still uniquely determined\footnote{There exist other candidates other than the Ricci scalar that results in second order field equations, see e.g. \cite{Burton97,Vitagliano10}.}.

The action is
\begin{equation}\label{eq:EHpalatini}
    S = \int d^4x \sg g^{\mu\nu}\sR_{\mu\nu}(\Gamma).
\end{equation}
The variation with respect to the metric is rather straightforward, since the variation of the Ricci tensor is zero; so, we have
\begin{equation}
    \delta(\sg g^{\mu\nu}\sR_{\mu\nu}(\Gamma)) = \delta \sg g^{\mu\nu}\sR_{\mu\nu}(\Gamma) + \delta(g^{\mu\nu})\sR_{\mu\nu}(\Gamma). 
\end{equation}
Using \eqref{eq:var_sg} the metric variation $\frac{\delta S}{\delta g^{\mu\nu}}=0$ gives
\begin{equation}
    \sR_{(\mu\nu)}-\half g_{\mu\nu}\sR=0.
\end{equation}
Notice that there were no appearances of any boundary terms. 

The variation with respect to the connection is more involved. Since the only object depending on the connection is the Ricci tensor we need to compute its variation $\delta \sR_{\mu\nu}$. For this we make use of the general Palatini identity
\begin{equation}\label{eq:palatiniID}
    \delta \sR\TI{^\rho_{\mu\lambda\nu}} = \cd_\lambda\delta\Gamma^\rho_{\nu\mu}-\cd_\nu\delta\Gamma^\rho_{\lambda\mu} + T\TI{^\sigma_{\lambda\nu}}\delta\Gamma^\rho_{\sigma\mu},
\end{equation}
see Appendix \ref{app:identities}. Taking the contraction of the first and third index gives the variation of $\delta \sR_{\mu\nu}$. So the variation of the action so far is
\begin{equation}\label{eq:EHPalVar1}
    \delta S = \int d^4x \sg g^{\mu\nu}\left( \cd_\lambda\delta\Gamma^\lambda_{\nu\mu}-\cd_\nu\delta\Gamma^\lambda_{\lambda\mu} + T\TI{^\sigma_{\lambda\nu}}\delta\Gamma^\lambda_{\sigma\mu} \right).
\end{equation}
The first and the second term inside the brackets need to be simplified further. We will focus on the first term in \eqref{eq:EHPalVar1}, since steps for the second term are the same. With the chain rule we have
\begin{equation} \label{eq:EHPalVar2}
    \sg g^{\mu\nu}\cd_\lambda\delta\Gamma^\lambda_{\nu\mu} = \cd_\lambda\left(\sg g^{\mu\nu}\delta\Gamma^\lambda_{\nu\mu}\right) - \cd_\lambda\left(\sg g^{\mu\nu}\right)\delta\Gamma^\lambda_{\nu\mu}.
\end{equation}
The first term on the right side of the equation resembles a surface term. We cannot, however, turn this into a surface term with the Stokes' theorem \eqref{eq:stokes}, since it is only valid for the Levi--Civita connection. To deal with this, we use a change of variables and write our general connection as
\begin{equation}\label{eq:con_C}
    \Gamma^\sigma_{\mu\nu} = \lcc^\sigma_{\mu\nu} + C\TI{^\sigma_{\mu\nu}},
\end{equation}
where $\lcc^\sigma_{\mu\nu}$ is the Levi--Civita connection and $C\TI{^\sigma_{\mu\nu}}$ is the difference from it and hence a tensor. In addition, the second term on the right hand side of equation \eqref{eq:EHPalVar2} is a covariant derivative acting on a tensor density of weight $+1$, which we will evaluate in order to get the desired form for the equation. For this we need the covariant derivative of a scalar density, which for a general scalar density weight $\omega$ is
\begin{equation} \label{eq:tensordensityCD}
    \cd_\mu \rho = \pd_\mu \rho -\omega\Gamma^\sigma_{\mu\sigma}\rho
\end{equation}
see Appendix \ref{app:identities} or \cite[ch. III]{Schouten54}. Note that it is important which of the contractions is taken from the connection coefficients in \eqref{eq:tensordensityCD} and this depends on the convention for the covariant derivative \eqref{eq:defCD}. Expanding \eqref{eq:EHPalVar2} further
\begin{equation}\label{eq:EHPalVar3}
    \sg g^{\mu\nu}\cd_\lambda\delta\Gamma^\lambda_{\nu\mu} =\sg\cd_\lambda(g^{\mu\nu}\delta\Gamma^\lambda_{\mu\nu})+g^{\mu\nu}(\cd_\lambda\sg)\delta\Gamma^\lambda_{\mu\nu}-\cd_\lambda\left(\sg g^{\mu\nu}\right)\delta\Gamma^\lambda_{\nu\mu}.
\end{equation}
Now using the change of variables \eqref{eq:con_C} and writing out the derivative in the second term we have

\begin{equation}\label{eq:EHPalVar4}
    \begin{aligned}
        \sg g^{\mu\nu}\cd_\lambda\delta\Gamma^\lambda_{\nu\mu} =&\sg\lcd_\lambda(g^{\mu\nu}\delta\Gamma^\lambda_{\mu\nu}) + \sg C\TI{^\lambda_{\lambda\sigma}}g^{\mu\nu}\delta\Gamma^\sigma_{\mu\nu}\\
        &+ g^{\mu\nu}\left[\lcd_\lambda\sg - C\TI{^\sigma_{\lambda\sigma}}\right]\delta\Gamma^\lambda_{\mu\nu}-\cd_\lambda\left(\sg g^{\mu\nu}\right)\delta\Gamma^\lambda_{\nu\mu}.
    \end{aligned}
\end{equation}
The first term in this expression can now be turned into a surface term using Stokes' theorem \eqref{eq:stokes} and it will vanish. The terms containing the $C\TI{^\sigma_{\mu\nu}}$ tensors can be combined and written using the torsion as $C\TI{^\lambda_{\lambda\sigma}}-C\TI{^\lambda_{\sigma\lambda}}=T\TI{^\lambda_{\lambda\sigma}}$ and finally we have $\lcd_\lambda\sg=0$. With these \eqref{eq:EHPalVar4} simplifies to
\begin{equation}\label{eq:EHPalVar5}
    \sg g^{\mu\nu}\cd_\lambda\delta\Gamma^\lambda_{\nu\mu} = \left[\sg g^{\mu\nu}T\TI{^\lambda_{\lambda\sigma}} - \cd_\sigma(\sg g^{\mu\nu})\right]\delta\Gamma^\sigma_{\mu\nu},
\end{equation}
up to the vanishing surface terms. The second term in \eqref{eq:EHPalVar1} goes similarly giving finally the equations of motion for the connection

\begin{equation}\label{eq:EHconnectionEOM}
    \begin{aligned}
        &\cd_\lambda(\sg g^{\nu\lambda})\delta^\mu_\sigma - \cd_\sigma(\sg g^{\mu\nu}) + \\
        &\sg\left[g^{\mu\nu}T\TI{^\lambda_{\lambda\sigma}}-g^{\nu\lambda}T\TI{^\rho_{\rho\lambda}}\delta^\mu_\sigma+g^{\nu\lambda}T\TI{^\mu_{\sigma\lambda}}\right]=0.
    \end{aligned}
\end{equation} 

At this point we can consider the case in which a symmetric connection is assumed. The obtained equation \eqref{eq:EHconnectionEOM} simplifies considerably since the torsion vanishes and only the symmetric part of the upper indices contributes, we thus have
\begin{equation}
    \cd_\lambda(\sg g^{\lambda(\nu})\delta^{\mu)}_\sigma - \cd_\sigma(\sg g^{\mu\nu}) = 0.
\end{equation}
Taking the trace of this with indices $\mu$ and $\sigma$ we end up with
\begin{equation}
    \cd_\sigma(\sg g^{\mu\nu}) = 0,
\end{equation}
which is equivalent to the metric compatibility condition. Thus if we initially assume a symmetric connection the connection dynamically becomes the Levi--Civita connection. Then the equations of motion for the metric reduces to the Einstein field equations, without needing to introduce additional YGH-boundary terms to deal with the boundaries.

There are several good reasons to impose restrictions to the general connection. For example, the equivalence principle in GR states that the connection coefficients can be locally set to zero; however this is only possible with a completely symmetric connection. Moreover, having a space with torsion means that infinitesimal parallelograms do not close. Additionally, autoparallel curves and the extremal curves of the metric\footnote{Autoparallel curves are curves where vectors are parallel transported with respect to themselves, i.e. straight lines described by the connection. Extremal curves of the metric are curves that tell the shortest or longest path between two points.} are equivalent if and only if the torsion tensor is totally antisymmetric $T_{[\sigma\mu\nu]}=T_{\sigma\mu\nu}$ and the connection metric compatible \cite{Hehl76}. Let us now inspect the most general case.

In the general case it can be shown that the most general solution to \eqref{eq:EHconnectionEOM} is given by \cite{NJ10}
\begin{equation}\label{eq:EHconnectionSolution}
    \Gamma^\sigma_{\mu\nu} = \lcc^\sigma_{\mu\nu} + \delta^\sigma_\nu V_\mu ,
\end{equation}
where $V_\mu$ is an arbitrary vector field. This says that the general solution for the connection differs from the Levi--Civita connection by a projective transformation, which acts as 
\begin{equation}\label{eq:ProjectiveT}
    \Gamma^\sigma_{\mu\nu}\rightarrow\Gamma^\sigma_{\mu\nu}+\delta^\sigma_\nu V_\mu . 
\end{equation}
It seems that to get the standard Einstein field equations one has to make an additional requirement of setting this arbitrary vector field to zero which is equivalent to saying that the trace of the torsion vanishes $T\TI{^\lambda_{\mu\lambda}}=0$.\footnote{This is obvious if one substitutes the solution \eqref{eq:EHconnectionSolution} to the definition of torsion.} There are different types of methods to impose this requirement e.g. where the constrain is introduced in terms of Lagrange multipliers \cite{hehl78}. However, it is easy to see that the Einstein--Hilbert action is in fact invariant under the projective transformation in the case of a symmetric metric. This can be seen by applying the transformation for the Ricci tensor which straightforwardly yields
\begin{equation}
    \sR_{\mu\nu} \rightarrow \sR_{\mu\nu} + 2\pd_{[\mu}V_{\nu]}.
\end{equation}
The projective transformation has the effect of adding an antisymmetric part to the Ricci tensor which vanishes when we take the contraction with symmetric metric, thus the action is invariant under projective transformations. This implies that at least the equations of motion are identical with the general solution \eqref{eq:EHconnectionSolution} and the Levi--Civita connection. However, geodesics described by the general solution could still be different from Levi--Civita geodesics. It can be shown that the geodesics of the general solution are pre-geodesics
\footnote{Pre-geodesic of a connection is a smooth curve on a manifold, which has a reparametrization that is a geodesic of the connection; i.e. the curves describe the same trajectory with different parametrizations.}
of the Levi--Civita ones, and so the geodesics coincide due to reparametrization invariance. The geodesics of the affine connection are solutions of the equation
\begin{equation}
    \dot{x}^\lambda\cd_\lambda \dot{x}^\mu = 0,
\end{equation}
where $x^\mu(\tau)$ is a curve and the dot denotes a derivative with respect to the curve parameter $\tau$. Using the solution \eqref{eq:EHconnectionSolution} this can be written as
\begin{equation}
    \dot{x}^\lambda\lcd_\lambda \dot{x}^\mu + V_\lambda \dot{x}^\lambda \dot{x}^\mu = 0. 
\end{equation}
The pre-geodesics are given by extremizing the arc-length functional 
\begin{equation}
    s(\lambda)=\int_0^\lambda \sqrt{g_{\mu\nu}\dot{x}^\mu\dot{x}^\nu}d\lambda,
\end{equation}
which has the extremum 
\begin{equation}
    \dot{x}^\lambda\cd_\lambda \dot{x}^\mu = ({\ddot{s}}/{\dot{s}})\dot{x}^\mu. 
\end{equation}
Now we can see that with the reparametrization
\begin{equation}
    \frac{dx^\mu(\lambda)}{d\lambda} = \frac{dx^\mu(\tau)}{d\tau}\frac{d\tau}{d\lambda} \quad\text{with}\quad \frac{d\tau}{d\lambda} = \exp \left( -\int_0^\lambda V_\mu \frac{dx^\mu}{d\lambda}d\lambda \right)
\end{equation}
the geodesics of the general connection and the Levi--Civita one coincide. In these two cases, however, parallel transport of a vector will differ by a homothetic transformation (the resulting vectors differ by a scaling factor which depends on the parametrization). It has been argued that the difference between these transports is unobservable and the arbitrary vector field in the general solution is thus also unobservable \cite{Bernal16}. These considerations rely on the assumption that the matter part of the Lagrangian is independent of the connection. This is not necessarily the case if one considers fermionic matter. This brings us to consider matter parts of the Lagrangian in the Palatini formulation.

\subsection{Matter in the Palatini formulation}

Under the assumption that the matter Lagrangian depends only on the metric, the matter fields and their derivatives $\Lagr_M=\Lagr_M(g,\Psi_i)$ the equations of motion for the connection do not change and we recover the Einstein field equations with a source term similarly to \eqref{eq:EFE}. This is straightforward if the matter Lagrangian is naturally independent of the connection, most notably in the case of a canonical scalar field. For a scalar field $\phi$ we have $\cd_\mu\phi = \pd_\mu\phi$, and thus it is naturally not coupled to the connection. In general the matter Lagrangian depends on the metric and the covariant derivative. Thus assuming that the matter Lagrangian is independent of the general connection amounts to using the metric compatible covariant derivative in the matter part of the Lagrangian. This leads to the general connection being an auxillary field not related to the geometrical structure of the spacetime, since the geodesics are now defined through the Levi--Civita connection a priori. Under this assumption the theory ends up being equivalent to a metric theory of spacetime \cite{SL07}.

In the literature the Palatini formulation is often defined by assuming that the matter part is independent of the connection and the general case where this is allowed $\Lagr_M = \Lagr_M(g,\Gamma,\Psi_i)$ is called the metric-affine formulation. In the case where the matter is coupled to the connection, which comes naturally when considering fermionic fields, the situation is more complicated. Since now the equations of motion for the connection will also have a source term usually in the literature called the hypermomentum 
\begin{equation}\label{eq:hypermomentum}
    \Delta\TI{_\sigma^{\mu\nu}} = \frac{-2}{\sg}\frac{\delta S_M}{\delta\Gamma^\sigma_{\mu\nu}}. 
\end{equation}
This formulation also has its problems. In the case of Einstein--Hilbert action taking the trace of the equations of motion \eqref{eq:EHconnectionEOM}, with respect to the indices $\nu$ and $\sigma$, the left side of the equation vanishes identically. Thus if we have a source term this leads to a constraint for the hypermomentum $\Delta\TI{_\sigma^{\mu\sigma}}=0$. This means that if we have a matter Lagrangian that does not satisfy this constraint we get inconsistent field equations. This problem can be traced back to the fact that the Einstein--Hilbert action is invariant under projective transformations, but in general the matter action is not, see \cite[Sec. 4.2]{SL07} and references therein. To get consistent field equations with general matter Lagrangians one is required to: impose some restrictions on the connections with e.g. the Lagrange multiplier method \cite{hehl78}, consider a higher-order action in the curvature or consider a non-symmetric metric i.e. Einstein-Strauss theories \cite[chapter XII]{schrodinger85}. However, we consider only scalar field matter and in this case the Palatini and metric-affine formulations are equivalent since the scalar field Lagrangian is naturally independent of the connection, (see e.g. \cite{Vitagliano10,Hehl76,SL07,SF10} for more details of metric-affine theories).
\\

We have shown that Palatini formulation in the case of the Einstein--Hilbert action, makes the connection dynamically the Levi--Civita connection and the Einstein field equations with a source term are obtained, if matter is not coupled to the connection. To be precise, we cannot state that the metric and Palatini formulations are completely equivalent in the case of Einstein--Hilbert action, since they differ by a surface term and additionally the solution of the general connection includes an extra vector field degree of freedom. It seems to be only a coincidence that the formulations produce the same dynamics with the Einstein--Hilbert action and in general the formulations produce different equations of motion. We will see this explicitly in chapters \ref{sec:HI} and \ref{sec:higherG} where we consider modified gravity sectors. Before that we review the case where scalar field drives inflation.

\chapter{Inflationary paradigm}\label{sec:Inflation}

    GR enabled us to start analyzing the evolution of the universe and together with the SM of particle physics it gives rise to the hot Big Bang model of the universe. The existence of the cosmic microwave background (CMB) and the abundances of light elements are nicely explained by the hot Big Bang model. However, there are problems that the standard hot Big Bang is unable to address.

    Inflation  can explain many features left unaddressed by the hot Big Bang model. These include: why the universe is so homogeneous and isotropic (the horizon problem) and why it is spatially flat. Most notably, it can explain the origin of the primordial perturbations that seed the structure formation of the universe.

    In this chapter we review cosmic inflation keeping emphasis on the generation of primordial perturbations, for more complete introduction see e.g. \cite[part IV]{Lyth09}\cite{Baumann09}.

    \section{Friedmann-Lemaître-Robertson-Walker model}

    Inflation quickly makes the universe spatially homogeneous and isotropic to a good approximation. Imposing these two properties implies that we can foliate our spacetime into spacelike slices that are maximally symmetric, decomposing spacetime into a time and a spacelike three dimensional manifold $\mathbb{R}\times \Sigma_t$. The spatial slices are equipped with an induced maximally symmetric metric. There exist a projection tensor $h_{\mu\nu}$ that projects vectors to the spatial slices, which satisfies
    \begin{equation}
        h_{\mu\nu}u^\nu=0,
    \end{equation}
    where $u^\mu$ is the normal vector of the hypersurface $\Sigma_t$, which can be thought as the time direction and is normalized by $g_{\mu\nu}u^\mu u^\nu=-1$. With these the full metric can be decomposed into a useful form
    \begin{equation}\label{eq:decomposeMetric}
        g_{\mu\nu} = h_{\mu\nu}-u_\mu u_\nu, \quad g^{\mu\nu} = h^{\mu\nu}-u^\mu u^\nu.
    \end{equation}
    The above costraints yields the Friedmann-Lemaître-Robertson-Walker (FLRW) metric \cite[ch. 8.2]{Carroll04}
    \begin{equation}
        ds^2 = -dt^2+a^2(t)\left[ \frac{dr^2}{1-k r^2}+r^2 (d\theta^2 +sin^2\theta d\phi^2) \right],
    \end{equation}
    where $t$ is the time coordinate, $a(t)$ the scale factor describing the expansion of space, $\{r,\theta,\phi\}$ are the spherical spatial coordinates and $k$ describes the curvature of space. Since inflation drives the universe to be spatially flat, we can further simplify and set $k=0$. This enables us to write the FLRW metric in a simple form in Cartesian coordinates
    \begin{equation}\label{eq:FLRWflat}
        ds^2 = -dt^2 + a^2(t)\left[dx^2+dy^2+dz^2\right].
    \end{equation}

    The symmetries of the FLRW model restrict the matter content of the universe to be described by an ideal fluid, for which the stress-energy tensor can be written as
    \begin{equation}\label{eq:perfectFluid}
        T_{\mu\nu} = p g_{\mu\nu} +(\rho + p)u_\mu u_\nu,
    \end{equation}
    where $u_\mu=(-1,0,0,0)$ is the four-velocity of the fluid, $\rho$ and $p$ are the energy density and pressure of the fluid, respectively. With \eqref{eq:FLRWflat} and \eqref{eq:perfectFluid} plugged into the Einstein equation \eqref{eq:EFE} gives the Friedmann equations
    \begin{equation}
        H^2 = \frac{\rho}{3}, \quad \frac{\ddot{a}}{a}=-\frac{1}{6}(\rho+3p),
    \end{equation}
    where $H=\dot{a}/a$ is the Hubble parameter and the dot denotes the time derivative. Thus for accelerated expansion $\ddot{a}>0$ we need $\rho+3p<0$. Next we will see how a scalar field can drive inflation.

    \section{Slow-Roll inflation}

    Adding a canonical scalar field to \eqref{eq:EH}, the action reads 
    \begin{equation}
        S = \int d^4x\sg \left[\half \sR -\half g^{\mu\nu}\pd_\mu\varphi\pd_\nu\varphi-V(\varphi) \right].
    \end{equation}
    Assuming that the scalar field is homogeneous and isotropic, the Friedmann equations turn out to be
    \begin{equation}\label{eq:FriedmanScalar}
        H^2 = \frac{1}{3}\left[ \half \dot{\varphi}^2+V(\varphi)\right],\quad \frac{\ddot{a}}{a}=-\frac{1}{3}\left[\dot{\varphi}^2-V(\varphi)\right],
    \end{equation}
    and the equations of motion for the scalar field become
    \begin{equation}\label{eq:scalarEOM}
        \ddot{\varphi}+3H\dot{\varphi}=-V_{,\varphi}(\varphi).
    \end{equation}
    Thus the condition for inflation is satisfied when $\dot{\varphi}^2<V$. In slow-roll (SR) approximation we assume $\dot{\varphi}^2\ll V(\varphi)$. Additionally we assume that the derivative of the field $\dot{\varphi}$ does not change quickly in one Hubble time. This is needed for a sufficiently long period of inflation to occur. The slow-roll conditions are 
    \begin{align}
        &\dot{\varphi}^2 \ll V(\varphi),\label{eq:sr1}\\
        &|\ddot{\varphi}|\ll|3H\dot{\varphi}|.\label{eq:sr2}
    \end{align}
    With these conditions satisfied the Friedmann equations \eqref{eq:FriedmanScalar} and \eqref{eq:scalarEOM} give the slow-roll equations
    \begin{equation}
        H^2 = \frac{1}{3}V(\varphi), \quad 3H\dot{\varphi}=-V_{,\varphi}(\varphi).
    \end{equation}
    Another useful way of writing the requirement for inflation $\ddot{a}>0$ is
    \begin{equation}
        \epsilon_H = -\frac{\dot{H}}{H^2} < 1.
    \end{equation}
    With the slow-roll equations this is approximately
    \begin{equation}\label{eq:epsV}
        \epsilon_H\approx\epsilon_V = \half \left( \frac{V_{,\varphi}}{V} \right) \overset{SR}{\ll} 1,
    \end{equation}
    where $\epsilon_V$ is the first slow-roll parameter, which describes the quality of the slow-roll approximation. The second slow-roll parameter is
    \begin{equation}\label{eq:etaV}
        |\eta_V| = \left|\frac{V_{,\varphi\varphi}}{V}\right| \overset{SR}{\ll} 1,
    \end{equation}
    describing the change of the first slow-roll parameter. It can be shown that $\epsilon_V\ll 1$ implies the first slow-roll condition \eqref{eq:sr1} and $|\eta_V|\ll 1$ implies \eqref{eq:sr2}. Smallness of the second and higher-order parameters ($\zeta_V = V_{,\varphi\varphi\varphi}V_{,\varphi}/V^2$ etc.) are needed for long period of inflation. It also turs out that slow-roll is an attractor, meaning that wide range of initial conditions rapidly tends to the slow-roll region, see e.g. \cite[sec 18.6.2]{Lyth09}.

    The amount of inflation is described by the number of $e$-folds N, which can be written as
    \begin{equation}\label{eq:efold}
        N(t) = \ln \frac{a(t_{end})}{a(t)} = \int_t^{t_{end}} Hdt= \int^\varphi_{\varphi_{end}}\frac{H}{\dot{\varphi}}d\varphi\overset{SR}{\approx} \int^\varphi_{\varphi_{end}}\frac{d\varphi}{\sqrt{2\epsilon_V}},
    \end{equation}
    where the subscript 'end' denotes the end of inflation. The end of slow-roll, and to a good approximation the end of inflation, is defined as the point where the first or the second slow-roll parameter becomes of the order of unity, which ever occurs first. Thus the field value at the end of inflation can be solved either from $\epsilon(\varphi_{end})=1$ or $|\eta(\varphi_{end})|=1$. This then allows us to solve the field in terms of $N$ from \eqref{eq:efold}. This is useful when evaluating the inflationary observables, which we shall discuss next.

    \section{Perturbations and inflationary observables}\label{sec:Pert}

    Inflation gives us a way of predicting the statistical properties of the primordial perturbations. The perturbations are generated by quantum vacuum fluctuations, which get amplified and stretched to large scales during inflation.
    
    The full calculation, of the statistical properties of the perturbations, uses quantum field theory in curved spacetime. The fields are expanded around the FLRW solution as $\phi \to \phi+\delta \phi$, $g_{\mu\nu}\to g_{\mu\nu}+\delta g_{\mu\nu}$. The important fact is that perturbations in the metric and Palatini formulations coincide in the case of the Einstein--Hilbert action. For a review of this computation see e.g. \cite[Part III]{Baumann09}. We will only give a high level overview of the main points and present the results. 
    
    Since the perturbations are gauge dependent, care has to be taken over fixing a gauge or constructing gauge invariant variables before proceeding with quantization of the perturbations. For gauge transformation in perturbation theory of gravity see e.g. \cite[ch. 5]{Weinberg08}. Then canonical quantization procedure can be applied: promoting the fields into operators and imposing canonical commutation relations. Due to the curved background, complications arise when constructing the vacuum state, see e.g. \cite[ch. 3]{BirrellDavies}. Approximating the background to be close to the static de Sitter space, which is the case during inflation, the vacuum turns out to be the Bunch-Davies vacuum, see e.g. \cite[ch. 5]{BirrellDavies}. The perturbations are then assumed to be initially in this adiabatic Bunch-Davies vacuum state. A crucial fact that enables inflationary theories to make predictions about the form of the perturbations is that initially adiabatic perturbations are conserved after Hubble crossing in slow-roll; by Hubble crossing we mean the instance when the wave numbers of the Fourier modes of the perturbations become larger than the Hubble parameter $k=aH$, see e.g. \cite[ch. 5.4]{Weinberg08}\cite{Lyth04}. This is also the point where the quantum fluctuations get transformed to look like a classical field, see e.g. \cite{Albrecht92}, which allows us to treat the perturbations as classical. In first order perturbation theory scalar, vector and tensor degrees of freedom do not mix; this is sometimes called the decomposition theorem, see e.g. \cite[ch. 5]{Weinberg08}. This allows us to separately calculate the spectrum for scalar and tensor perturbations. It turns out that vector perturbations quickly decay away and can be ignored here.
    
    The gauge invariant quantity describing the scalar perturbations is the so called Sasaki-Mukhanov variable $\nu$, which in the spatially flat gauge is $\nu = a\delta\phi$. The Fourier mode functions $\nu_k$ of this variable obey the equations of motion \cite{Mukhanov88}
    \begin{equation}
        \nu^{\prime\prime}_k+\left(k^2-\frac{z^{\prime\prime}}{z}\right)\nu_k=0,
    \end{equation}
    where $z\equiv a\dot{\phi}/H$, the prime denotes a derivative with respect to the conformal time $d\tau = dt/a$ and $k$ is the comoving wave number of the Fourier mode. This is solved in the slow-roll approximation with the Bunch-Davies vacuum state as the initial condition. These solutions are used to compute the correlation functions that describe the statistical properties of the perturbations. To the first order the perturbations are Gaussian and completely described by the power spectrum (two point correlation function)
    \begin{equation}
        \mathcal{P}_{\delta\phi}=\frac{1}{a^2}\mathcal{P}_\nu = \frac{k^3}{2\pi^2 a^2}\lvert u_k\rvert^2 = \left. \left(\frac{H}{2\pi}\right)^2\right|_{k=aH}.
    \end{equation}
    The comoving curvature perturbation $\mathcal{R}$ turns out to be an useful quantity that can be related to the spectrum measured from the CMB \footnote{Sometimes in the literature another useful quantity is used, the curvature perturbation on uniform-density hypersurfaces $\zeta$. It can be shown that these quantities are equal outside the horizon and during slow-roll.}. It is proportional to the Sasaki-Mukhanov variable $\mathcal{R} = \nu/z$ \cite{Baumann09}. Thus the power spectum for the curvature perturbation is
    \begin{equation}
        \mathcal{P}_{\mathcal{R}} = \frac{1}{z^2}\mathcal{P}_\nu = \left. \frac{1}{2\epsilon}\left(\frac{H}{2\pi}\right)^2 \right|_{k=aH},
    \end{equation}
    where $\epsilon$ is the first slow-roll parameter. It is customary to parametrize the spectrum in a nearly scale invariant form
    \begin{equation}
        \mathcal{P}_{\mathcal{R}}=A_s\left(\frac{k}{k_*}\right)^{n_s-1},
    \end{equation}
    where $k_*$ is some reference scale, $A_s$ is the amplitude and $n_s$ is the spectral tilt, which describes the scale dependence of the spectrum. Thus when $n_s = 1$ the spectrum is scale invariant. The amplitude in the slow-roll aproximation can be written as
    \begin{equation}
        A_s = \frac{1}{24\pi^2}\frac{V}{\epsilon_V}
    \end{equation}
    and to first order the spectral tilt becomes 
    \begin{equation}
        n_s-1\equiv \frac{d \ln\mathcal{P}_{\mathcal{R}}(k)}{d \ln k} \overset{SR}{\approx} -6\epsilon_V+2\eta_V.
    \end{equation}

    Inflation also predicts the form of tensor perturbations. Similar calculation as for the scalar perturbations gives the spectrum for the tensor perturbations
    \begin{equation}
        \mathcal{P}_t = 2\left(\frac{2}{a}\right)^2 \mathcal{P}_\nu = \left. 8\left(\frac{H}{2\pi}\right)^2 \right|_{k=aH}.
    \end{equation}
    This can also parametrized in a nearly scale invariant form as
    \begin{equation}
        \mathcal{P}_t = A_t\left(\frac{k}{k_*}\right)^{n_t}.
    \end{equation}
    The amplitude in the slow-roll approximation reads
    \begin{equation}
        A_t = \frac{2}{3\pi^2}V.
    \end{equation}
    The spectral tilt for this to the first order becomes 
    \begin{equation}
        n_t \equiv \frac{d \ln\mathcal{P}_t}{d \ln k} \overset{SR}{\approx} -2\epsilon_V.
    \end{equation}
    The amplitude of tensor spectrum is not amplified by the slow-roll parameters, in contrast to the spectrum of the scalar perturbations, and is thus harder to observe. The important quantity that the current observations can set bounds on is the tensor-to-scalar ratio, which in the slow-roll becomes
    \begin{equation}
        r \equiv \frac{\mathcal{P}_\mathcal{R}}{\mathcal{P}_t} \overset{SR}{\approx} 16\epsilon_V.
    \end{equation}
    We have seen that when a single scalar field drives inflation inflationary predictions can be computed from the inflaton potential alone.

    \chapter{Higgs inflation}\label{sec:HI}

    The Standard Model of particle physics (SM) contains one scalar field: the Higgs boson. For this reason it is tempting to identify the Higgs field as the inflaton. However, the Higgs self-coupling ($\lambda\simeq 0.129$) is too large to generate an amplitude for the primordial scalar perturbations that is consistent with current observations. At tree level the calculated amplitude is too large. This remains the case when taking into account quantum corrections to the potential \cite{Isidori07,Hamada13,Fairbairn14}. 
    
    Adding a non-minimal coupling between the Higgs and gravity can generate a spectrum compatible with current observations. The non-minimal coupling term is well motivated. It is the only new dimension four operator when the Einstein--Hilbert and SM actions are combined. Furthermore, even if the term is absent in the action at classical level this kind of non-minimal coupling term is generated when a scalar field is quantized in a curved background; more specifically when renormalizing the energy-momentum tensor \cite[ch. 3, 6]{BirrellDavies}.
    
    The feature that only one new parameter is introduced makes this one of the most interesting models to study. Furthermore, the couplings between the Higgs and the rest of the SM are known experimentally, which is advantageous since the reheating period after inflation can be calculated in detail. This is important since the length of the reheating period affects the inflationary observables. In contrast, many models introduce new unknown couplings between the inflaton and the SM.

    In Higgs inflation the SM Higgs boson non-minimally coupled to gravity is identified as the inflaton field. The action reads
    \begin{equation}
        S = \int d^4x\sqrt{-g}\left[ (\tfrac{1}{2} M^2 + \xi \mathcal{H}^\dagger \mathcal{H})\sR -(D_\mu \mathcal{H})^\dagger(D^\mu \mathcal{H}) + m_h \mathcal{H}^\dagger \mathcal{H}-\lambda (\mathcal{H}^\dagger \mathcal{H})^2 + \mathcal{L}_{SM}  \right],
    \end{equation}
    where $M$ is a mass parameter, $\xi$ is positive dimensionless non-minimal coupling parameter to be fixed by observations, $\mathcal{H}$ is the Higgs doublet, $D_\mu$ is the gauge covariant derivative and $\Lagr_{SM}$ is the rest of the standard model. During inflation the rest of the SM acts as spectator fields and do not affect the evolution of the universe. We can write the Higgs doublet in the unitary gauge as $\mathcal{H} = 1/\sqrt{2}\left(\begin{smallmatrix} 0\\v+h\end{smallmatrix}\right)$, where $v=246\quad\text{GeV}$ is the vacuum expectation value (vev) of the Higgs field $h$. We can thus write the relevant part of the action as
    \begin{equation}\label{eq:toy}
        S = \int d^4x\sg \left[ \half(M^2+\xi h^2)\sR - \half g^{\mu\nu}\pd_\mu h \pd_\nu h - V(h) \right],
    \end{equation}
    where $V(h)$ is the symmetry breaking potential
    \begin{equation}\label{eq:Higgspot}
        V(h) = \frac{\lambda}{4}(h^2-v^2)^2.
    \end{equation}
    The non-minimal coupling to gravity changes the strength of gravitational interaction by effectively changing Newton's gravitational constant, or equivalently the square of Planck mass
    \begin{equation}
        M_p^2 = M^2+\xi h^2.
    \end{equation}
    The Planck mass we measure today in the Jordan frame (the definition of the Jordan frame is discussed in the next section) would then be $M_p^2 = M^2+\xi v^2$. With a non-minimal coupling in the range $1\ll \xi \ll M^2/v^2$ we can approximate $M\simeq M_p$, which is what we will do from now on. Additionally we again set $M_p$ to unity. Differences occur only in the large field regime $h \gg M_p/\xi$, which turns out to be the inflationary regime.

    The non-minimal coupling makes the computation of inflationary observables more involved. However, we can make use of conformal transformations to remove the non-minimal coupling. Then determining the spectrum of the perturbations will follow the standard computation outlined in section \ref{sec:Pert}. Let us next discuss these transformations more closely.
    
    \section{Conformal transformations}\label{sec:conformal}

    Conformal transformation is a local rescaling of the metric
    \begin{equation}
        g_{\mu\nu}\to \accentset{\sim}{g}_{\mu\nu}=\Omega^2(x)g_{\mu\nu},
    \end{equation}
    where $\Omega^2(x)$ is a positive smooth function called the conformal factor. This is thus a map between two pseudo-Riemannian spaces. The causal structure is unchanged. Space(Time)like vectors remain space(time)like and null vectors remain as null vectors. Thus null geodesics are left invariant under conformal transformations and light cones are unchanged. Angles between vectors are also unchanged. What conformal transformations do change is the geometry. For example the time-like geodesics generically differ.

    To be able to perform this transformation to the action \eqref{eq:toy} we need to compute how the relevant quantities transform. First of all for the inverse metric we have $\accentset{\sim}{g}_{\lambda\nu}\accentset{\sim}{g}^{\mu\lambda} = \delta^\mu_\nu$ and thus $\accentset{\sim}{g}^{\mu\nu}=\Omega^{-2}(x)g^{\mu\nu}$. The metric determinant transforms as 
    \begin{equation}
        \det (\accentset{\sim}{g}_{\mu\nu}) = \det( \Omega^2 g_{\mu\nu})=\Omega^{2d} \det( g_{\mu\nu}),
    \end{equation}
    where $d$ is the number of spacetime dimensions. So we have altogether
    \begin{equation}
        \begin{aligned}
            g_{\mu\nu} &= \Omega^{-2}\accentset{\sim}{g}_{\mu\nu}\\
            g^{\mu\nu} &= \Omega^{2}\accentset{\sim}{g}^{\mu\nu}\\
            \sg &= \Omega^{-d}\sqrt{-\accentset{\sim}{g}}.
        \end{aligned}
    \end{equation}
    This is all that is needed for the Palatini formulation. For the metric formulation we also need to know how the Ricci scalar transforms under the conformal transformations. First transforming the Levi--Civita connection we find
    \begin{equation}
        \lcc^\sigma_{\mu\nu} = \accentset{\sim}{\lcc}^\sigma_{\mu\nu} - \left[ 2\delta^\sigma_{(\nu} \ln\Omega_{,\mu)}-\accentset{\sim}{g}_{\mu\nu}\accentset{\sim}{g}^{\sigma\lambda}\ln\Omega_{,\lambda} \right].
    \end{equation}
    We can plug this into the definition of the Riemann tensor \eqref{eq:defRie} and see how it transforms
    \begin{equation}
        \begin{aligned}
            {\lcR} \TI{^\sigma_{\mu\lambda\nu}}=&\accentset{\sim}{\lcR}\TI{^\sigma_{\mu\lambda\nu}} + 2\accentset{\sim}{g}_{\mu[\nu}\accentset{\sim}{g}^{\sigma\rho}\accentset{\sim}{\cd}_{\lambda]}\accentset{\sim}{\cd}_\rho \ln\Omega + 2 \delta^\sigma_{[\lambda}\accentset{\sim}{\cd}_{\nu]}\accentset{\sim}{\cd}_\mu\ln\Omega + 2\delta^\sigma_{[\lambda}\ln\Omega_{,\nu]}\ln\Omega_{,\mu}\\
            &2\accentset{\sim}{g}_{\mu[\lambda}\delta^\sigma_{\nu]}\accentset{\sim}{g}^{\rho\tau}\ln\Omega_{,\rho}\ln\Omega_{,\tau} + \accentset{\sim}{g}_{\mu[\nu}\accentset{\sim}{g}^{\rho\sigma}\ln\Omega_{,\lambda]}\ln\Omega_{,\rho} ,
        \end{aligned}
    \end{equation}
    where quantities with a tilde are constructed from the metric $\accentset{\sim}{g}_{\mu\nu}$. Taking the contraction we get how the Ricci tensor transforms
    \begin{equation}
        \begin{aligned}
            \lcR_{\mu\nu} = &\accentset{\sim}{\lcR}_{\mu\nu} + \accentset{\sim}{g}_{\mu\nu}\accentset{\sim}{\Box}\ln\Omega + (d-2)\accentset{\sim}{\cd}_\nu\accentset{\sim}{\cd}_\mu \ln\Omega\\
            & + (d-2)\ln\Omega_{,\mu}\ln\Omega_{,\nu} + (2-d)\accentset{\sim}{g}_{\mu\nu} \accentset{\sim}{g}^{\rho\tau}\ln\Omega_{,\rho}\ln\Omega_{,\tau}.
        \end{aligned}
    \end{equation}
    Contracting with the metric $g^{\mu\nu}$ we get how the Ricci scalar transforms
    \begin{equation}
        \lcR = \Omega^2\left[ \accentset{\sim}{\lcR} + 2(d-1)\accentset{\sim}{\Box}\ln\Omega + (d-1)(2-d)\accentset{\sim}{g}^{\rho\tau}\ln\Omega_{,\rho}\ln\Omega_{,\tau}\right].
    \end{equation}
    It is also straightforward to derive how higher-order curvature terms (discussed later in chapter \ref{sec:higherG}) transform, but with the higher-order terms we focus on the Palatini formulation. Considering higher-order terms in metric formulation is out of the scope of this thesis. In contrast in the Palatini formulation the situation is simple, since the Riemann tensor does not depend on the metric and is thus unchanged. All the different invariants that are second order in curvature, that we will encounter in chapter \ref{sec:higherG}, are invariant under conformal transformations (in 4 dimensions). For example 
    \begin{equation}
        \begin{aligned}
            \sqrt{-g}g^{\mu\lambda}g^{\nu\sigma}\sR_{\mu\nu}\sR_{\lambda\sigma} &= \left[{-\Omega^{-2\times 4}\accentset{\sim}{g}}\right]^{1/2}\Omega^4 \accentset{\sim}{g}^{\mu\lambda}\accentset{\sim}{g}^{\nu\sigma}\sR_{\mu\nu}\sR_{\lambda\sigma}\\
             &= \sqrt{-\accentset{\sim}{g}}\accentset{\sim}{g}^{\mu\lambda}\accentset{\sim}{g}^{\nu\sigma}\sR_{\mu\nu}\sR_{\lambda\sigma}.
        \end{aligned}
    \end{equation}
    
    The usefulness of these transformations is due to the fact that two different-looking actions related by a conformal transformation are physically equivalent at the classical level\footnote{When quantizing the theory this point is less clear.}. By performing a conformal transformation we thus find a different representation for the theory. Customarily in the literature these different representations that are related by conformal transformations are called different frames. There are two specific important frames that have their own names. The so called Jordan frame, which is characterized by the fact that there is a non-minimal coupling between a scalar field and the curvature term. The other important frame is the so called Einstein frame, where matter is minimally coupled to the curvature. The physical equivalence of these two frames is clear from the fact that the transformation is required to be invertible and thus it gives a one-to-one correspondence between these two frames. The transformation can also be thought simply as a local change of units \cite{Flanagan04}. When comparing quantities between two frames attention has to be paid to appropriately account for the conformal factor.
    
    The inflationary observables, both the curvature and tensor perturbations, can be shown to be invariant under conformal transformations \cite{Tsujikawa04,Postma14}.
    The inflationary observables can thus be calculated in any frame. In the case of Higgs inflation inflationary observables have also been computed in both frames and the results agree, see \cite{Rubio18} and references therein.

    Transforming the action \eqref{eq:toy} to the Einstein frame is different in the metric and Palatini formulation due to the distinct metric dependence of the actions. Let's now perform this transformation in both cases. 
    %
    \subsection{Einstein frame in metric formulation}
    Let's first consider the metric formulation, where $\sR(\Gamma) = \lcR(g)$ in \eqref{eq:toy}. From the action \eqref{eq:toy} we see that we can remove the non-minimal coupling by choosing the conformal factor to be 
    \begin{equation}\label{eq:conformalfactor}
        \Omega^2(x) = 1 + {\xi h^2}.
    \end{equation}
    With the previously obtained transformations the action \eqref{eq:toy} becomes
    \begin{equation}
        \begin{aligned}
            S = \int d^4x \sqrt{-\accentset{\sim}{g}}\Omega^{-4} &\left\{\vphantom{\half}\right. \frac{1}{2}\Omega^2 \Omega^2 \left[ \accentset{\sim}{\lcR} + 3\accentset{\sim}{\Box}\ln\Omega^2 - \frac{3}{2}\accentset{\sim}{g}^{\rho\lambda}\ln\Omega^2_{,\rho}\ln\Omega^2_{,\lambda}  \right]\\
            & - \half \Omega^2 \accentset{\sim}{g}^{\mu\nu}\pd_\mu h\pd_\nu h - V(h) \left.\vphantom{\half}\right\},
        \end{aligned}
    \end{equation}
    where the second term inside the square brackets is a total derivative. This term can be converted to a surface term using the Stokes' theorem \eqref{eq:stokes} and thus vanishes. Writing 
    \begin{equation}
        \ln\Omega^2_{,\lambda} = \frac{(\Omega^2)_{,\lambda}}{\Omega^2} = \frac{\Omega^2_{,h}}{\Omega^2}\pd_\lambda h,
    \end{equation}
    the action simplifies to
    \begin{equation}
        S = \int d^4x \sqrt{-\accentset{\sim}{g}} \left\{\vphantom{\half}\right. \frac{1}{2}\accentset{\sim}{\lcR} - \half \frac{ \tfrac{3}{2}  (\Omega^2)_{,h}^2+\Omega^2}{\Omega^4} \accentset{\sim}{g}^{\mu\nu}\pd_\mu h\pd_\nu h  \\
             - \Omega^{-4}V(h) \left.\vphantom{\half}\right\}.
    \end{equation}
    We see that we have been able to remove the non-minimal coupling between gravity and the scalar field. The cost for this is a modified matter sector with a non-canonical kinetic term. The kinetic term can be brought into a canonical form by a field redefinition
    \begin{equation}\label{eq:CTredef}
        \frac{d\chi}{d h}=\sqrt{\frac{ \tfrac{3}{2} (\Omega^2)_{,h}^2+\Omega^2}{\Omega^4}}.
    \end{equation}
    With this the action becomes
    \begin{equation}\label{eq:actionSTCT}
        S=\int d^4x \sqrt{-\accentset{\sim}{g}} \left\{ \frac{1}{2}\accentset{\sim}{\lcR}-\half\accentset{\sim}{g}^{\mu\nu}\pd_\mu\chi\pd_\nu\chi - U(\chi) \right\},
    \end{equation}
    where we defined the modified potential
    \begin{equation}\label{eq:EPot}
        U(\chi) \equiv \frac{V(h(\chi))}{\Omega^4(h(\chi))}.
    \end{equation}
    Note that the $h$ field has to be solved in terms of the new field from the redefinition \eqref{eq:CTredef}, we will come to this later.

    \subsection{Einstein frame in the Palatini formulation}
    In the Palatini formulation this is considerably easier since the Ricci tensor is unchanged. We can again get rid of the non-minimal coupling by the same conformal factor \eqref{eq:conformalfactor}. The action \eqref{eq:toy} becomes
    \begin{equation}\label{eq:actionCTpal}
    S=\int d^4x \sqrt{-\accentset{\sim}{g}}\left[ \frac{1}{2}\accentset{\sim}{g}^{\mu\nu}\sR_{\mu\nu} -\half\frac{1}{\Omega^2}\accentset{\sim}{g}^{\mu\nu}\pd_\mu h\pd_\nu h-\frac{V(h)}{\Omega^4} \right].
    \end{equation}
    The non-canonical kinetic term can be again brought in to canonical form by a field redefinition, which now simply reads 
    \begin{equation}\label{eq:CTredef}
        \frac{d\chi}{d h}=\sqrt{\frac{1}{\Omega^2}}.
    \end{equation}
    So the action can be written
    \begin{equation}
        S=\int d^4x \sqrt{-\accentset{\sim}{g}} \left\{ \frac{1}{2}\accentset{\sim}{\sR}-\half\accentset{\sim}{g}^{\mu\nu}\pd_\mu\chi\pd_\nu\chi - U(\chi) \right\},
    \end{equation}
    with
    \begin{equation}\label{eq:ModPot}
        U(\chi) \equiv \frac{V(h(\chi))}{\Omega^4(h(\chi))}.
    \end{equation}
    We see that the metric and Palatini formulations differ by the form of the potential coming from the fact that the field redefinitions differ. We will see this difference more explicitly in the following section.
\\

    \section{Tree-level inflationary observables}\label{sec:TreeObsHI}

    The tree-level inflationary observables are now straightforward to compute in the Einstein frame. As we saw in chapter \ref{sec:Inflation}, these can be computed from the potential, which are different between the metric and Palatini formulations. To get the potential we need to solve the new field variable $\chi$ from the field redefinition. Let's again start with the metric formulation.

    \subsection{Metric formulation}
    It is possible to solve $\chi$ from the field redefinition  \eqref{eq:CTredef} analytically \cite{Rubio18}. However, the form of the solution is cumbersome and not particularly illuminating. We can solve it approximately by assuming that $\xi\gg 1$, which we shall later see is the case in Higgs inflation. Additionally in the inflationary regime the field value is large $h \gg 1/\xi$. The field redefinition \eqref{eq:CTredef} is then approximately \footnote{When this was first done in the literature \cite{Bezrukov07} the approximation $h \gg 1/\sqrt{\xi}$ was made which results in a different form for the potential; the overall approximate form is different but the asymptotic behavior is the same. The first slow-roll parameter turns out to be the same but second and higher order slow-roll parameters are different.}
    \begin{equation}
        \begin{aligned}
            \frac{d\chi}{d h}&= \sqrt{\frac{1+(1+6\xi)\xi h^2}{\left(1+\xi h^2\right)^2}}\\
            &\simeq \frac{\sqrt{6}\xi h}{1+ \xi h^2}.
        \end{aligned}
    \end{equation}
    Which when solved gives
    \begin{equation}\label{eq:h2solved}
        h^2 = \frac{1}{\xi}\left(e^{\frac{2}{\sqrt{6}}\chi}-1\right).
    \end{equation}
    Solving for $\chi$ gives 
    \begin{equation}
        \chi = \sqrt{\frac{3}{2}} \ln\left(1+{\xi}h^2\right) = \sqrt{\frac{3}{2}}\ln\Omega(h)^2.
    \end{equation}
    We also see that in the small field regime $h \ll 1/\xi$ we simply have $h\simeq\chi$. So the field redefinition \eqref{eq:CTredef} is approximately solved in different asymptotic regions as
    \begin{equation}
        \chi \simeq \begin{cases}
            h \quad&,\text{when}\quad h \ll 1/\xi\\
            \sqrt{\frac{3}{2}}\ln\Omega(h)^2\quad &,\text{when}\quad h \gg 1/\xi
        \end{cases}
    \end{equation}
    In the large field region the Higgs vev is much smaller than the field value and we can approximate the Jordan frame potential \eqref{eq:Higgspot} as
    \begin{equation}\label{eq:JordanPotAprx}
        V(h) \simeq \frac{\lambda}{4}h^4.
    \end{equation}
    With these, in the large field regime, the potential \eqref{eq:EPot} becomes
    \begin{equation}\label{eq:ApproxPot}
        U(\chi) = \frac{\lambda }{4\xi^2}\left( 1-e^{-\frac{2}{\sqrt{6}}\chi} \right)^2.
    \end{equation}
    From this we see that the potential is asymptotically flat: $U(\chi) \to \lambda /(4\xi^2)$ when $\chi \to \infty$. Thus there is a plateau in the large field regime where slow-roll inflation can occur. This also implies that the action has an asymptotic shift symmetry, i.e. it is symmetric under the shift of the inflaton field ($\chi\to\chi+\text{constant}$) in the limit $\chi\to\infty$.\footnote{In the Jordan frame this asymptotic symmetry manifests itself as a asymptotic scale symmetry: $x^\mu\to\alpha x^\mu$ and $h(x) \to \alpha^{-1}h(\alpha x)$ where $\alpha$ is a constant.} This approximate symmetry turns out to be important when considering quantum corrections \cite{Bezrukov10}.

    We can now calculate the inflationary observables in the slow-roll approach from the potential. The first slow-roll parameters are
    \begin{align}
            \epsilon_U &= \frac{1}{2}\left(\frac{U_{,\chi}}{U}\right)^2 = \frac{4}{3}\left(e^{\frac{2}{\sqrt{6}}\chi}-1\right)^{-2}=\frac{4}{3 \xi^2 h^4}\label{eq:SReps}\\
            \eta_U &= \frac{U_{,\chi\chi}}{U} = \frac{4}{3}\left(2-e^{\frac{2}{\sqrt{6}}\chi}\right)\left(e^{\frac{2}{\sqrt{6}}\chi}-1\right)^{-2}=\frac{4}{3\xi^2 h^4}\left(1-{\xi h^2}\right).\label{eq:SReta}
    \end{align}
    We can get the field value at the end of inflation defined by the usual condition, which in this case is $\epsilon(\chi_{end})=1$. Equating \eqref{eq:SReps} to unity and solving for $\chi$ gives the field value at the end of inflation
    \begin{equation}\label{eq:inf_end_metric}
        \chi_{end} = \sqrt{\frac{3}{2}}\ln\left(1+\sqrt{\frac{3}{4}}\right).
    \end{equation}
    As described in the section \ref{sec:Pert} the observables are to be evaluated at a horizon crossing. We can do this by solving the field in terms of the number of e-folds $\chi(N_{*})$, where $N_{*}$ is the number of e-folds when some reference scale $k_{*}$ crosses the horizon $k_{*}=a_* H_*$. From \eqref{eq:efold} and \eqref{eq:SReps} we have
    \begin{equation}
        N_* = \int_{\chi_{end}}^{\chi_*}\frac{d\chi}{\sqrt{2\epsilon}} = \int_{\chi_{end}}^{\chi_*}\frac{U}{ U_{,\chi}}d\chi = \int_{\chi_{end}}^{\chi_*} \frac{1}{2}\sqrt{\frac{3}{2}}\left(e^{\frac{2}{\sqrt{6}}\chi}-1\right)d\chi,
    \end{equation} 
    giving us
        \begin{equation}
            \begin{aligned}
                N_* &= \frac{3}{4}\left[\exp\left(\frac{2}{\sqrt{6}}\chi_*\right)-\exp\left(\frac{2}{\sqrt{6}}\chi_{end}\right)\right] - \frac{\sqrt{6}}{4}\left[\chi_*-\chi_{end}\right].\\
            \end{aligned}
        \end{equation}
    Plugging in the field value at the end of inflation \eqref{eq:inf_end_metric} we get 
    \begin{equation}
            N_* = \frac{1}{4}\left[3\exp\left(\frac{2}{\sqrt{6}}\chi_*\right)-\sqrt{6}\chi_*\right] - \frac{3}{4}\left[ \ln\left(1+\sqrt{\frac{3}{4}}\right)-\left(1+\sqrt\frac{3}{4}\right) \right],
    \end{equation}
    where the numerical factor is of the order of unity. Assuming that the exponential term dominates we finally get
    \begin{equation}
        N_* \simeq \frac{3}{4}\exp\left({\frac{2}{\sqrt{6}}\chi_*}\right).
    \end{equation}
    It is also possible solve $\chi_*$ in terms of $N_*$ and $\chi_{end}$ analytically. The approximated form is easier to work with, which gives
    \begin{equation}
        \chi_*(N_*) = \sqrt{\frac{3}{2}}\ln\left(\frac{4}{3}N_*\right).
    \end{equation}
    With this we can write the inflationary observables in terms of the number of e-folds. The slow-roll parameters are
    \begin{align}
        \epsilon(N_*) &= \frac{12}{(4 N_*-3)^2}\simeq\frac{3}{4N_*^2}\\
        \eta(N_*) &= 8\frac{3-2N_*}{(4N_*-3)^2}\simeq -\frac{1}{N_*},
    \end{align}
    where we approximated further by assuming $4N_*\gg 1$. The amplitude of the scalar perturbations is
    \begin{equation}\label{eq:amp}
        A_s = \frac{1}{24\pi^2}\frac{\lambda(4N_*-3)^4}{ 768\xi^2 N_*^2}\simeq \frac{1}{24\pi^2}\frac{\lambda N_*^2}{ 3\xi^2}.
    \end{equation}
    The spectral tilt of scalar perturbations is
    \begin{equation}
        n_s = 1 -\frac{8(3+4N_*)}{(4N_*-3)^2}\simeq 1-\frac{2}{N_*}.
    \end{equation}
    The spectral tilt of tensor perturbations is
    \begin{equation}
        n_t = \frac{-24}{(4N_*-3)^2}\simeq-\frac{3}{2N_*^2}.
    \end{equation}
    The tensor-to-scalar ratio is
    \begin{equation}
        r = \frac{192}{(4N_*-3)^2}\simeq\frac{12}{N_*^2}.
    \end{equation}
    Analytical expressions, with out the approximations made here, for these observables can be found in \cite{Rubio18}. The precise number of e-folds depends on the heating period after inflation, which advantageously can be calculated in detail since the SM parameters are experimentally known. These computations have been carried out in the literature \cite{GarciaBellido08,Repond16,Bezrukov08}. We use the simplifying assumption that the reheating period was instant, which gives us an estimate of $N_*\simeq 55$ when the usual reference scale $k_*=0.05\quad\text{Mpc}^{-1}$ passes the horizon. We can now find an estimate for the non-minimal coupling by matching the amplitude \eqref{eq:amp} with the observed value \cite{Akrami18}
    \begin{equation}\label{eq:observedAmp}
        \ln\left(10^{10} A_s\right)\simeq 3.094\pm 0.034.
    \end{equation}
    This gives a relation between the non-minimal coupling and the Higgs self-coupling. With the previous estimate for the number of e-folds we have
    \begin{equation}
        \xi \simeq 800 N_*\sqrt{\lambda}\simeq 44000\sqrt{\lambda}
    \end{equation}
    We see that the non-minimal coupling is required to be quite large since $\lambda\simeq 0.129$, but still $\xi \ll M_p^2/v^2\sim 10^{32}$ as we previously required. It is also smaller than the upper bound $\xi<2.6\times10^{15}$ set by considering observations about the Higgs at the LHC \cite{Atkins12}.

    We can now also get numerical values for the important observables
    \begin{equation}
        n_s \simeq 0.964, \quad r \simeq 0.00396,
    \end{equation}
    which are in good agreement with the current observed values \cite{Akrami18}.
    
    However, this model is not without problems. The non-minimal coupling to gravity makes this model non-renormalizable in the Jordan frame. Written in terms of canonically normalized variables $g_{\mu\nu} = \eta_{\mu\nu} + M_p^{-1} h_{\mu\nu}$ in the weak field regime $\phi^2 R$ results in a dimension $5$ term and is thus non-renormalizable. In the Einstein frame the non-renormalizability comes from the nonlinear form of the interactions in the potential. Non-renormalizability implies that there is some cutoff scale where perturbation theory breaks down, and in the effective field theory point of view one would have to add higher dimension terms in the scalar field to the action suppressed by the cutoff. These higher-order terms could spoil the inflationary plateau. Hence, it is important to estimate this cutoff in order to figure out whether the theory is consistent in the inflationary region or does it need to be modified. One way of estimating this cutoff is to see where tree-level unitarity is violated, see e.g. \cite[ch. 24]{Schwartz14}. Estimating the cutoff by expanding only the metric around a background leads to a cutoff which is right around the inflationary scale, and thus seems to lead to inconsistencies \cite{Barbon09}. However, during inflation both metric and the scalar are naturally expanded around a background. In \cite{Bezrukov10} the cutoff was estimated by expanding both the metric and the scalar field around a background, in both Einstein and Jordan frames. This results in a field dependent cutoff, which is parametrically much larger than the Hubble scale during inflation and is equal to the Planck scale at small field values. These considerations suggest that our tree-level analysis is consistent, however this is still an open question.

    \subsection{Palatini formulation}

    In the Palatini formulation the analysis follows the same steps. The field redefinition is now simpler
    \begin{equation}
        \frac{d\chi}{dh} = \frac{1}{\sqrt{\Omega^2}} = \frac{1}{\sqrt{1 + {\xi h^2}}},
    \end{equation}
    which is easily integrated to give
    \begin{align}
        \chi(h) &= \frac{1}{\sqrt{\xi}}\sinh^{-1}\left({\sqrt{\xi}h}\right),\label{eq:ChiToH}\\
        h(\chi) &= \frac{1}{\sqrt{\xi}}\sinh\left({\sqrt{\xi}\chi}\right).
    \end{align}
    Again the Higgs vev is negligible in the large field regime and we take the Jordan frame potential to be \eqref{eq:JordanPotAprx}. Plugging \eqref{eq:ChiToH} into the potential \eqref{eq:ModPot} gives
    \begin{equation}
        U(\chi) = \frac{\lambda }{4\xi^2}\tanh^4\left({\sqrt{\xi}\chi}\right),
    \end{equation}
    which is asymptotically flat: $\chi\to\infty$, $U(\chi)\to\lambda /(4\xi^2)$ and has a plateau in the large field regime where slow-roll can occur. The slow-roll parameters can be approximated in the large field regime $\sqrt{\xi}\chi\gg 1$ as
    \begin{align}
        \epsilon &\simeq 128\xi \exp\left({{-4\sqrt{\xi}\chi}}\right),\\
        \eta &\simeq -32\xi\exp\left({-2\sqrt{\xi}\chi}\right),
    \end{align}
    The second slow-roll parameter becomes of order of unity before the first parameter. Thus we get the field value at the end of inflation from $|\eta(\chi_{end})|=1$. This gives
    \begin{equation}\label{eq:palEndInf}
        \chi_{end} = \frac{1}{2\sqrt{\xi}}\ln\left(32\xi\right) .
    \end{equation}
    Next we need to solve the field in terms of e-folds. Starting with
    \begin{equation}
        \begin{aligned}
            N_* &= \int_{\chi_{end}}^{\chi_*}\frac{U}{ U_{,\chi}}d\chi =  \int_{\chi_{end}}^{\chi_*} \frac{1}{8\sqrt{\xi}}\sinh\left({2\sqrt{\xi}\chi}\right)d\chi\\ 
            &= \left.\frac{1}{16\xi}\cosh\left({2\sqrt{\xi}\chi}\right)\right|^{\chi_*}_{\chi_{end}},
        \end{aligned}
    \end{equation}
    which gives
    \begin{equation}
        N_* = \frac{1}{32\xi}\left[ \exp\left({2\sqrt{\xi}\chi_*}\right)-\exp\left({2\sqrt{\xi}\chi_{end}}\right) \right].
    \end{equation}
    Plugging in the field value at the end of inflation \eqref{eq:palEndInf} into the previous expression
    \begin{equation}
        N_* = -1 + \frac{1}{32\xi}\exp\left({2\sqrt{\xi}\chi_*}\right),
    \end{equation}
    then with the assumption that $N_* \gg 1$ we get\footnote{In the ref. \cite{Bauer08} the assumption $\chi_*\gg \chi_{end}$ is made, which in fact result in the same approximate form \eqref{eq:Naprx} and thus does not change the results that we obtain. However, when looking at \eqref{eq:palEndInf} and \eqref{eq:chiN} we see that $\chi_*,\chi_{end}$ are of the same order. In the reference \cite{Bauer08} the end of inflation is defined to be at $\epsilon=1$, which leads $\chi_*$ to be about twice the value of $\chi_{end}$.}
    \begin{equation}\label{eq:Naprx}
        N_* \simeq \frac{1}{32\xi}\exp\left({2\sqrt{\xi}\chi}\right).
    \end{equation}
    Allowing us to solve the field in terms of the e-folds
    \begin{equation}\label{eq:chiN}
        \chi(N_*) = \frac{1}{2\sqrt{\xi}}\ln\left(32\xi N_*\right).
    \end{equation}
    The slow-roll parameters in terms of e-folds become
    \begin{align}
        \epsilon(N_*)\simeq\frac{1}{8N_*^2 \xi}\\
        \eta(N_*)\simeq -\frac{1}{N_*}.
    \end{align}
    Which allows us to write the observables in terms of the e-folds
    \begin{align}
        A_{s} &\simeq \frac{1}{24\pi^2}\frac{2\lambda N_*^2}{\xi},\\
    %
        n_s &\simeq 1 -\frac{2}{N_*},\\
    %
        n_t &\simeq -\frac{1}{4N_*^2\xi},\\
    %
        r &\simeq \frac{2}{N_*^2\xi},
    \end{align}
    where we again approximated $N_*\gg 1$. The reheating stage after inflation is also different in the Palatini case and it has been studied in \cite{RubioEemeli19}. There it was found that the reheating stage is almost instant, which increases the required number of e-folds. The number of e-folds is estimated to be $N_*\simeq 50$. The number of e-folds depends on the value of the non-minimal coupling $\xi$. With the range $\xi=10^6 \dots 10^9$ we have $N_* \simeq 51 \dots 50$. We can now match the observed amplitude of the scalar perturbations \eqref{eq:observedAmp}. This now requires a much larger non-minimal coupling
    \begin{equation}\label{eq:XiPal}
        \xi \simeq 3.8 \times 10^{6}N_*^2 \lambda \simeq 3.9\times 10^9 \lambda,
    \end{equation}
    which is many orders of magnitude larger than in the metric formulation. However, it is still much smaller than what we required $\xi\ll M_p^2/v^2\sim 10^{32}$ and smaller than the LHC upper bound $\xi<2.6\times10^{15}$.

    We see that the tilt of the scalar spectrum has the same form to the first order as in the metric case, but with the smaller estimated number of e-folds has a slightly smaller value. The main difference comes in the tensor-to-scalar ratio, which is suppressed by the non-minimal coupling. Future planned experiments could thus rule out Palatini-Higgs inflation if significant tensor-to-scalar ratio is observed. With our estimates the important observables have the values
    \begin{equation}
        n_s \simeq 0.960, \quad r\simeq 2.1 \times 10^{-13}\lambda^{-1},
    \end{equation}
    which are still compatible with observations.\\

    The unitarity violation is less of a problem in Palatini formulation. A straightforward estimation of the cutoff turns out to be parametrically larger than the scale during inflation, for details see \cite{Bauer11} and references therein.

    Now that we have reviewed the Higgs inflation in Palatini formulation we are ready to consider higher-order curvature terms and Higgs inflation. We shall begin by reviewing some basic properties of higher-order gravity.

    \chapter{Higher-order gravity} \label{sec:higherG}

GR has been well tested on different scales, since its formulation, and it has been able to predict new phenomena. Despite the unquestioned success of the theory there are good reasons to consider modified theories of gravity. From a theoretical standpoint the lack of theory of quantum gravity suggests that this is not the full story. Moreover, the non-renormalizability of GR (see e.g.\cite[Ch. 22.4]{Schwartz14}), further suggest that GR is a effective field theory of some underlying more complete theory.

There are many ways to start modifying GR (for a exhaustive survey in view of cosmology see e.g. \cite{Clifton11}). Here we are interested in higher-order gravity theories, where terms that are higher-order in the curvature are added to the action. Existence of these terms can be motivated in several ways: considering relativistic QFT in classical curved background generates terms to the action which are quadratic in the curvature tensor \cite{DeWitt62}\cite{BirrellDavies}. Additionally, Weinberg and Deser suggested that adding quadratic curvature terms to the action makes the theory renormalizable\footnote{Renormalization is out of the scope of this thesis. For an introduction see e.g. \cite[Part III]{Schwartz14}\cite[Part II]{PnS}}, which was later proven \cite{Stelle76}. Also these kind of terms are generated in string theories as low energy effective actions at classical level, see \cite{Boulware85,Deser87}\cite[Ch. 7]{Tong09}.

However, adding these kinds of terms introduces new kinds of problems. In the case of metric formulation the equations of motion contain higher-order derivative terms of the form, which by the Ostrogradsky theorem \cite[Ch 2.5]{Woodard09}, lead to instabilities. By instabilities we mean that there exists solutions for which the energies are not bounded from below and can thus have arbitrary negative energies\footnote{It has been suggested that with higher-order terms all the unstable solutions are unphysical and all the physical solutions remain stable in semiclassical limit \cite{Parker93}.}. These negative energy modes are often called ghosts in the literature. These lead into serious difficulties when trying to interpret the theory.

In the Palatini formulation appearance of ghosts is more unclear. Naively one might think that since in Palatini formulation the resulting equations of motion are only second order, and thus there are no Ostrogradsky instabilities, there would not be any ghosts. However, by analyzing the degrees of freedom of some quadratic curvature terms it has been shown that ghost modes can still appear \cite{BJ19}. The question, what is the most general form of the Lagrangian where ghosts do not appear, requires further research.

In this chapter we will consider the addition of quadratic curvature invariants in both the metric and Palatini formulations and derive the equations of motion for these terms.

\section{$f(R)$ gravity} \label{sec:fR}

To illustrate the differences between the metric and Palatini formulations arising from adding higher-orders curvature terms to the action, we first consider adding higher powers of the Ricci scalar. We can do this in general by considering a general function of the Ricci scalar $f(R)$. This is called $f(R)$ gravity, for a review see e.g. \cite{SF10}. It is well known that in both the metric and Palatini formulations these kind of theories are equivalent to GR coupled to a scalar field known as scalar-tensor theories\footnote{Note that this equivalence does not apply in metric-affine formulation.} \cite{Sotiriou06}. It is also known that with the condition $f''(R)\geq 0$ (prime denotes the derivative with respect to the Ricci scalar $R$) these theories are ghost free \cite[Ch. V, and references therein]{SF10}.

Let's see how the differences between the metric and Palatini formulation appear.

\subsection{In the metric formulation}
The action under consideration is 
\begin{equation}\label{eq:FR}
    S = \int d^4x \sg f(\lcR).
\end{equation}
The steps for taking the variation of this action are very similar as those we skimmed over in section \ref{sec:EH-metric-var}. To deal with the boundaries without imposing extra constraints YGH-term needs to be added similarly as in section \ref{sec:YGH-term}
\footnote{The exact form of the term differs from the one in the section \ref{sec:YGH-term}. The term needed is 
\begin{equation*}\label{}
    S_{YGH} = 2\int_{\pd\Sigma}d^3x\epsilon\sqrt{\gamma}f'(\lcR)\lcd_\mu n^\mu,
\end{equation*}
for a detailed derivation see \cite{GCT10}.}. 
The differences occur when performing partial integrations, since this now introduces second order derivative terms of $f'(\lcR)$. The resulting equations of motion are 
\begin{equation}\label{eq:fR_EOM}
    f'(\lcR)\lcR_{\mu\nu}-\half f(\lcR)g_{\mu\nu} - \lcd_\mu \lcd_\nu f'(\lcR) + g_{\mu\nu}\lBox f'(\lcR) = 0,
\end{equation}
where $\lBox \equiv \lcd_\mu\lcd^\mu$ is the d'Alembertian operator. These are fourth order partial differential equations of the metric, if $f'(\lcR)$ is dynamical; and reduce to second order equations, if $f'(\lcR)$ is a constant.

The difference between the metric and Palatini formulation is more easily seen when moving to the Einstein frame. If we assume that $f''(\lcR)\neq 0$ we can introduce a new scalar field $\phi$ and write the action as
\begin{equation}\label{}
    S = \int d^4x\sg \left[f(\phi)+f'(\phi)(\lcR-\phi)\right].
\end{equation}
We see that the variation with respect to the field $\phi$ leads to a constraint $\lcR=\phi$ and thus we recover equivalent equations of motion to the original action. Now with a simple change of variables $\Phi(\phi) = f'(\phi)$, and with the requirement that the inverse $\phi(\Phi)$ exists, we can write $V(\Phi) = \phi(\Phi)-f(\phi(\Phi))$, which brings the action to the form
\begin{equation}\label{eq:fRmetricJ}
    S = \int d^4x \sg \left[\Phi \lcR - V(\Phi)\right].
\end{equation}
There now appears a non-minimal coupling between the scalar and gravity. We can now remove this non-minimal coupling by performing a conformal transformation, as we have done before. With
\begin{equation}\label{eq:fRconf}
    g_{\mu\nu} \to \Phi g_{\mu\nu},
\end{equation}
and using relations from section \ref{sec:conformal}, the action can be written in Einstein frame 
\begin{equation}\label{eq:fRmetricE}
    S=\int d^4x\sg\left[ \lcR -\frac{3}{2\Phi^2}\pd^\mu\Phi\pd_\mu\Phi - \Phi^{-2}V(\Phi)\right],
\end{equation}
where a surface term was dropped. Thus we see that in the metric formulation there appears a new gravitational degree of freedom.
\subsection{In the Palatini formulation}

Next we will see how the equations of motion turn out in the Palatini formulation. Variation of the action \eqref{eq:FR} with respect to the metric gives
\begin{equation}
    f'(\sR)\sR_{(\mu\nu)} - \half f(\sR)g_{\mu\nu} = 0.
\end{equation}
Variation of the action with respect to the connection follows the same steps as we did in section \ref{sec:EH-Palatini}. Thus the equations of motion for the connection are
\begin{equation} \label{eq:fR-pal}
    \begin{aligned}
        &\cd_\lambda(\sg f'(\sR) g^{\nu\lambda})\delta^\mu_\sigma - \cd_\lambda(\sg f'(\sR) g^{\mu\nu})\\ &+ \sg f'(\sR)\left[g^{\mu\nu}T\TI{^\lambda_{\lambda\sigma}}-g^{\nu\lambda}T\TI{^\rho_{\rho\lambda}}\delta^\mu_\sigma + g^{\nu\lambda}T\TI{^\mu_{\sigma\lambda}}\right] = 0.
    \end{aligned}
\end{equation}

Let us now move to the Einstein frame. Often in the literature the torsion is assumed to vanish; this is not, however, necessary for finding the Einstein frame. We start by noticing that the equations of motion for the connection \eqref{eq:fR-pal} with the change of variables $q^{\mu\nu} = f'(\sR)g^{\mu\nu}$ are the same as for the Einstein--Hilbert action. We know the general solution for this equation, namely \eqref{eq:EHconnectionSolution}. Thus \eqref{eq:fR-pal} has the general solution
\begin{equation}
    \Gamma^\sigma_{\mu\nu} = \lcc^\sigma_{\mu\nu}(q) + \delta^\sigma_\nu V_\mu,
\end{equation}
where $\lcc^\sigma_{\mu\nu}(q)$ is the Levi--Civita connection for the metric $q^{\mu\nu}$ and $V_\mu$ is an arbitrary vector field. The action is invariant under the projective transformations \eqref{eq:ProjectiveT}, and thus the arbitrary vector field does not come into the action. We can thus say that $\sR_{\mu\nu}$ is equivalent to the Ricci scalar with Levi--Civita connection of the metric $q^{\mu\nu}$, $\sR_{\mu\nu} = \lcR_{\mu\nu}(q)$. So the action becomes
\begin{equation}
    S = \int d^4x \sg \left[\Phi g^{\mu\nu}\lcR_{\mu\nu}(q) - V(\Phi)\right].
\end{equation}
We can now perform a conformal transform to get rid of the $q^{\mu\nu}$ dependence, since $g^{\mu\nu} = \Phi^{-1}q^{\mu\nu}$. Using the relations from section \ref{sec:conformal} and performing the conformal transformation the action can be written as
\begin{equation}\label{eq:fRPalJ}
    S = \int d^4x \sg \left[\Phi \lcR + \frac{3}{2\Phi}\pd_\mu\Phi\pd^\mu\Phi-V(\Phi)\right],  
\end{equation}
where a surface term was ignored. There is still a non-minimal coupling present, so we do a another conformal transformation to the Einstein frame. With \eqref{eq:fRconf} the action becomes 
\begin{equation}
    S = \int d^4x \sg \left[ \lcR + \frac{3}{2\Phi^2}\pd_\mu\Phi\pd^\mu\Phi- \frac{3}{2\Phi^2}\pd_\mu\Phi\pd^\mu\Phi-\Phi^{-2}V(\Phi) \right].
\end{equation}
The kinetic term vanishes and the action becomes
\begin{equation}\label{eq:fRPalE}
    S = \int d^4x \sg \left[ \lcR -\Phi^{-2}V(\Phi) \right] + S_m\left[\Phi g_{\mu\nu},\Psi_i\right],
\end{equation}
where we wrote explicitly the matter part of the action to illustrate that in the Palatini formulation there is no new degrees of freedom, since the scalar $\Phi$ does not have a kinetic term, but the relation between existing degrees of freedom is altered.

From the Einstein frame action \eqref{eq:fRmetricE} in the metric formulation we observe there is one extra degree of freedom. In contrast from the Einstein frame action \eqref{eq:fRPalE} in the Palatini formalism we see that there are no new degrees of freedom\footnote{Comparing the Jordan frame actions, the metric case \eqref{eq:fRmetricJ} is of the form of Brans-Dicke action with $\omega_0=0$ and the Palatini case $\omega_0 = -3/2$}.

\section{Quadratic gravity in metric formulation}
Let us next investigate the other possible quadratic curvature terms, starting with the metric formulation.

Due to the high number of symmetries of the Riemann tensor, in the metric formulation, there are only a few independent terms that can be written down, namely: $\lcR^2$, $\lcR_{\mu\nu}\lcR^{\mu\nu}$ and $\lcR_{\mu\nu\sigma\lambda}\lcR^{\mu\nu\sigma\lambda}$ (there is one more possible term $\epsilon^{\mu\nu\sigma\rho}\lcR\TI{^{\alpha\beta}_{\mu\nu}}\lcR_{\alpha\beta\sigma\rho}$, but this turns out to be a total derivative).

Equations of motion for the $\lcR^2$ term can be easily read from the $f(\lcR)$ gravity equations of motion \eqref{eq:fR_EOM}. Derivation of the equations of motion for the last two terms is straightforward but somewhat lengthy. We give some needed tricks and just state the results. Keep in mind that when raising and lowering the indices of the variation: $\delta g_{\mu\nu} = -g_{\mu\alpha}g_{\nu\beta}\delta g^{\alpha\beta}$. We make use of the Bianchi identities $\lcR_{\mu\nu[\sigma\lambda;\rho]}=0$ (more specifically the contracted Bianchi identity) and the definition of how the commutator of two covariant derivatives $[\lcd_\mu,\lcd_\nu]$ act on a general tensor (see e.g. \cite[p. 123]{Carroll04}).
With these in mind and ignoring surface terms the equations of motion turn out to be\footnote{This derivation gives rise to boundary terms that can be canceled by adding YHG-terms to the action as done before; these terms are ignored here.}
    \begin{flalign}\label{}
            &\frac{\delta(\lcR^2)}{\sg \delta g^{\mu\nu}} = -\half \lcR^2 g_{\mu\nu}+2\lcR\lcR_{\mu\nu} - 2\lcd_\nu \lcd_\mu \lcR + 2g_{\mu\nu}\lBox \lcR=0,&&
    \end{flalign}
    \begin{flalign}\label{}
        \begin{aligned}
            \frac{\delta(\lcR_{\alpha\beta}\lcR^{\alpha\beta})}{\sg\delta g^{\mu\nu}}=&-\half g_{\mu\nu}\lcR_{\alpha\beta}\lcR^{\alpha\beta} + \lBox \lcR_{\mu\nu} + \half g_{\mu\nu}\lBox \lcR\\ &- \lcd_\nu \lcd_\mu \lcR + 2\lcR^{\lambda\sigma}R_{\mu\lambda\nu\sigma}=0,\\
        \end{aligned}&&
    \end{flalign}
    \begin{flalign}\label{}
    \begin{aligned}
        \frac{\delta(\lcR_{\alpha\beta\sigma\lambda}\lcR^{\alpha\beta\sigma\lambda})}{\sg\delta g^{\mu\nu}} =& -\half g_{\mu\nu}\lcR_{\alpha\beta\sigma\lambda}\lcR^{\alpha\beta\sigma\lambda}+2\lcR\TI{_{\mu}^{\alpha\beta\sigma}}\lcR_{\nu\alpha\beta\sigma}+4\lBox \lcR_{\mu\nu}\\ 
        &- 2\lcd_\nu\lcd_\mu \lcR + 4\lcR^{\lambda\sigma}\lcR_{\lambda\mu\sigma\nu}-4\lcR_{\mu\lambda}\lcR\TI{_\nu^\lambda}=0.
    \end{aligned}&&
    \end{flalign}
These are again seen to be fourth order differential equations. Notice that the sum of these terms with the right coefficients leads to the higher-order derivative terms to cancel out. Namely the combination 
\begin{equation}\label{eq:GB1}
    \Lagr_{GB} = \lcR^2 - 4\lcR_{\mu\nu}R^{\mu\nu} + \lcR_{\mu\nu\sigma\lambda}\lcR^{\mu\nu\sigma\lambda},
\end{equation}
called the Gauss-Bonnet term. The variation of this term can be written in a compact form using the Weyl tensor \eqref{eq:weyl}. This is identically zero and it is called the Bach-Lanczos identity \cite{Lanczos38}
\begin{equation}\label{eq:bach-lanczos}
    \frac{\delta\Lagr_{GB}}{\delta g^{\mu\nu}}=W\TI{_\mu^{\alpha\beta\sigma}}W_{\nu\alpha\beta\sigma} - \frac{1}{4}g_{\mu\nu}W_{\alpha\beta\sigma\lambda}W^{\alpha\beta\sigma\lambda}=0,
\end{equation}
for a simple proof see e.g. \cite[Appendix]{Alvarez16}. The combination \eqref{eq:GB1} does not contribute to the equations of motion in 4 dimensions. The term \eqref{eq:GB1} is a topological invariant (Euler characteristic) of the manifold. The generalized Gauss-Bonnet theorem in 4 dimensions states that
\begin{equation}
    \frac{1}{32\pi^2}\int_M d^4x \sg \left[\lcR^2 - 4\lcR_{\mu\nu}\lcR^{\mu\nu} + \lcR_{\mu\nu\sigma\lambda}\lcR^{\mu\nu\sigma\lambda}\right] = \chi(M),
\end{equation}
where $\chi(M)$ is a topological invariant of the manifold M. This implies that only two of the three different higher-order terms are independent. We can solve out one of the three terms (choosing the most complicated one including the Riemann tensor) and write the general action quadratic in the curvature as (ignoring the constant term)
\begin{equation}\label{}
    S = \int d^4x\sg \left(\alpha \lcR_{\mu\nu}\lcR^{\mu\nu}+\beta \lcR^2+\gamma \lcR\right),
\end{equation}
where $\alpha,\beta$ and $\gamma$ are constants. This is exactly the form of the Lagrangian that has been proven to be renormalizable in all orders of perturbation theory \cite{Stelle76}. This, however, comes with the cost of massive spin-2 ghost \cite{Salvio18}\cite{Stelle78}. Next we will consider Lagrangians that do not lead to ghosts in the metric formulation and see that this requirement restricts the form of the Lagrangian considerably.

\subsection{Lovelock gravity} \label{sec:Lovelock}
Lovelock constructed the most general Lagrangian in metric formulation, that leads to second order field equations, and do not contain ghosts, in arbitrary number of spacetime dimension \cite{Lovelock69,lovelock71}. The restrictions that Lovelock imposed were: the general Einstein tensor can only depend on the metric and its first two derivatives and it should be divergence free. It turns out that in four spacetime dimensions the only such tensors are the metric and the Einstein tensor of GR. Thus the Lovelock action is the Einstein--Hilbert action; this is also called the Lovelock's theorem \cite[Theorem 5]{Lovelock69}. The Lovelock Lagrangian in $d$ dimensions is
\begin{equation}\label{eq:lovelock}
    \Lagr^d_{Lovelock} = \sum_{n=0}^{n<d/2}c_n\Lagr_n,
\end{equation}
where $c_n$ are constants with dimensionality $2n-d$ and
\begin{equation}
    \Lagr_n = \frac{1}{2^n}\delta^{\alpha_1\beta_1...\alpha_{n}\beta_{n}}_{\mu_1\nu_1...\mu_{n}\nu_n}\prod_{m=1}^{n}\lcR\TI{^{\mu_m\nu_m}_{\alpha_m\beta_m}},
\end{equation}
where $\delta^{...}_{...}$ is the generalized Kronecker delta \eqref{eq:genKronecker}. Notice that if $n > d/2$ then $\Lagr_n$ will be zero, since there are more indices in the Kronecker delta than there are dimensions. The first few terms of $\Lagr_n$ are
\begin{subequations}
    \begin{align}
            \Lagr_0 &= \Lambda,\\
            \Lagr_1 &= R,\\
            \Lagr_2 &= \lcR^2 - 4\lcR_{\mu\nu}R^{\mu\nu} + \lcR_{\mu\nu\sigma\lambda}\lcR^{\mu\nu\sigma\lambda}.\label{eq:GaussBonnet}
    \end{align}
    \end{subequations}
The term with $n=d/2$ turns out to be a topological invariant also in general and this justifies why it is not included in the sum \eqref{eq:lovelock}.

It turns out that for Lovelock gravity every solution in the metric formulation is also a solution in the Palatini formulation, but not the other way around. Not every solution of the Palatini formulation is a solution of the metric formulation. Thus in the case of Lovelock gravity metric formulation is contained in the Palatini formulation. There are also other forms of higher-order curvature Lagrangians that have this relation between these two formulations \cite{BJB08}. 
\\
    
\section{Quadratic gravity in the Palatini formulation}\label{sec:QGPAL}

Let us now consider adding quadratic curvature invariants to the action in the Palatini formulation. In contrast to the metric case there are more different invariants that we can construct out of two Riemann tensors. We can think of the most general Lagrangian second order in the Riemann tensor, that does not violate parity (i.e. ignoring terms with the Levi--Civita symbol) as \footnote{There might be some redundancy in these terms, if there exists some equivalent more general version of the Gauss-Bonnet theorem in metric affine space.}
\begin{equation}
    S = \int d^4x \sg \left(\text{all contractions of }\sR\TI{^\sigma_{\rho\mu\nu}}\sR\TI{^\alpha_{\beta\tau\delta}}\right).
\end{equation}
There are $16$ possible independent contractions due to the Riemann tensor having only one symmetry. These contractions can be written in terms of the different possible Ricci tensors found in \eqref{eq:Ricci}--\eqref{eq:antisymRicci}. With these the action can be written as
\begin{equation}\label{eq:PAL-RR-lag}
    \begin{aligned}
        S = \int &d^4x \sg  \left[ 
        \alpha   \sR^2 +
        \beta_1  \sR_{\mu\nu}\sR^{\mu\nu} + 
        \beta_2  \sR_{\mu\nu}\sR^{\nu\mu} + 
        \beta_3  \sR_{\mu\nu}\sRc^{\mu\nu} + 
        \beta_4  \sR_{\mu\nu}\sRc^{\nu\mu} +\right. \\
       &\beta_5  \sRc_{\mu\nu}\sRc^{\mu\nu} + 
        \beta_6  \sRc_{\mu\nu}\sRc^{\nu\mu} + 
        \beta_7  \sRc_{\mu\nu}\sR^{\prime\mu\nu} +
        \beta_8  \sR^{\prime}_{\mu\nu}\sR^{\prime\mu\nu} +
        \beta_9  \sR_{\mu\nu}\sR^{\prime\mu\nu} + \\
       &\gamma_1 \sR\TI{_{\mu\nu\sigma\lambda}}\sR\TI{^{\mu\nu\sigma\lambda}} +
        \gamma_2 \sR\TI{_{\mu\nu\sigma\lambda}}\sR\TI{^{\mu\sigma\nu\lambda}} +
        \gamma_3 \sR\TI{_{\mu\nu\sigma\lambda}}\sR\TI{^{\nu\mu\sigma\lambda}} +
        \gamma_4 \sR\TI{_{\mu\nu\sigma\lambda}}\sR\TI{^{\nu\sigma\mu\lambda}} + \\
&\left. \gamma_5 \sR\TI{_{\mu\nu\sigma\lambda}}\sR\TI{^{\sigma\nu\mu\lambda}} +
        \gamma_6 \sR\TI{_{\mu\nu\sigma\lambda}}\sR\TI{^{\sigma\lambda\mu\nu}} \right].
        \end{aligned}
\end{equation}
Derivation of the equations of motion for these terms follows the similar procedure as that we did for the Einstein--Hilbert action in the section \ref{sec:EH-Palatini}. We will derive the equations of motion for one of these terms for clarity and present the rest of the equations of motion for each term in the appendix \ref{app:EOM_Palatini}.

We will consider the term $\sR_{\mu\nu}\sR^{\mu\nu}$. Starting with the metric variation
\begin{equation}
    \delta\int d^4x\sg \sR_{\mu\nu}\sR^{\mu\nu} = \int d^4x \left[  \sR_{\mu\nu}\sR^{\mu\nu}\delta\sg + \sg \delta ( \sR_{\mu\nu}\sR^{\mu\nu}) \right].
\end{equation}
Taking the variation of $\sR_{\mu\nu}\sR^{\mu\nu}$ gives
\begin{equation}\label{eq:RRterm}
    \begin{aligned}
        \delta\left(\sR_{\mu\nu}\sR^{\mu\nu}\right) &= \delta\left(g^{\tau\mu}g^{\lambda\nu}\sR\TI{^\sigma_{\mu\sigma\nu}}\sR\TI{^\rho_{\tau\rho\lambda}}\right)\\
        &=(\delta g^{\tau\mu} )g^{\lambda\nu}\sR\TI{_{\mu\nu}}\sR\TI{_{\tau\lambda}} + g^{\tau\mu}(\delta g^{\lambda\nu})\sR\TI{_{\mu\nu}}\sR\TI{_{\tau\lambda}}\\
        &= \left( \sR\TI{_{\mu}^\lambda}\sR\TI{_{\nu\lambda}} + \sR\TI{^{\lambda}_\mu}\sR\TI{_{\lambda\nu}} \right)\delta g^{\mu\nu}
    \end{aligned}
\end{equation}
With the previous identity and \eqref{eq:var_sg} the equation of motion turn out to be 
\begin{equation}
    \sR\TI{_{\mu}^\lambda}\sR\TI{_{\nu\lambda}} + \sR\TI{^{\lambda}_\mu}\sR\TI{_{\lambda\nu}} - \half g_{\mu\nu} \sR_{\alpha\beta}\sR^{\alpha\beta} = 0.
\end{equation}
Next let's look at the variation with respect to the connection. First of all we have
\begin{equation}
    \begin{aligned}
        \int d^4x\sg \delta(\sR_{\mu\nu}\sR^{\mu\nu}) &= \int d^4x\sg g^{\tau\mu}g^{\lambda\nu}\left[\sR_{\tau\lambda}\delta\sR_{\mu\nu}+\sR_{\mu\nu}\delta\sR_{\tau\lambda}\right]\\
        &=\int d^4x \sg 2\sR^{\mu\nu}\delta\sR_{\mu\nu}.
    \end{aligned}
\end{equation}
Then using the Palatini identity \eqref{eq:palatiniID} this becomes
\begin{equation}\label{eq:RRvarP}
    \int d^4x\sg 2\sR\TI{^\mu^\nu}\left[ \cd_\sigma\delta\Gamma^\sigma_{\nu\mu} - \cd_\nu\delta\Gamma^\sigma_{\sigma\mu} + T\TI{^\sigma_{\lambda\nu}}\delta\Gamma^\lambda_{\sigma\mu} \right].
\end{equation}
Like previously the first and the second term inside the brackets have to be simplified further. This computation is very similar to the one we performed in section \ref{sec:EH-Palatini}. First we expand out the first term in \eqref{eq:RRvarP}
\begin{equation}
    \begin{aligned}
        2\sR\TI{^\mu^\nu} \cd_\sigma\delta\Gamma^\sigma_{\nu\mu}&=2\sg\cd_\sigma(\sR\TI{^\mu^\nu}\delta\Gamma^\sigma_{\nu\mu}) +2 \sR\TI{^\mu^\nu}\cd_\sigma\sg\delta\Gamma^\sigma_{\nu\mu}\\ 
        &-2\cd_\sigma(\sg \sR\TI{^\mu^\nu})\delta\Gamma^\sigma_{\nu\mu},
    \end{aligned}
\end{equation}
using the change of variables $\Gamma^\sigma_{\mu\nu} = \lcc^\sigma_{\mu\nu} + C\TI{^\sigma_{\mu\nu}}$ this becomes
\begin{equation}
    \begin{aligned}
        2\sR\TI{^\mu^\nu} \cd_\sigma\delta\Gamma^\sigma_{\nu\mu}&=2\sg\lcd_\sigma(\sR\TI{^\mu^\nu}\delta\Gamma^\sigma_{\nu\mu}) + 2\sg C\TI{^\lambda_{\lambda\sigma}} (\sR\TI{^\mu^\nu}\delta\Gamma^\sigma_{\nu\mu}) +\\
        &2\sR\TI{^\mu^\nu}\left[ \lcd_\nu\sg - C\TI{^\lambda_{\sigma\lambda}} \right]\delta\Gamma^\sigma_{\nu\mu} - 2\cd_\sigma(\sg \sR\TI{^\mu^\nu})\delta\Gamma^\sigma_{\nu\mu}.
    \end{aligned}
\end{equation}
The first term in this expression is now a total divergence and can be turned into a surface term by Stokes' theorem \eqref{eq:stokes} and it will vanish. The terms containing the $C\TI{^\sigma_{\mu\nu}}$ tensors can be again written in terms of the torsion tensor. With these this simplifies to 
\begin{equation}
    2\sR\TI{^\mu^\nu} \cd_\sigma\delta\Gamma^\sigma_{\nu\mu}=\left[ 2\sg \sR\TI{^\mu^\nu} T\TI{^\lambda_{\lambda\sigma}} - 2\cd_\sigma(\sg \sR\TI{^\mu^\nu})\right]\delta\Gamma^\sigma_{\nu\mu},
\end{equation}
up to the surface terms. The second term in \eqref{eq:RRvarP} goes similarly and the equations of motion turn out to be
\begin{equation}\label{eq:EOM_connection_RR}
    \begin{aligned}       
        &2\cd_\lambda(\sg \sR\TI{^\mu^\lambda})\delta^\nu_\sigma - 2\cd_\sigma(\sg \sR\TI{^\mu^\nu})\\
        &+2\sg\left[\sR\TI{^\mu^\nu}T\TI{^\lambda_{\lambda\sigma}} - \sR\TI{^\mu^\rho}T\TI{^\lambda_{\lambda\rho}} + \sR\TI{^\mu^\lambda}T\TI{^\nu_{\sigma\lambda}}\right] = 0.
    \end{aligned}
\end{equation}
The non-metricity nature of these equations becomes more transparent if we write it in terms of the non-metricity tensor, defined by
\begin{equation}\label{eq:defNonMet}
    Q_{\sigma\mu\nu} \equiv \cd_\sigma g_{\mu\nu}.
\end{equation}
Note that it is symmetric in the last two indices. It is important to also note that $\cd_\sigma g^{\mu\nu} \neq Q\TI{_\sigma^{\mu\nu}}$. A quick computation will give us the right relation,
\begin{equation}
    \cd_\sigma g^{\mu\nu} = \cd_\sigma( g^{\mu\alpha}g^{\nu\beta}g_{\alpha\beta} ) = Q\TI{_\sigma^{\mu\nu}} + 2\cd_\sigma g^{\mu\nu},    
\end{equation}
and thus
\begin{equation}
    \cd_\sigma g^{\mu\nu} = - Q\TI{_\sigma^{\mu\nu}}.
\end{equation}
With the definition of non-metricity and torsion it is straightforward to verify that a general connection can be written as \footnote{Plug in the definition of non-metricity \eqref{eq:defNonMet} into \eqref{eq:ConDecompose} and write out the covariant derivative of the metric. With a few steps of algebra the left and right hand side of the equation will match.}
\begin{equation}\label{eq:ConDecompose}
    \Gamma^\sigma_{\mu\nu} = \lcc^\sigma_{\mu\nu}+\half\left[ -Q\TI{_\mu^\sigma_\nu}+Q\TI{^\sigma_{\nu\mu}}-Q\TI{_{\nu\mu}^\sigma} \right]+\half\left[T\TI{^\sigma_{\mu\nu}}-T\TI{_{\mu\nu}^\sigma}+T\TI{_\nu^\sigma_\mu}\right].
\end{equation}
Using this we can write $\cd_\mu \sg$ in terms of the non-metricity, since
\begin{equation}
    \cd_\mu\sg = \lcd_\mu\sg-C\TI{^\lambda_{\mu\lambda}}\sg.
\end{equation}
Taking the trace of \eqref{eq:ConDecompose} we get 
\begin{equation}
    \begin{aligned}
        C\TI{^\lambda_{\mu\lambda}} &= \half\left[ -Q\TI{_\mu^\lambda_\lambda}+Q\TI{^\lambda_{\lambda\mu}}-Q\TI{_{\lambda\mu}^\lambda} \right]+\half\left[T\TI{^\lambda_{\mu\lambda}}-T\TI{_{\mu\lambda}^\lambda}+T\TI{_\lambda^\lambda_\mu}\right]\\
        &= -\half Q\TI{_\mu^\lambda_\lambda},
    \end{aligned}
\end{equation}
where the torsion tensors vanish due to them being antisymmetric in the last two indices and two of the non-metricity terms vanish due to the symmetry of the last two indices. Thus we have
\begin{equation}
    \cd_\mu\sg = \half Q\TI{_\mu^\lambda_\lambda}\sg.
\end{equation}
Using this we can write the equations of motion \eqref{eq:EOM_connection_RR} as 
\begin{equation}
    \begin{aligned}
        &2\sR^{\mu\lambda}\left[\half Q\TI{_\lambda^\rho_\rho}\delta^\nu_\sigma-T\TI{^\rho_\rho_\lambda}\delta^\nu_\sigma+T\TI{^\nu_\sigma_\lambda}\right]+2\sR^{\mu\nu}\left[-\half Q\TI{_\sigma^\rho_\rho}+T\TI{^\lambda_\lambda_\sigma}\right]\\
        &-Q\TI{_\lambda^\mu^\lambda}\delta^\nu_\sigma+Q\TI{_\sigma^\mu^\nu}+2\delta^\nu_\sigma\cd_\lambda\sR^{\mu\lambda}-2\cd_\sigma\sR^{\mu\nu} = 0.
    \end{aligned}
\end{equation} 
The non-metricity and torsion dependence of the last two terms containing covariant derivatives of the Ricci tensor can also be written out using \eqref{eq:ConDecompose}. This form for the equation is useful when trying to find solutions in the general case. However, as we can see even for this one quadratic term $\sR_{\mu\nu}\sR^{\mu\nu}$ the equations of motion turn out to be very complicated. Furthermore, the fact that some of the extra degrees of freedom coming from the higher-order terms turn out to be ghost degrees of freedom complicate the picture \cite{Alvarez18,Alvarez-Gaume15,BJ19}. To be able to say something about higher-order terms and Higgs inflation we will restrict our action such that no new degrees of freedom appear, and thus avoid any inconsistencies. In the next chapter we shall see how this is accomplished.

    \chapter[Higher-order gravity and Higgs inflation]{Higher-order gravity and Higgs\\ inflation}\label{sec:QGHI}
    
    In this chapter we present a new result. We analyze Higgs inflation with higher-order curvature terms in the Palatini formulation. We have to make several simplifying assumptions to make the calculations analytically tractable. The first simplification that we make is to assume vanishing torsion, i.e. our general connection is assumed to be symmetric $\Gamma^{\sigma}_{\nu\mu}=\Gamma^{\sigma}_{\mu\nu}$. We will only add terms constructed from the Ricci tensor and additionally we will consider an action invariant under projective transformations, which does not introduce extra gravitational degrees of freedom, that would complicate the picture. The action that we will consider in the Jordan frame reads 
    \begin{equation}\label{eq:START}
        \begin{aligned}
            S = \int d^4x \sg \left[\vphantom{\half}\right. &\half \left\{\vphantom{\half}  \left( 1 + \xi h^2 \right) \sR + \beta \sR^2 + \alpha \sR_{(\mu\nu)}\sR^{(\mu\nu)}  \right\}\\
             &-\half g^{\mu\nu} \pd_\mu h \pd_\nu h - V(h)\left.\vphantom{\half}\right].
        \end{aligned}
    \end{equation}
    Like in the case of pure Higgs inflation one could work with this action and compute the inflationary observables. However we can again simplify our life by transforming our action into an Einstein--Hilbert like action; we will again call this the Einstein frame. However, this is now a more involved task due to the higher-order curvature terms. We will also present an alternative method of computing the slow-roll equations without going into the Einstein frame. 
    
    First we will carry out the computation in the Einstein frame. We start by introducing a general procedure how to bring an projective invariant Ricci based theory, like we have, into the Einstein frame. We then use this method to compute the slow-roll equations and parameters with our action \eqref{eq:START}. Next we will derive the slow-roll equations without going to the Einstein frame and see that our results agree. Finally we will show how the inflationary observables are modified by the higher-order curvature terms.

    \section{Finding the Einstein frame representation}

    Here we will outline the general procedure how to transform an projective invariant Ricci based theory to the Einstein frame. This general procedure was also described in \cite{Afonso17}. We will then use this procedure in our special case to obtain our new result. We write our general action as
    \begin{equation}\label{eq:genericRicci}
        S = \int d^4x\sg \half\left[ F\left( g_{\mu\nu}, \sR_{(\mu\nu)},\psi_i,\pd_\mu \psi_i\right) + 2\Lagr_m\left(g_{\mu\nu},\psi_i,\pd_\mu\psi_i\right)\vphantom{\half}\right],
    \end{equation}
    where $F$ is a general function of the symmetric part of the Ricci tensor $\sR_{\mu\nu}$, the metric $g_{\mu\nu}$ and matter fields $\psi_i$; and $\Lagr_m$ is the matter part of the action. We can perform a Legendre transformation so that the action becomes
    \begin{equation}\label{eq:Legendre}
        \begin{aligned}
            S =  \int d^4x\sg\half&\left[\vphantom{\half}\right. F\left( g_{\mu\nu}, \Sigma_{\mu\nu},\psi_i,\pd_\mu \psi_i\right)+\frac{\pd F}{\pd \Sigma_{\mu\nu}}\left(\sR_{(\mu\nu)}-\Sigma_{\mu\nu}\right)\\
            & + 2\Lagr_m\left[g_{\mu\nu},\psi_i,\pd_\mu\psi_i\right] \left.\vphantom{\half}\right],
        \end{aligned}
    \end{equation}
    where $\Sigma_{\mu\nu}$ is an auxillary field, which is symmetric by definition. Taking the variation of the action \eqref{eq:Legendre} with respect to $\Sigma_{\mu\nu}$ gives the constraint $\Sigma_{\mu\nu}=\sR_{(\mu\nu)}$ and thus the actions \eqref{eq:genericRicci} and \eqref{eq:Legendre} describe the same theory. The Legendre transformed action is now linear in the Ricci tensor and we are one step closer to finding the Einstein frame. We next make a field redefinition 
    \begin{equation}\label{eq:redefq}
        \sqrt{-q}q^{\mu\nu}=\sg \frac{\pd F}{\pd \Sigma_{\mu\nu}},
    \end{equation}
    where $q\equiv \det q_{\mu\nu}$ and $q_{\mu\nu}$ is the inverse of $q^{\mu\nu}$, which means that $q^{\mu\lambda}q_{\nu\lambda}=\delta^\mu_\nu$. Note that the new variable is also symmetric $q_{\mu\nu}=q_{\nu\mu}$. From this redefinition we can solve the auxillary field $\Sigma_{\mu\nu}$ in terms of the metric $g_{\mu\nu}$, the matter fields $\psi_i$ and the new variable $q_{\mu\nu}$, i.e. $\Sigma_{\mu\nu}=\Sigma_{\mu\nu}(g_{\mu\nu},q_{\mu\nu},\psi_i)$. This enables us to write the action as
    \begin{equation}\label{eq:legendre2}
        \begin{aligned}
            S =  \int d^4x \half&\left\{ \vphantom{\half}\right. \sqrt{-q} q^{\mu\nu}\sR_{\mu\nu}\\
            &\left.-\sg\left[ \frac{\pd F}{\pd \Sigma_{\mu\nu}}\Sigma_{\mu\nu}(q_{\mu\nu},g_{\mu\nu},\psi_i)- F(q_{\mu\nu},g_{\mu\nu},\psi_i)-2\Lagr_m(g_{\mu\nu},\psi_i) \right] \right\}.
        \end{aligned}
    \end{equation}
    From this action it is apparent that the metric $g_{\mu\nu}$ does not have a kinetic term and thus its equations of motion are algebraic equations. We can thus solve the original metric $g_{\mu\nu}$ in terms of the new field $q_{\mu\nu}$ and the matter fields $\psi_i$, that is $g_{\mu\nu}=g_{\mu\nu}(q_{\mu\nu},\psi_i)$. Plugging this back into the \eqref{eq:legendre2} gives us an action that depends only on the independent connection $\Gamma^\sigma_{\mu\nu}$, the new field $q_{\mu\nu}$ and the matter fields $\psi_i$
    \begin{equation}\label{eq:actionq}
        \begin{aligned}
            S =  \int d^4x \half\left\{\vphantom{\half}\right. \sqrt{-q}q^{\mu\nu}\sR_{\mu\nu}
            \left.-\sg \left[ \frac{\pd F}{\pd \Sigma_{\mu\nu}}\Sigma_{\mu\nu}(q_{\mu\nu},\psi_i)- F(q_{\mu\nu},\psi_i)-2\Lagr_m(q,\psi) \right] \right\},
        \end{aligned}
    \end{equation}
    where $\sg$ is also solved in terms of $q_{\mu\nu}$ and $\psi_i$. Thus we have managed to transform our action into the Einstein frame where the gravitational sector is the Einstein--Hilbert action for the metric $q_{\mu\nu}$.

    In general, if we do not assume a symmetric Ricci tensor the new metric $q_{\mu\nu}$ is not symmetric. Even if a symmetric connection is assumed the Ricci tensor can have an antisymmetric part: $\sR_{[\mu\nu]} = -\cd_{[\nu}Q_{\mu]}$. and thus from the field redefinition \eqref{eq:redefq} it is clear that also $q^{\mu\nu}$ has an antisymmetric part. The non-symmetric part will then contain new gravitational degrees of freedom and in general these can lead to instabilities in the theory \cite{Moffat}\cite{Damour92}. In fact in the non-symmetric case the action can be seen to be equivalent to a non-symmetric gravity theory, which has been shown to contain instabilities. However, it is possible to construct a non-projective invariant theory in Ricci based gravity without instabilities by requiring that the torsion will vanish by means of Lagrange multipliers. Then the resulting new gravitational degrees of freedom from the antisymmetric part will result in one massive vector field \cite{BJ19}. In fact, in a cosmological setting with the FLRW universe the Ricci tensor is restricted to be symmetric since the only available tensors are the projection tensor for the spacelike slices $h_{\mu\nu}$ and the timelike vector $u_\mu$. Thus the background evolution will be the same with a general Ricci tensor, but the perturbations will include these extra degrees of freedom. Considering these extra degrees of freedom are out of the scope of this thesis.
    
    If the action is projective invariant there are no new gravitational degrees of freedom and $q^{\mu\nu}$ is symmetric. The action \eqref{eq:actionq} is then the Einstein--Hilbert action for the metric $q_{\mu\nu}$ with modified matter sector. The connection is thus the Levi--Civita connection for the metric $q_{\mu\nu}$. 

    In what follows we shall carry out this procedure in the case of Higgs inflation with higher-order curvature terms.
    
    \section{Einstein frame}
    
    The general procedure works for non-minimally coupled actions. However, we will simplify our action \eqref{eq:START} by removing the non-minimal coupling with a conformal transformation like previously
    \begin{equation}
        g^{\mu\nu}\rightarrow \Omega^2(x) g^{\mu\nu},\quad \Omega^2 = 1+\xi h^2.
    \end{equation}
    Since the higher-order terms are invariant under conformal transformations in the Palatini formulation, as we noted in section \ref{sec:conformal}, only the matter sector is modified. The action becomes
    \begin{equation}\label{eq:hActionCT}
        \begin{aligned}
            S = \int d^4x \sg \left[\vphantom{\half}\right. \half \left\{\sR + \beta \sR^2 + \alpha \sR_{\mu\nu}\sR^{\mu\nu} \right\}
            \left. -\half \Omega^{-2} g^{\mu\nu} \pd_\mu h \pd_\nu h - \Omega^{-4}V(h)\right].
        \end{aligned}
        \end{equation}
    The kinetic term is brought to the canonical form with the same field redefinition as before \eqref{eq:CTredef}. And the potential $U(\chi)$ is defined as \eqref{eq:ModPot}.
    With these the action simplifies to
    \begin{equation}\label{eq:chiAction}
        S = \int d^4x \sg \left[ \half\left\{\sR + \beta \sR^2 + \alpha \sR_{\mu\nu}\sR^{\mu\nu} \right\} -\half g^{\mu\nu} \pd_\mu \chi \pd_\nu \chi - U(\chi)\right].
    \end{equation}
    We will call the action in this form the higher-order gravity frame. We will also later derive the slow-roll equations in this frame to compare with the Einstein frame results. First, let us carry out the procedure outlined above and transform \eqref{eq:chiAction} into the Einstein frame.

    With our action \eqref{eq:chiAction} the function $F$ is 
    \begin{equation}\label{eq:F}
        F(g_{\mu\nu},\sR_{\mu\nu}) = \sR + \beta \sR^2 + \alpha \sR_{\mu\nu}\sR^{\mu\nu},
    \end{equation}
    and the matter sector is
    \begin{equation}\label{eq:Lm}
        \Lagr_m(g_{\mu\nu},\chi) = -\half g^{\mu\nu} \pd_\mu \chi \pd_\nu \chi - U(\chi).
    \end{equation}
    To bring our action explicitly to the form of \eqref{eq:actionq} we start by solving the auxillary field $\Sigma_{\mu\nu}$ in terms of $q^{\mu\nu}$ and $g^{\mu\nu}$. We will write out all the metrics explicitly when performing contractions to see more easily the metric dependence and to make it clear which metric is used when making the contractions. Writing out \eqref{eq:redefq} with \eqref{eq:F} we have 
    \begin{equation}\label{eq:redefq2}
        \sqrt{-q}q^{\mu\nu} = \sg \left[ (1+2\beta g^{\sigma\lambda}\Sigma_{\sigma\lambda} )g^{\mu\nu} + 2\alpha g^{\sigma\mu}g^{\lambda\nu}\Sigma_{\sigma\lambda} \right].
    \end{equation}
    Taking the trace of this we have
    \begin{equation}
        \sq q^{\mu\nu}g_{\mu\nu} = \sg\left[ 4(1+2\beta g^{\sigma\lambda}\Sigma_{\sigma\lambda} ) + 2\alpha g^{\sigma\lambda}\Sigma_{\sigma\lambda} \right],
    \end{equation}
    from which we can solve
    \begin{equation}
        g^{\sigma\lambda}\Sigma_{\sigma\lambda} = \frac{\sq q^{\mu\nu}g_{\mu\nu}}{\sg (8\beta+2\alpha)} - \frac{4}{8\beta+2\alpha}.
    \end{equation}
    Substituting this back into \eqref{eq:redefq2} and solving for $\Sigma_{\mu\nu}$ we get
    \begin{equation}
        \Sigma_{\mu\nu} = \frac{1}{2\alpha}\frac{\sq}{\sg}q^{\sigma\lambda}g_{\sigma\mu}g_{\lambda\nu}-\frac{1}{2\alpha+8\beta}\left[1 + \frac{\beta \sq}{\alpha \sg} q^{\sigma\lambda}g_{\sigma\lambda}\right]g_{\mu\nu}.
    \end{equation}
    Next we need to substitute this back to the Legendre transformed action \eqref{eq:Legendre} and derive the equations of motion for the original metric $g_{\mu\nu}$. Computing the variation is straightforward but lengthy, see appendix \ref{app:SigmaVariation} for a few more details, here we will just present the result
    \begin{equation}\label{eq:metricConstraint}
        \begin{aligned}
            &\frac{1}{\sg}\frac{\delta S}{\delta g^{\mu\nu}} = - \frac{1}{4(\alpha+4\beta)}\frac{\sq}{\sg}q^{\sigma\lambda}g_{\sigma\mu}g_{\lambda\nu} \\
            &+ \frac{1}{4\alpha}\frac{q}{g} \left[ q^{\sigma\lambda}q^{\rho\delta}g_{\lambda\delta}g_{\rho\nu}g_{\sigma\mu}-\frac{\beta}{(\alpha+4\beta)}q^{\delta\rho}g_{\delta\rho}q^{\sigma\lambda}g_{\sigma\mu}g_{\lambda\nu}\right]\\
            &+\frac{1}{2}g_{\mu\nu}\left[\frac{1}{\alpha+4\beta}\left(\frac{\beta}{8\alpha}\frac{q}{g}q^{\lambda\sigma}g_{\lambda\sigma}q^{\rho\delta}g_{\rho\delta}+ \half \right) -\frac{q}{g}\frac{1}{8\alpha} q^{\lambda\sigma}q^{\delta\rho}g_{\lambda\delta}g_{\sigma\rho}\right]\\
            &-\half \pd_\mu \chi \pd_\nu \chi + \half g_{\mu\nu}\left[U(\chi) + \half g^{\lambda\sigma}\pd_\lambda \chi \pd_\sigma \chi\right] = 0.
        \end{aligned}
    \end{equation}
    We now need to solve the original metric $g_{\mu\nu}$ in terms of the new metric $q_{\mu\nu}$ and the scalar field $\chi$. This turns out to be the most difficult part.
    We will solve this by introducing the following ansatz
    \begin{equation}\label{eq:metricAnsatz}
        g_{\mu\nu} = \gamma_1(\chi,X_q)q_{\mu\nu}+\gamma_2(\chi,X_q)\pd_\mu \chi \pd_\nu \chi,
    \end{equation}
    where $X_q = q^{\mu\nu}\pd_\mu \chi \pd_\nu \chi$. This is in the form of a disformal transformation. We are thus effectively finding the disformal transformation which brings the action to the Einstein frame. Let us briefly discuss these kind of metric transformations.

    \subsection{Disformal transformations}\label{sec:disformal}
    Another useful type of metric transformations that is able to bring more complicated actions into the Einstein frame, is disformal transformations. These are of the form
    \begin{equation}\label{eq:disformaltransform}
        q_{\mu\nu}=\Gamma_1(\chi,X_g)g_{\mu\nu}+\Gamma_2(\chi,X_g)\pd_\mu\chi\pd_\nu\chi,
    \end{equation}
    where $X_g = g^{\mu\nu}\pd_\mu \chi \pd_\nu \chi$ and $\Gamma_1, \Gamma_2$ are called the conformal and disformal factor respectively. We write the inverse of this transformation as
    \begin{equation}\label{eq:disformaltransform_inv}
        g_{\mu\nu}=\gamma_1(\chi,X_q)q_{\mu\nu}+\gamma_2(\chi,X_q)\pd_\mu\chi\pd_\nu\chi.
    \end{equation}
    Notice that both of the factors are allowed to depend on the kinetic term of the scalar field and thus these factors depend on the metric $g_{\mu\nu}$ and in the case of the inverse transformation they depend on the new metric $q_{\mu\nu}$. The factors between the transformation and its inverse have the following relations
    \begin{equation}\label{eq:func_rel}
        \Gamma_1(\chi,X_g) = \frac{1}{\gamma_1(\chi,X_q(X_g))},\quad \Gamma_2(\chi,X_g) = -\frac{\gamma_2(\chi,X_q(X_g))}{\gamma_1(\chi,X_q(X_g))},
    \end{equation}
    where the $X_q$ has to be solved in terms of $X_g$. We will get a more explicit form for $X_q(X_g)$ later. The factors $\Gamma_1, \Gamma_2$ are restricted by requiring that the new metric describes a well defined spacetime. For that we require the following four conditions:
    
    \begin{enumerate}
        \item The transformation should be invertible, so that there exists an inverse metric $q^{\mu\nu}$.
        \item The transformation does not change the Lorentzian signature of the metric.
        \item Causal trajectories have to remain causal.
        \item The transformation has to be invertible. 
    \end{enumerate}
    Sometimes in the literature also another condition is listed, which is that the volume element has to be non-zero. This is however redundant since this is already guaranteed by the first condition listed. These requirements assure that the transformation is a map between two pseudo-Riemannian spaces and the two representations are physically equivalent, for investigations of this see e.g. \cite{Domenech15,Takahashi17}.

    It is easier to compute the constraints in a cosmological setting and since we are interested in analyzing inflation this is enough for our purposes. Let's look at how these requirements restrict the two factors. Looking at the invariant line element
    \begin{equation}
        d\hat{s}^2 = \Gamma_1 g_{\mu\nu}dx^\mu dx^\nu + \Gamma_2\pd_\mu\chi\pd_\nu\chi dx^\mu dx^\nu = \Gamma_1 ds^2 + \Gamma_2 (\pd_\mu\chi dx^\mu)^2,
    \end{equation}
    we see that the light-cones get modified and are either stretched or squeezed depending on the sing of $\Gamma_2$. To ensure that causal trajectories, $ds^2<0$, are also causal with the transformed metric we must have $d\hat{s}^2<0$, i.e. $\Gamma_2<0$ everywhere.

    To maintain the Lorentzian signature of the transformed metric, i.e. $q_{00}<0$ with $q_{0i}=0$, we must have
    \begin{equation}
        q_{00} = \Gamma_1 g_{00} + \Gamma_2 \pd_0\chi\pd_0\chi < 0.
    \end{equation}
    Multiplying both sides with $g^{00}$, which is negative, we get
    \begin{equation}
        \Gamma_1 + \Gamma_2 g^{00}\pd_0\chi\pd_0\chi > 0.
    \end{equation}
    In a FLRW model $\pd_\mu\chi$ is time-like; thus, taking a frame where $\pd_\mu\chi=(\pd_0\chi,0,0,0)$ we can write $X_g = g^{00}\pd_0\chi\pd_0\chi$. So we have the frame independent requirement
    \begin{equation}\label{eq:DisformalCondition}
        \Gamma_1 + \Gamma_2 X_g > 0.
    \end{equation}
    The inverse of the transformed metric $q_{\mu\nu}$ can be obtained from $q_{\lambda\nu}q^{\mu\lambda} = \delta^{\mu}_\nu$. We introduce an ansatz for the inverse metric
    \begin{equation}\label{eq:gAnzInv}
        q^{\mu\nu} = \Gamma_3(\chi,X_g)g^{\mu\nu}+\Gamma_4(\chi,X_g)g^{\mu\lambda}g^{\nu\sigma}\pd_\lambda\chi\pd_\sigma\chi
    \end{equation}
    Substituting this and \eqref{eq:disformaltransform} into the identity $\delta^\mu_\nu = q^{\mu\lambda}q_{\nu\lambda}$ we get 
    \begin{equation}
        \begin{aligned}
            \delta^\mu_\nu &= \left( \Gamma_3 g^{\mu\lambda} + \Gamma_4 g^{\mu\rho}q^{\lambda\tau}\pd_\rho\chi\pd_\tau\chi \right)\left(\Gamma_1 g_{\mu\nu} + \Gamma_2 \pd_\mu\chi\pd_\nu\chi\right)\\
            &= \Gamma_1\Gamma_3 \delta^\mu_\nu + \left(\Gamma_2\Gamma_4 X_g +\Gamma_1\Gamma_4+\Gamma_2\Gamma_3\right)g^{\mu\lambda}\pd_\lambda\chi\pd_\nu\chi .
        \end{aligned}
    \end{equation}
    Requiring for the coefficients to match on the left and right side of the equation we get
    \begin{equation}
        \begin{dcases}
            \Gamma_1\Gamma_3 = 1 \\
            \Gamma_2\Gamma_4 X_g + \Gamma_1\Gamma_4+\Gamma_2\Gamma_3 = 0.\\
        \end{dcases}
    \end{equation}
    Solving for $\Gamma_3$ and $\Gamma_4$ we get
    \begin{equation}
        \begin{dcases}
            \Gamma_3 = \frac{1}{\Gamma_1}\\
            \Gamma_4 = \frac{-\Gamma_2}{\Gamma_1(\Gamma_1+X_g\Gamma_2)}.
        \end{dcases}
    \end{equation}
    The inverse metric is thus
    \begin{equation}\label{eq:ansatzInverse}
        q^{\mu\nu} = \frac{1}{\Gamma_1}g^{\mu\nu} - \frac{\Gamma_2}{\Gamma_1(\Gamma_1+X_g \Gamma_2)} g^{\lambda\mu}g^{\sigma\nu}\pd_\lambda\chi\pd_\sigma\chi,
    \end{equation}
    For this to be non-singular we must require $\Gamma_1>0$ and $\Gamma_1+X_q \Gamma_2 > 0$ everywhere. These two conditions are enough to assure the existence of the inverse metric and that the Lorentz signature is unchanged. We can now relate $X_q$ with $X_g$ by multiplying \eqref{eq:ansatzInverse} with $\pd_\mu\chi\pd_\nu\chi$ which yields
    \begin{equation}\label{eq:kin_qg}
        X_q = \frac{X_g}{\Gamma_1+X_g\Gamma_2}.
    \end{equation}
    The existence of the inverse transformation \eqref{eq:ansatzInverse} is guaranteed by the condition of non vanishing Jacobian determinant, which yields the condition \cite{Zumalacarregui13}
    \begin{equation}
        \Gamma_1\left(\Gamma_1- X_g\frac{\pd \Gamma_1}{\pd X_g}- X_g^2\frac{\pd\Gamma_2}{\pd X_g}\right)\neq 0.
    \end{equation}
    This constraint also guarantees that the equation \eqref{eq:kin_qg} is solvable. Next let's look at how the determinant of the metric transforms. We begin by taking \eqref{eq:disformaltransform} and contracting both sides with $g^{\nu\sigma}$ and taking the determinant
    \begin{equation}
        \det (q_{\mu\nu}g^{\nu\sigma}) = \det \left[ \Gamma_1 \delta_\mu^\sigma + \Gamma_2 g^{\nu\sigma} \pd_\mu\chi\pd_\nu\chi \right].
    \end{equation}
    Thinking in terms of matrices the left-hand side is a determinant of the product of two square matrices which is equal to the product of their determinants. Additionally $\det (M^{-1}) = (\det M)^{-1}$, so we have
    \begin{equation}
        q g^{-1} = \det \left[ \Gamma_1 \delta^\mu_\sigma + \Gamma_2 g^{\nu\sigma} \pd_\mu \chi \pd_\nu \chi \right].
    \end{equation}
    For the right hand side we write the determinant in terms of the completely antisymmetric Levi--Civita symbol $\epsilon^{\mu\nu\sigma\lambda}$ as
    \begin{equation}
        \begin{aligned}
            q g^{-1} = \epsilon^{\mu\nu\sigma\lambda} \left[\right.& \left( \Gamma_1 \delta^0_\mu+\Gamma_2 \pd_\mu\chi\pd_\rho\chi q^{0\rho} \right)\\
            &\times\left( \Gamma_1 \delta^1_\nu+\Gamma_2 \pd_\nu\chi\pd_\rho\chi g^{1\rho} \right)\\
            &\times\left( \Gamma_1 \delta^2_\sigma+\Gamma_2 \pd_\sigma\chi\pd_\rho\chi g^{2\rho} \right)\\
            &\times\left( \Gamma_1 \delta^3_\lambda+\Gamma_2 \pd_\lambda\chi\pd_\rho\chi g^{3\rho} \right) \left.\right].
        \end{aligned}
    \end{equation}
    Since the left hand side is a ratio of two scalar densities with the same weight it is a scalar and so we can pick any convenient frame to evaluate this expression. Thus, again picking a frame where $\pd_\mu\chi = (\pd_0\chi,0,0,0)$ makes all except few terms vanish giving us
    \begin{equation}
        \begin{aligned}   
            q g^{-1} &= \Gamma_1^4\epsilon^{\mu\nu\sigma\lambda}\delta^0_\mu\delta^1_\nu\delta^2_\sigma\delta^3_\lambda + \Gamma_1^3\Gamma_2 \pd_\mu\chi\pd_\rho\chi g^{0\rho} \delta^1_\nu\delta^2_\sigma\delta^3_\lambda \epsilon^{\mu\nu\sigma\lambda} \\
            &= \Gamma_1^4 + \Gamma_1^3\Gamma_2 \pd_0\chi\pd_0\chi g^{00}\\
            &= \Gamma_1^3\left( \Gamma_1 + X_g \Gamma_2 \right).
        \end{aligned}
    \end{equation}
    Finally the determinant of $q_{\mu\nu}$ in terms of $g_{\mu\nu}$ and $\chi$ becomes
    \begin{equation}\label{eq:detq}
        q = g \Gamma_1^3\left( \Gamma_1 + X_g \Gamma_2 \right).
    \end{equation}
    Thus, the volume element transforms as
    \begin{equation}
        d^4x\sq = d^4x\sg\sqrt{\Gamma_1^3( \Gamma_1 + X_g \Gamma_2)},
    \end{equation}
    which is already guaranteed to be non-zero by the previously found constraints. The invariance of inflationary observables under disformal transformations has been investigated in the literature \cite{Domenech15,Motohashi15,Watanabe15,Tsujikawa14,Minamitsuji14}.

    For our computation we need to perform the inverse transformation \eqref{eq:disformaltransform_inv} and thus we need relations for the inverse metric $g^{\mu\nu}$ and the metric determinant in terms of the factors $\gamma_1,\gamma_2$. We see that the computations can be performed with the exactly same steps as we did above and so we have
    \begin{equation}\label{eq:ansatzInverse_g}
        g^{\mu\nu} = \frac{1}{\gamma_1}g^{\mu\nu} - \frac{\gamma_2}{\gamma_1(\gamma_1+X_q \gamma_2)} q^{\lambda\mu}q^{\sigma\nu}\pd_\lambda\chi\pd_\sigma\chi,
    \end{equation}
    for the inverse metric $g^{\mu\nu}$ and
    \begin{equation}\label{eq:detg}
        g = q \gamma_1^3\left( \gamma_1 + X_q \gamma_2 \right),
    \end{equation}
    for the metric determinant. Similarly $X_g$ can be written in terms of $X_q$ as
    \begin{equation}\label{eq:Xg}
        X_g = \frac{X_q}{\gamma_1 + X_q \gamma_2}.
    \end{equation}
    Since we will be solving only the form of the inverse transformation it is easier to check the four conditions that we require in terms of the factors $\gamma_1$ and $\gamma_2$. Using the relations between the transformation and its inverse \eqref{eq:func_rel} and how the kinetic terms are related \eqref{eq:Xg} the requirements for the existence of the inverse metric become
    \begin{equation}
        \frac{1}{\gamma_1}>0 \iff \gamma_1>0,\quad \frac{1}{\gamma_1+X_q\gamma_2}>0\iff \gamma_1+X_q\gamma_2>0.
    \end{equation}
    The requirement for causal trajectories to remain causal becomes
    \begin{equation}
        -\frac{\gamma_2}{\gamma_1}<0\iff \gamma_2 > 0.
    \end{equation}
    Finally the invertibility condition becomes
    \begin{equation}
        \gamma_1\left( \gamma_1 - X_q \frac{\pd\gamma_1}{\pd X_q}- X_q^2\frac{\pd\gamma_2}{\pd X_q} \right) \neq 0.
    \end{equation}
    Now we have all the necessary tools to carry out the computation.

    \subsection{Solving the metric}
    In the previous section we found the inverse \eqref{eq:ansatzInverse_g} for the ansatz of the metric $g_{\mu\nu}$ \eqref{eq:metricAnsatz}, and additionally the metric determinant $g$ in terms of the new metric $q_{\mu\nu}$ and the scalar $\chi$ \eqref{eq:detg}.

    Now we can substitute the ansatz \eqref{eq:metricAnsatz},\eqref{eq:ansatzInverse} and the determinant \eqref{eq:detg} to the constraint equation for the metric $g_{\mu\nu}$ \eqref{eq:metricConstraint}. This will give us algebraic equations for the coefficients $\gamma_1$ and $\gamma_2$. After doing the substitutions and requiring that the coefficients of $q_{\mu\nu}$ and $\pd_\mu\chi\pd_\nu\chi$ vanish identically we get the system of equations:
    \begin{equation}\label{eq:gamma_eq1}
        \begin{aligned}
            &\frac{1}{16 \alpha (\alpha + 4 \beta) \gamma_1^2 (\gamma_1 + X_q \gamma_2)} \left[\vphantom{\half}\right. -4 \alpha \gamma_1^4 + 2 X_q (\alpha + 2 \beta) \gamma_1 \gamma_2 \\
            &+ 
            X_q^2 (\alpha + 3 \beta) \gamma_2^2 - 
            4 X_q \alpha \gamma_1^3 (\alpha + 4 \beta + 
             \gamma_2) + 
            4 \alpha \gamma_1 \sqrt{\gamma_1^3 (\gamma_1 + 
            X_q \gamma_2)} \left.\vphantom{\half}\right]\\
            & - \half\gamma_1 U(\chi) = 0,
        \end{aligned}
    \end{equation}
    from the coefficient of $q_{\mu\nu}$ and 
    \begin{equation}\label{eq:gamma_eq2}
        \begin{aligned}
            &\frac{1}{16\alpha (\alpha + 
              4 \beta) \gamma_1^3 (\gamma_1 + X_q \gamma_2)}\left[\vphantom{\half}\right. 
              -4 \alpha \gamma_1^4 (2\alpha + 8 \beta - \gamma_2) + 
        8 (\alpha + 2 \beta) \gamma_1^2 \gamma_2
        \\ 
        &-4 X_q \alpha \gamma_1^3 (\alpha + 4 \beta - 
            \gamma_2) \gamma_2 + 
        2 X_q (5 \alpha + 12 \beta) \gamma_1 \gamma_2^2 + 
        3 X_q^2 (\alpha + 3 \beta) \gamma_2^3 \\
        &-\left(8\alpha\gamma_1\gamma_2 + 4 X_q \alpha \gamma_2^2\right) \sqrt{\gamma_1^3 (\gamma_1 + 
        X_q \gamma_2)} \left.\vphantom{\half}\right] + \half\gamma_2 U(\chi)=0,
        \end{aligned}
        \end{equation}
        from the coefficient of $\pd_\mu\chi\pd_\nu\chi$. We can now check what happens in the vacuum limit. Taking $X_q\rightarrow 0$ ,$U(\chi)\rightarrow 0$ and $\chi\rightarrow 0$ the equation \eqref{eq:gamma_eq1} simplifies to $\gamma_1 - 1 = 0$, the second equation is nonexistent ($0=0$) since there is no kinetic term for the scalar. Substituting $\gamma_1=1$ to the modified scalar sector in the action \eqref{eq:actionq} gives zero. Thus, we obtain the Einstein equations in vacuum.

        In general it is hard to find solutions to this set of algebraic equations. However, in slow-roll approximation the square of the time derivative of the field is small $\dot{\chi}^2/H^2\ll 1$. In FLRW universe we have $X_q = -\dot{\chi}^2$, thus we can treat $X_q$ as a small parameter and find approximate solutions of the form
        \begin{equation}\label{eq:seriesAnz}
            \begin{aligned}
                &\gamma_1(\chi,X_q) = a_0 + a_1 X_q + a_2 X_q^2 + \dots \quad,\\
                &\gamma_2(\chi,X_q) = b_0 + b_1 X_q + b_2 X_q^2 + \dots \\
            \end{aligned}
        \end{equation}  
        We can then find solutions of this form by substituting \eqref{eq:seriesAnz} into the equations \eqref{eq:gamma_eq1} and \eqref{eq:gamma_eq2} and again expanding in terms of $X_q$ and requiring that each of the coefficients vanish identically. This gives us systems of equations from which we can solve the coefficients $\{a_i,b_i\}$.
        In the zeroth order we get the system of equations
        \begin{equation}
        \begin{aligned}
            &\frac{a_0 \left[1 + 2\alpha U(\chi) + 8 \beta U(\chi)\right]-1}{\alpha + 4 \beta} = 0\\
            &\frac{-2 \alpha (\alpha + 4 \beta) a_0^2 + (4 \beta + \alpha (2 + (-2 + a_0) a_0)) b_0}{4\alpha (\alpha + 4 \beta) a_0^2} + \half b_0 U(\chi)=0,
        \end{aligned}
        \end{equation}
        which are easily solved giving
        \begin{equation}
            \begin{aligned}
                &a_0 = \frac{1}{1 + (2\alpha  + 8 \beta) U(\chi)}\\
                &b_0 = \frac{2\alpha }{\left[1 + (2\alpha + 8 \beta) U(\chi)\right] \left[1 + (4 \alpha + 
                8 \beta) U(\chi)\right]}.
            \end{aligned}
        \end{equation}
        These expressions quickly become very cumbersome to write down and are not particularly illuminating. So we will not present any more terms here; a few more coefficients can be found from the appendix \ref{app:coeffs}. It can be now verified that the requirements that we required for the transformation to be physical are met. To this first order the requirements reduce to $a_0>0,b_0>0$ and $a_0 + b_0 X_q>0$.
        
        Now substituting the solution of the form \eqref{eq:seriesAnz} with the solved coefficients $\{a_i, b_i\}$ to the action \eqref{eq:actionq} and writing this again in series of $X_q$ we get the modified matter sector
        \begin{equation}
            \begin{aligned}
                \Lagr_m =& -\frac{U}{1+(2\alpha + 8\beta) U} - \frac{\half q^{\mu\nu}\pd_\mu\chi\pd_\nu\chi}{1+(2\alpha  +8 \beta) U}\\
                &+ \frac{\left[\alpha^2 U + 
                \beta (1 + 8 \beta U) + \alpha (1 + 6 \beta U)\right]}{2 \left[1 + (2\alpha + 8 \beta) U\right] \left[1 + (4 \alpha + 8 \beta) U\right]}(q^{\mu\nu}\pd_\mu\chi\pd_\nu\chi)^2\\ 
                &+ \dots
            \end{aligned}
        \end{equation}
        As before we can bring the kinetic term to a canonical form by another field redefinition
        \begin{equation}\label{eq:redefXi}
            \frac{d\chi}{d \phi} = \sqrt{1+ (2\alpha + 8\beta) U}.
        \end{equation}
        The potential of the new field $\phi$ reads
        \begin{equation}\label{eq:UtPot}
            \Ut(\phi) = \frac{U(\chi(\phi))}{1+(2\alpha+8\beta)U(\chi(\phi))}.
        \end{equation}
        This brings the matter Lagrangian to the form
        \begin{equation}\label{eq:matterSec:redef}
            \begin{aligned}
                \Lagr_m &= - \Ut -\half q^{\mu\nu}\pd_\mu\phi\pd_\nu\phi\\
                &+ \frac{\left[\alpha^2 U + 
                \beta (1 + 8 \beta U) + \alpha (1 + 6 \beta U)\right]\left[1 + (2\alpha + 8 \beta) U\right]}{2  \left[1 + (4 \alpha + 8 \beta) U\right]}(q^{\mu\nu}\pd_\mu\phi\pd_\nu\phi)^2\\
                &+ \dots
            \end{aligned}
        \end{equation}
        For later convenience we will write this as
        \begin{equation}\label{eq:matterSec:redef_c}
            \begin{aligned}
                \Lagr_m &= - \Ut -\half q^{\mu\nu}\pd_\mu\phi\pd_\nu\phi\\
                & + c_2(U(\phi))(q^{\mu\nu}\pd_\mu\phi\pd_\nu\phi)^2+\dots+c_n(U(\phi))(q^{\mu\nu}\pd_\mu\phi\pd_\nu\phi)^n+ \dots
            \end{aligned}
        \end{equation}
        At this point we can also note that when taking the limit $\alpha \to 0$ we obtain the result of \cite{EERW18}, where non-minimally coupled scalar with the addition of only $\sR^2$ term was analyzed. This is, however, with the assumption that all the higher than second order kinetic terms vanish when the limit is taken. This is expected to be the case, but should still be explicitly proven. Easier way of seeing this is to take the limit $\alpha \to 0$ from the start. This simplifies the computations and with some effort the same result is obtained. 
        
        When the $\alpha \to 0$ limit is taken we have to require $\beta > 0$ to have the correct sign for the kinetic terms in \eqref{eq:matterSec:redef} and avoid instabilities. We see that the correct signs for the kinetic terms are preserved in the general case if both $\alpha,\beta > 0$. There might exist more of the parameter space where this is the case, but this needs further investigation. Thus here we require that $\alpha,\beta > 0$.

        We are now ready to derive the equations of motion, which now is an easy task since the gravitational sector is nothing but the Einstein--Hilbert action.
        \subsection{Equations of motion}

        We have managed to bring our action to the form
        \begin{equation}
            S = \int d^4x \sq \left[\half q^{\mu\nu}\sR_{\mu\nu} -\Ut-\half X_q + c_2(U(\phi)) X_q^2 + \dots \right],
        \end{equation}
        where $c_n(\phi)$ are functions of the field defined in \eqref{eq:matterSec:redef_c}. Thus as previously shown the connection is solved to be the Levi--Civita connection of the metric $q^{\mu\nu}$. The equations of motion for the metric thus are 
        \begin{equation}
            \begin{aligned}
                G^{(q)}_{\mu\nu} \equiv& R^{(q)}_{\mu\nu}-\half R^{(q)} q_{\mu\nu} =\\
                &- q_{\mu\nu}\left[ \Ut +\half \left(X_q - c_2 X_q^2 +\dots\right) \right]\\
                &+\pd_\mu\phi\pd_\nu\phi \left[ 1 -4 c_2 X_q + \dots )\right],
            \end{aligned}
        \end{equation}
        and for the scalar $\phi$ we have
        \begin{equation}
            \begin{aligned}
                &\Box\phi \left[1 - 4c_2 X_q -6c_3 X_q^2 + \dots  \right]\\ 
                &-\Ut_{,\phi}\left[1+3{c_{2}}_{,U} X_q^2 + \dots \right]\\
                &-\left[8c_2+24c_3 X_q + \dots \right](\cd_\mu\cd_\nu\phi)\cd^\mu\phi\cd^\nu\phi = 0.
            \end{aligned}
        \end{equation}
        With the assumption that the kinetic term for the field is sufficiently small such that we can ignore all terms of order higher than one, we get the Einstein equations with a scalar field. With the FLRW solution with zero spatial curvature the temporal component of the metric field equations becomes 
        \begin{equation}\label{eq:Fried00EF}
            3 H^2 = \Ut + \half\dot{\phi}^2,
        \end{equation}
        and the equations of motion for the scalar field become
        \begin{equation}\label{eq:ScalarEOMEF}
            \ddot{\phi}+3H\dot{\phi}=-\Ut_{,\phi}
        \end{equation}
        With the slow-roll conditions $|\ddot{\phi}| \ll |3H\dot{\phi}|$ and $\dot{\phi}^2 \ll \Ut$ the equations reduce to the familiar slow-roll equations with the modified potential $\Ut$
        \begin{equation}
        \begin{aligned}
            &3H^2 =  \Ut\\
            &3H\dot{\phi} = -\Ut^\prime,
        \end{aligned}
        \end{equation}
        The first few slow-roll parameter are thus
        \begin{equation}\label{eq:SRparamQG}
        \begin{aligned}
            \tilde{\epsilon}  =& \half\left(\frac{\Ut^\prime}{\Ut}\right)^2\\        
            \tilde{\eta} =& \frac{\Ut^{\prime\prime}}{\Ut}\\
            \tilde{\zeta} =& \frac{\Ut^\prime \Ut^{\prime\prime\prime}}{\Ut^2}.
        \end{aligned}
        \end{equation}
        Now it is clear that $\dot{\phi}^2 = 2/3 \Ut \epsilon$ and with \eqref{eq:redefXi} we get that $\dot{\chi}^2 = 2/3 U \epsilon$ thus validating our assumption that $\dot{\chi}^2$ is small in slow-roll. What is left to determine is when does this approximation break down and when the higher-order terms in $X_q$ should be taken in to account.


        \section{Higher-order gravity frame representation}

        Now we return to the second method of arriving to these slow-roll equations. We derive the equations of motion for the action \eqref{eq:chiAction} and show that by performing the disformal transformation they are equal to the Einstein frame equations. We derived the equations of motion for the higher-order terms in the most general case in the chapter \ref{sec:QGPAL}. In the case of the action \eqref{eq:START} and vanishing torsion these simplify considerably. For the metric we obtain
        \begin{equation}\label{eq:gEOMnoTsymR}
            \begin{aligned}
                &\left(\half+\beta \sR\right)\sR_{\mu\nu}+\alpha \sR\TI{_\mu^\lambda}\sR_{\nu\lambda} -\half \pd_\mu\chi\pd_\nu\chi\\
                &-\half g_{\mu\nu} \left[\half(1+\beta\sR)\sR +\frac{\alpha}{2}\sR_{\sigma\lambda}\sR^{\sigma\lambda}+\Lagr_m\right]=0,
            \end{aligned}
        \end{equation}
        where $\Lagr_m$ is defined in \eqref{eq:Lm} as before. For the connection we get
        \begin{equation}
            \cd_\sigma\left\{ \sg\left[ \left(1+2\beta\sR\right)g^{\mu\nu}+2\alpha \sR^{\mu\nu} \right] \right\}=0.
        \end{equation}
        We see that this is a metric compatibility condition for a metric defined with
        \begin{equation}\label{eq:q_met_def2}
            \sq q^{\mu\nu} = \sg\left[ \left(1+2\beta\sR\right)g^{\mu\nu}+2\alpha \sR^{\mu\nu} \right].
        \end{equation}
        Thus, the connection is the Levi--Civita connection for the metric $q_{\mu\nu}$. We need to solve $q_{\mu\nu}$ in terms of $g_{\mu\nu}$ and the scalar field $\chi$. This is accomplished by solving $\sR$ and $\sR_{\mu\nu}$. We can start by solving $\sR$, which is easily solved by taking the trace of \eqref{eq:gEOMnoTsymR}. This gives
        \begin{equation}
            \sR = g^{\mu\nu}\pd_\mu\chi\pd_\nu\chi + 4U(\chi).
        \end{equation}
        Solving $\sR_{\mu\nu}$ is more complicated. In a FLRW universe the Ricci tensor must be of the form 
        \begin{equation}\label{eq:RicciAnz}
            \sR_{\mu\nu} = R_1(g_{\lambda\sigma},\chi)g_{\mu\nu} + R_2(g_{\lambda\sigma},\chi)u_\mu u_\nu.
        \end{equation}
        We can plug in this ansatz into the the equations of motion for the metric \eqref{eq:gEOMnoTsymR} and obtain a system of equations for $R_1$ and $R_2$ ($\pd_\mu\chi = \dot{\chi}u_\mu$):
        \begin{equation}
            \begin{aligned}
                \left(\half + \beta \sR\right) R_1 + &\alpha R_1^2 - 
                \frac{1}{4} \left(\sR + \beta \sR^2 + 2\Lagr_m\right) \\
                &- \frac{\alpha}{4}\left(4 R_1^2 +  R_2^2 - 2 R_1 R_2\right) = 0,
            \end{aligned}
        \end{equation}
        \begin{equation}
            \left(\half +  \beta \sR\right) R_2 -  \alpha R_2^2 + 2 \alpha R_1 R_2 - 
            \half \dot{\chi}^2 = 0,
        \end{equation}
        This system of equations is solved by 
        \begin{equation}\label{eq:Rsols}
            \begin{aligned}
                R_1^{\pm} &= \frac{1 + 12 \alpha\Lagr_m + 2\sR(3\alpha+2\beta)(1+\beta\sR) - 3\alpha \dot{\chi}^2\pm A}{8 (\alpha + 2\alpha \beta\sR)}
                \\
                \\
                R_2^{\pm} &= \frac{1 + 4 \alpha\Lagr_m + 2\sR(\alpha+2\beta)(1+\beta\sR) - \alpha \dot{\chi}^2\pm A}{
  2 \alpha  (1 + 2 \beta \sR)},
            \end{aligned}
        \end{equation}
        with
        \begin{equation}
            A\equiv\sqrt{
                -4 \alpha (1 + 2 \beta\sR)^2 \dot{\chi}^2 + \left[1 + 4\alpha \Lagr_m + 
                   2 \sR (\alpha + 2 \beta) (1 + \beta\sR) - 
                    \alpha \dot{\chi}^2\right]^2}.
        \end{equation}
        We can now solve the metric $q^{\mu\nu}$ from \eqref{eq:q_met_def2} giving
        \begin{equation}\label{eq:qSol}
            \begin{aligned}
                q^{\mu\nu} &= \frac{(1+2\beta\sR+2\alpha R_1)g^{\mu\nu} + 2\alpha R_2 u^\mu u^\nu}{\sqrt{(1+2\beta\sR+2\alpha R_1)^3\left[(1+2\beta\sR+2\alpha R_1)- 2\alpha R_2\right]}}\\
                &\equiv \frac{1}{q_1}g^{\mu\nu}+q_3u^\mu u^\nu
            \end{aligned}
        \end{equation} 
        where we have used the previously found identity \eqref{eq:detg} to write the determinant of $q_{\mu\nu}$ in terms of $g_{\mu\nu}$ and $\chi$. We can use the previously obtained identity for the inverse metric \eqref{eq:ansatzInverse} and write the inverse, for later convenience, as 
        \begin{equation}\label{eq:qSolInv}
            q_{\mu\nu}\equiv q_1 g_{\mu\nu} + q_2 u_\mu u_\nu,\quad q_2 = \frac{q_3}{\frac{1}{q_1}\left(\frac{q_3}{q_1}-\frac{1}{q_1}\right)}
        \end{equation}
        The solution \eqref{eq:Rsols} with the minus signs will correspond to the solution found with the earlier method and it is in fact the inverse of the transformation found before. The second solution \eqref{eq:Rsols} with the plus signs turns out to be non-invertible, at least in the inflationary region, and thus was not found by the earlier method. Additionally, the resulting disformal transformation does not preserve the Lorentzian signature of the metric $q_{\mu\nu}$ in the regime where $\dot{\chi}^2\ll U(\chi)$. With the positive signed solution the constraint for preserving the Lorentzian signature \eqref{eq:DisformalCondition}, to zeroth order in $\dot{\chi}^2$, simplifies to 
        \begin{equation}
            -\left[1 + 8 (\alpha + \beta) U(\chi)\right] > 0,
        \end{equation}
        which is not satisfied.
        In contrast with the negatively signed solution the constraint to zeroth order becomes
        \begin{equation}
            1 + (2\alpha + 8 \beta) U(\chi) > 0,
        \end{equation}
        which is satisfied. The positively signed solution \eqref{eq:Rsols} is thus dismissed as unphysical.

        Now let us continue deriving the equations of motion. We can start by computing the connection explicitly by plugging in the solution of $q_{\mu\nu}$ into the definition of Levi--Civita connection. After a straightforward simplification, with the abbreviations \eqref{eq:qSol},\eqref{eq:qSolInv} and making use of \eqref{eq:decomposeMetric}, we obtain
        \begin{equation}\label{eq:connQ}
            \begin{aligned}
                \lcc^\sigma_{\mu\nu}(q) = \lcc^\sigma_{\mu\nu}(g) &+\left[ \half\dot{q_1}\left(q_3-\frac{1}{q_1}\right) + q_3 q_1 H \right]u^\sigma h_{\mu\nu}\\
                & + \half \left[ \frac{1}{q_1}(\dot{q_2}-\dot{q_1}) + q_3(\dot{q_1}-\dot{q_2}) \right]u^\sigma u_\mu u_\nu + \half\frac{\dot{q_1}}{q_1}h\TI{^\sigma_{(\mu}}u_{\nu)}
                \\                
                & \equiv  \lcc^\sigma_{\mu\nu}(g) + L^\sigma_{\mu\nu},
            \end{aligned}
        \end{equation}
        where $H=\dot{a}/a$ is the Hubble parameter and $h_{\mu\nu}$ is the metric of the spacelike slices. We also used the fact that $u_\lambda \lcc^\lambda_{\mu\nu}(g) = H h_{\mu\nu}$ in FLRW universe. Now we can plug this into the definition of the Riemann tensor and see how it is modified, this gives 
        \begin{equation}\label{eq:RicciQ}
            \begin{aligned}
                R^{(q)}_{\mu\nu} &= R^{(g)}_{\mu\nu} + \lcd^{(g)}_\lambda L^\lambda_{\mu\nu}-\lcd^{(g)}_\nu L^\lambda_{\mu\lambda}+L^\lambda_{\lambda\sigma}L^\sigma_{\mu\nu}-L^\lambda_{\nu\sigma}L^\sigma_{\lambda\mu}\\
                & \equiv R^{(g)}_{\mu\nu} + L_1 g_{\mu\nu} + L_2 u_\mu u_\nu.
            \end{aligned}
        \end{equation}
        Now instead of plugging this into the full equations of motion for the metric we can plug this into \eqref{eq:RicciAnz} and we get 
        \begin{equation}\label{eq:eomGRic}
            R^{(g)}_{\mu\nu} = (R_1 - L_1)g_{\mu\nu} + (R_2 - L_2)u_\mu u_\nu,
        \end{equation}
        where $L_1$ and $L_2$ have to be explicitly computed from \eqref{eq:connQ} and \eqref{eq:RicciQ}. This is a straightforward but tedious task and so we shall not do it explicitly here. When $L_1,L_2$ are computed using the solutions \eqref{eq:Rsols} for $R_1^-,R_2^-$ and plugged into \eqref{eq:eomGRic} we get the Friedmann equations. The full equation is cumbersome to write down and we will not need it for our purposes, thus we will again expand in powers of $\dot{\chi}$ and do not write down terms $\mathcal{O}(\dot{\chi}^2)$. For the negatively signed solution $\{R_1^-,R_2^-\}$ we get 
        \begin{equation}\label{eq:FriedJordan}
         3H^2 = U - \frac{12H\dot{\chi}U^\prime\left[\alpha+3\beta+3(\alpha^2+6\alpha\beta+8\beta^2)\right]}{\left[1+(4\alpha+8\beta)U\right]\left[1+(2\alpha+8\beta)U\right]} + \mathcal{O}(\dot{\chi}^2).
        \end{equation}
        The equations of motion for the scalar field are not modified and are simply 
        \begin{equation}
            \ddot{\chi}+3H\dot{\chi} = -U_{,\chi}.
        \end{equation}
        These are the equations of motion in the higher-order gravity frame. To see that these are indeed equivalent to the Einstein frame counterparts, that we obtained earlier, we need to know how the Hubble parameter and the time derivatives change under the disformal transformation. 
        
        These transformation rules can be seen by looking at the invariant line element. With $g_{\mu\nu} = \gamma_1 q_{\mu\nu}+\gamma_2 u_\mu u_\nu$:
        \begin{equation}
            \begin{aligned}
                ds^2 &= g_{\mu\nu}dx^\mu dx^\nu = \gamma_2 u_\mu u_\nu dx^\mu dx^\nu+ \gamma_1 q_{\mu\nu}dx^\mu dx^\nu\\
                &=-(\gamma_1-\gamma_2)dt^2 + \gamma_1 a^2(t)\delta_{ij}dx^i dx^j,
            \end{aligned}
        \end{equation}
        where we used a frame $u_\mu = (-1,0,0,0)$ and $q_{\mu\nu} = \text{diag}(-1,a^2,a^2,a^2)$. From this we see that the scale factor and time coordinate transform as
        \begin{equation}
            \begin{aligned}
                &dt \to \sqrt{\gamma_1-\gamma_2}dt\\
                &a(t) \to \sqrt{\gamma_1}a(t).
            \end{aligned}
        \end{equation}
        Thus the Hubble parameter transforms like 
        \begin{equation}
            H \to \frac{1}{\sqrt{\gamma_1-\gamma_2}}\left[H + \half \frac{\dot{\gamma_1}}{\gamma_1}\right].
        \end{equation}
        Applying these rules to the scalar field equations of motion (with our previous definitions $\gamma_1 = 1/q_1$ and $\gamma_2 = -q_2/(q_1-q_2)$) and keeping only first order terms in $\dot{\chi}$ we get 
        \begin{equation}
            \begin{aligned}
                \left[1+(\alpha+4\beta)U\right] \left( \ddot{\chi} + 3H\dot{\chi} \right) = -U_{,\chi}.
            \end{aligned}
        \end{equation}
        Now making the field redefinition \eqref{eq:redefXi} we get the same result as before \eqref{eq:ScalarEOMEF}. The equation \eqref{eq:FriedJordan} is also straightforward. Plugging in the transformations and keeping only linear terms we get
        \begin{equation}
            \begin{aligned}
                &3H^2\left[1+(2\alpha+8\beta)U\right] - \frac{12H\dot{\chi}U^\prime\left[\alpha+3\beta+3(\alpha^2+6\alpha\beta+8\beta^2)\right]}{1+(4\alpha+8\beta)U}\\ 
                &= U - \frac{12H\dot{\chi}U^\prime\left[\alpha+3\beta+3(\alpha^2+6\alpha\beta+8\beta^2)\right]}{1+(4\alpha+8\beta)U} + \mathcal{O}(\dot{\chi}^2),
            \end{aligned}
        \end{equation}
        which simplifies to the previous result \eqref{eq:Fried00EF}. This was just to show another way of obtaining the slow-roll equations in the Einstein frame. We will not pursue the analysis in the higher-order gravity frame further. 

        With the slow-roll equations of motions in the Einstein frame we are now ready to analyze how the inflationary behavior is modified due to the higher-order terms. 

        \section{Changes to inflationary observables}
        We can now see how the inflationary observables are modified by the higher-order curvature terms. It is easier to write the slow-roll parameters \eqref{eq:SRparamQG} with the field variable $\chi$. Using \eqref{eq:redefXi} and \eqref{eq:UtPot}, simple manipulations yield
        \begin{equation}
            \begin{aligned}
                \tilde{\epsilon}_U  =& \half\left(\frac{\Ut^\prime}{\Ut}\right)^2 
                =\frac{\epsilon_U}{1+(2\alpha+8\beta)U(\chi)}\\       
                \tilde{\eta}_U=& \frac{\Ut^{\prime\prime}}{\Ut} 
                = \eta -3\frac{(2\alpha+8\beta)U(\chi)}{1+(2\alpha+8\beta)U(\chi)}\epsilon,
            \end{aligned}
        \end{equation}
        where $\epsilon_U=1/2(U_{,\chi}/U)^2$, $\eta_U=U_{,\chi\chi}/U$ are the slow-roll parameters that we obtained previously without the higher-order curvature terms in chapter \ref{sec:HI}. The field value at the end of inflation is again obtained from $|\tilde{\eta}(\chi_{end})|=1$, which in the large field limit $\sqrt{\xi}\chi \gg 1$ results in the same value as obtained earlier
        \begin{equation}
            \chi_{end} = \frac{1}{2\sqrt{\xi}}\ln\left(32\xi\right).
        \end{equation}
        We also find that in the slow-roll approximation the formula for the number of e-folds is unchanged
        \begin{equation}
            N_* \simeq\int_{\phi_{end}}^{\phi_*}\frac{d\phi}{\sqrt{2\tilde{\epsilon}}}=\int_{\chi_{end}}^{\chi_*}\frac{d\chi}{\sqrt{2\epsilon_U}}.
        \end{equation}
        We thus obtain the same relation between the field and the number of e-folds
        \begin{equation}\label{}
            \chi(N_*) = \frac{1}{2\sqrt{\xi}}\ln\left(32\xi N_*\right).
        \end{equation}
        This holds assuming that our expansion in terms of the kinetic term of the field is still valid when approaching the end of inflation. This needs further investigations. 
        
        Assuming our expansion to be valid we see that the scalar spectrum amplitude is unchanged
        \begin{equation}
            A_s = \frac{1}{24\pi^2}\frac{\Ut}{\tilde{\epsilon}}=\frac{1}{24\pi^2}\frac{1+(2\alpha+8\beta)U}{1+(2\alpha+8\beta)U}\frac{U}{\epsilon_U}=\frac{1}{24\pi^2}\frac{U}{\epsilon_U}.
        \end{equation}
        Thus matching this to the observed amplitude gives the same requirement for the non-minimal coupling \eqref{eq:XiPal}. The tilt of the scalar spectrum is also unchanged to first order in the slow-roll parameters
        \begin{equation}
            \begin{aligned}
                n_s-1 &= -6\tilde{\epsilon}+2\tilde{\eta}=-6\frac{\epsilon_U}{1+(2\alpha+8\beta)U(\chi)}+ 2\eta_U -6\frac{(2\alpha+8\beta)U(\chi)}{1+(2\alpha+8\beta)U(\chi)}\epsilon_U\\
                &=-6\epsilon_U+2\eta_U.
            \end{aligned}
        \end{equation}
        What changes is the spectrum of tensor perturbations. The amplitude 
        \begin{equation}
            A_t = \frac{2}{3\pi^2}\frac{U}{1+(2\alpha+8\beta)U},
        \end{equation}
        tilt of the spectrum
        \begin{equation}
            n_t = -2\tilde{\epsilon}=\frac{-2\epsilon_U}{1+(2\alpha+8\beta)U},
        \end{equation}
        and the tensor-to-scalar ratio
        \begin{equation}
            r = 16\tilde{\epsilon} = \frac{16\epsilon_U}{1+(2\alpha+8\beta)U}
        \end{equation}
        are all suppressed by the same factor. The observables in terms of the number of e-folds (approximating with $\xi\gg 1$) become
        \begin{align}
            A_t&\simeq \frac{\lambda}{\lambda(2\alpha+8\beta)+4\xi^2}\\
            n_t &\simeq -\frac{\xi}{2N_*^2\left[\lambda(\alpha+4\beta)+2\xi^2\right]}\\
            r &\simeq -\frac{4\xi}{N_*^2\left[\lambda(\alpha+4\beta)+2\xi^2\right]}.
        \end{align}
        Thus we see that both higher-order curvature terms $\beta \sR^2$ and $\alpha \sR_{(\mu\nu)}\sR^{(\mu\nu)}$ have the same effect of suppressing the tensor perturbation spectrum but keeping the scalar perturbation spectrum unchanged. Again when we take the limit $\alpha\to 0$ we recover the results of \cite{EERW18}.
        \\

        Higgs inflation and higher-order gravity in the metric formulation has been studied in the literature. In ref. \cite{Enckell18} Higgs inflation with the addition of $\lcR^2$ term was analyzed. As we saw in section \ref{sec:fR} in the metric formulation there is additional degrees of freedom of gravity and the inflationary model becomes a multi field model. It was found out that in the case of metric formulation the pure Higgs inflation, where the inflation is driven completely by the Higgs field, is ruined.

    \chapter{Conclusions}\label{sec:conclusion}

    \vspace{1cm}
    We have reviewed the metric and Palatini formulations in a comparative manner and discussed their subtle differences in the case of the Einstein--Hilbert action and more obvious differences when considering higher-order theories of gravity. 
    
    We reviewed Higgs inflation in both the metric and Palatini formulations and obtained the inflationary observables in terms of the number of e-folds. We computed the numerical values for the observables, and saw that they are compatible with current observational bounds. The tensor-to-scalar ratio is suppressed by the large non-minimal coupling parameter $\xi$ in the Palatini formulation $r\simeq 2.1\times 10^{-13}\lambda^{-1}$, compared to $r\simeq 0.00396$ in metric formulation. The spectral tilt is also slightly smaller in Palatini formulation due to the different reheating period. These allow us to distinguish Palatini Higgs inflation from the metric Higgs inflation. This could possibly be accomplished by the next generation of planned experiments \cite{LiteBIRD20}.
    
    We briefly discussed the problems of ghost modes appearing in both metric and Palatini formulations when considering higher-order gravity. This problem requires further research. We derived the equations of motion for all the different parity preserving quadratic curvature invariants in the Palatini formulation. 
    \\

    When we considered adding higher-order curvature terms to the Higgs inflation scenario we simplified our action to contain only the symmetric part of the Ricci tensor. Actions constructed this way do not suffer from ghosts, due to the fact that no new gravitational degrees of freedom are present. This also allows us to transform our action to the Einstein frame. By adding only terms $\beta\sR^2+\alpha\sR_{(\mu\nu)}\sR^{(\mu\nu)}$ we were able to find the Einstein frame representation in the inflationary region by finding an approximate solution in powers of the kinetic term of the inflaton field, i.e. the small parameter $\dot{\chi}^2/H^2$. This method could possibly be applied to more general Lagrangians than the one we considered here. However, the region of validity of this approximation needs further investigation. Furthermore, the two parameters $\alpha$ and $\beta$ were assumed to be positive to assure correct signs for the kinetic terms. However, there could be other valid regions for these parameters that give the correct sings, which were not explored in this thesis. 
    
    With the assumption that we can extrapolate the approximate solution to the end of inflation the number of e-folds turns out to be unchanged when comparing to the case where no higher-order terms are present. This results to the same relation between the number of e-folds and the inflation field. The scalar amplitude and tilt of the scalar spectrum are also unchanged. The changes appeared in the tensor perturbation spectrum. The amplitude, spectral tilt and the tensor to scalar ratio are all suppressed by the combination of the couplings of the higher-order terms and the Higgs self coupling: $1+(2\alpha+8\beta)U$. We conclude that in this simple case the higher-order curvature terms do not ruin the Palatini Higgs inflation.

    \begin{appendices}    
        
        \chapter{Useful identities} \label{app:identities}

    \section{Palatini identity}
    Start from the definition of the Riemann tensor \eqref{eq:defRie} and take the variation with respect to the connection
    \begin{equation}
        \delta\sR\TI{^\rho_{\mu\lambda\nu}} = \pd_\lambda\delta\Gamma^\rho_{\nu\mu}+\Gamma^\rho_{\lambda\sigma}\delta\Gamma^\sigma_{\nu\mu} + \delta\Gamma^\rho_{\lambda\sigma}\Gamma^\sigma_{\nu\mu} - (\lambda \leftrightarrow \nu).
    \end{equation}
    Notice that difference of connections is a tensor, hence $\delta\Gamma^\rho_{\nu\mu}$ is a tensor, so the covariant derivative is well defined
    \begin{equation}
        \cd_\lambda(\delta\Gamma^\rho_{\nu\mu}) = \pd_\lambda\delta\Gamma^\rho_{\nu\mu}+\Gamma^\rho_{\lambda\sigma}\delta\Gamma^\sigma_{\nu\mu} - \Gamma^\sigma_{\lambda\nu}\delta\Gamma^\rho_{\sigma\mu} - \Gamma^\sigma_{\lambda\mu}\delta\Gamma^\rho_{\nu\sigma}.
    \end{equation}
    Now gathering terms 
    \begin{equation}
        \begin{aligned}
            \delta\sR\TI{^\rho_{\mu\lambda\nu}} &=  \left[\pd_\lambda\delta\Gamma^\rho_{\nu\mu} + \Gamma^\rho_{\lambda\sigma}\delta\Gamma^\sigma_{\nu\mu} - \Gamma^\sigma_{\lambda\mu}\delta\Gamma^\rho_{\nu\sigma}\right]\\
            &-\left[\pd_\nu\delta\Gamma^\rho_{\lambda\mu} + \Gamma^\rho_{\nu\sigma}\delta\Gamma^\sigma_{\lambda\mu} - \Gamma^\sigma_{\nu\mu}\delta\Gamma^\rho_{\lambda\sigma}\right].
        \end{aligned}
    \end{equation}
    Notice the terms inside the brackets almost equal the covariant derivative of the variation of the connection, we can make it equal by adding and subtracting the missing terms 
    \begin{equation}
        \begin{aligned}
            \delta\sR\TI{^\rho_{\mu\lambda\nu}} &=  \left[\pd_\lambda\delta\Gamma^\rho_{\nu\mu} + \Gamma^\rho_{\lambda\sigma}\delta\Gamma^\sigma_{\nu\mu} - \Gamma^\sigma_{\lambda\mu}\delta\Gamma^\rho_{\nu\sigma} - \Gamma^\sigma_{\lambda\nu}\delta\Gamma^\rho_{\sigma\mu}\right]\\
            &-\left[\pd_\nu\delta\Gamma^\rho_{\lambda\mu} + \Gamma^\rho_{\nu\sigma}\delta\Gamma^\sigma_{\lambda\mu} - \Gamma^\sigma_{\nu\mu}\delta\Gamma^\rho_{\lambda\sigma} - \Gamma^\sigma_{\nu\lambda}\delta\Gamma^\rho_{\sigma\mu}\right] + T\TI{^\sigma_{\lambda\nu}}\delta\Gamma^\rho_{\sigma\mu}
        \end{aligned}
    \end{equation}
    and we get the result
    \begin{equation}
        \delta \sR\TI{^\rho_{\mu\lambda\nu}} = \cd_\lambda(\delta\Gamma^\rho_{\nu\mu})-\cd_\nu(\delta\Gamma^\rho_{\lambda\mu}) + T\TI{^\sigma_{\lambda\nu}}\delta\Gamma^\rho_{\sigma\mu}.
    \end{equation}

    \section{Tensor densities}

    A general tensor density $\mathfrak{T}$ of weight $\omega$ is an object that transforms as
    \begin{equation} 
        {\mathfrak{T}^\prime}\TI{^{\mu_1 ... \mu_n}_{\nu_1 ... \nu_n}} = \ \left|\frac{\pd x^{\alpha}}{\pd x^\prime_{\beta}}\right|^\omega\frac{\pd x^{\prime\mu_1}}{\pd x^{\sigma_1}}\dots\frac{\pd x^{\prime\mu_n}}{\pd x^{\sigma_n}}\dots\frac{\pd x^{\tau_1}}{\pd x^{\nu_1}}\dots\frac{\pd x^{\tau_n}}{\pd x^{\nu_n}} \mathfrak{T}\TI{^{\sigma_1 ... \sigma_n}_{\tau_1 ... \tau_n}},
    \end{equation}
    where $||$ denotes the determinant. Scalar density of weight $\omega$ transforms as 
    \begin{equation} 
        \rho^\prime = \left|\frac{\pd x^{\alpha}}{\pd x^\prime_{\beta}}\right|^\omega\rho.
    \end{equation}
    A way of looking at a scalar density of weight one is as a single independent component of a covariant antisymmetric tensor $\rho = \rho_{[\mu_1...\mu_n]}=\rho_{\mu_1...\mu_n}$, see e.g. \cite[ch. II]{schrodinger85}. It is easy to see that a single component of this transforms as a scalar density. Now 
    \begin{equation}
        \cd_\mu(\rho_{\nu_1...\nu_n}) = \pd_\mu\rho_{\nu_1...\nu_n} - \Gamma^\sigma_{\mu \nu_1}\rho_{\sigma\nu_2...\nu_n} - \Gamma^\sigma_{\mu \nu_2}\rho_{\nu_1\sigma...n} - ... - \Gamma^\sigma_{\mu \nu_n}\rho_{\nu_1...\sigma} .
    \end{equation}
    Since all the non zero components are equal up-to a sign we can focus on one of them 
    \begin{equation}
        \cd_\mu(\rho_{0...n}) = \pd_\mu\rho_{0...n} - \Gamma^\sigma_{\mu 0}\rho_{\sigma 1...n} - \Gamma^\sigma_{\mu 1}\rho_{0\sigma...n} - ... - \Gamma^\sigma_{\mu n}\rho_{0...\sigma} ,
    \end{equation}
    we see that this equals
    \begin{equation}
        \cd_\mu\rho = \pd_\mu\rho-\Gamma^\sigma_{\mu\sigma}\rho .
    \end{equation}
    This is the result for a scalar density of weight +1. From this it is easy to generalize to arbitrary scalar density of weight $\omega$ by requiring that the covariant derivative of such a density transforms as a $(0,1)-$tensor density. One finally obtains
    \begin{equation}
        \cd_\mu\rho = \pd_\mu\rho-\omega\Gamma^\sigma_{\mu\sigma}\rho,
    \end{equation}
    for a scalar density $\rho$ of weight $\omega$.

    \chapter{Equations of motion in the Palatini formulation}\label{app:EOM_Palatini}
    Here we list the equations of motion for the different curvature invariants in the action \eqref{eq:PAL-RR-lag} in Palatini formulation.\\
    \\

\noindent$\bullet$ $\Lagr=\sR^2$
    \begin{equation}
        2\sR\sR_{(\alpha\beta)} -\half g_{\alpha\beta}\sR^2=0.
    \end{equation}
    \begin{equation}
        \begin{aligned}
            &2\cd_\lambda(\sg\sR g^{\nu\lambda})\delta^\mu_\sigma - 2\cd_\sigma(\sg \sR g^{\mu\nu}) \\
            &+ 2\sg \sR\left[g^{\mu\nu}T\TI{^\lambda_{\lambda\nu}} - g^{\nu\rho}T\TI{^\lambda_{\lambda\rho}}\delta^\mu_\sigma + g^{\nu\lambda}T\TI{^\mu_{\sigma\lambda}}\right]=0.
        \end{aligned}
    \end{equation}
\noindent$\bullet$ $\Lagr=\sR^{\mu\nu}\sR_{\mu\nu}$
    \begin{equation}
        \sR\TI{_\alpha^\lambda}\sR_{\beta\lambda} + \sR\TI{^\lambda_\alpha}\sR_{\lambda\beta} - \half g_{\alpha\beta}\sR^{\mu\nu}\sR_{\mu\nu} = 0.
    \end{equation}
    \begin{equation}
        \begin{aligned}
            &2\cd_\lambda(\sg\sR^{\mu\lambda})\delta^\nu_\sigma - 2\cd_\sigma(\sg \sR^{\mu\nu})\\ 
            &+ 2\sg\left[ \sR^{\mu\nu}T\TI{^\lambda_{\lambda\sigma}} - \sR^{\mu\rho}T\TI{^\lambda_{\lambda\rho}}\delta^\nu_\sigma + \sR^{\mu\lambda}T\TI{^\nu_{\sigma\lambda}} \right] = 0.
        \end{aligned}
    \end{equation}
    %
\noindent$\bullet$ $\Lagr=\sR^{\mu\nu}\sR_{\nu\mu}$
    \begin{equation}
        \sR\TI{_\alpha^\lambda}\sR_{\lambda\beta} + \sR\TI{^\lambda_\alpha}\sR_{\beta\lambda} - \half g_{\alpha\beta}\sR^{\mu\nu}\sR_{\nu\mu} = 0.
    \end{equation}
    \begin{equation}
        \begin{aligned}
            &2\cd_\lambda(\sg \sR^{\lambda\mu})\delta^\nu_\sigma - 2\cd_\sigma(\sg \sR^{\nu\mu})\\
            &+2\sg \left[ \sR^{\nu\mu}T\TI{^\lambda_{\lambda\sigma}} - \sR^{\rho\mu}T\TI{^\lambda_{\lambda\rho}}\delta^\nu_\sigma + \sR^{\lambda\mu}T\TI{^\nu_{\sigma\lambda}} \right] = 0.
        \end{aligned}
    \end{equation}
    %
\noindent$\bullet$ $\Lagr=\sRc^{\mu\nu}\sRc_{\mu\nu}$
    \begin{equation}
        2\sR_{\mu(\alpha|\nu|\beta)}\sRc^{\mu\nu} + \sRc_{\mu\alpha}\sRc\TI{^\mu_\beta} - \sRc_{\alpha\nu}\sRc\TI{_\beta^\nu} - \half g_{\alpha\beta}\sRc^{\mu\nu}\sRc_{\mu\nu}= 0.
    \end{equation}
    \begin{equation}
        \begin{aligned}
            &2\cd_\lambda(\sg g^{\mu\lambda}\sRc\TI{_\sigma^\nu})-2\cd_\lambda(\sg g^{\mu\nu}\sRc\TI{_\sigma^\lambda})\\
            &+2\sg\left[ g^{\mu\nu}\sRc\TI{_\sigma^\lambda}T\TI{^\rho_{\rho\lambda}} - g^{\mu\lambda}\sRc\TI{_\sigma^\nu}T\TI{^\rho_{\rho\lambda}} + g^{\mu\lambda}\sRc\TI{_\sigma^\rho}T\TI{^\nu_{\rho\lambda}} \right] = 0.
        \end{aligned}
    \end{equation}
\noindent$\bullet$ $\Lagr=\sRc^{\mu\nu}\sRc_{\nu\mu}$
    \begin{equation}
        2\sR_{\mu(\alpha|\nu|\beta)}\sRc^{\nu\mu} -\half g_{\alpha\beta}\sRc^{\mu\nu}\sRc_{\nu\mu} = 0.
    \end{equation}
    \begin{equation}
        \begin{aligned}
            &2\cd_\lambda(\sg g^{\mu\lambda}\sRc\TI{^\nu_\sigma})-2\cd_\lambda(\sg g^{\mu\nu}\sRc\TI{^\lambda_\sigma})\\
            &+2\sg\left[ g^{\mu\nu}\sRc\TI{^\lambda_\sigma}T\TI{^\rho_{\rho\lambda}} - g^{\mu\lambda}\sRc\TI{^\nu_\sigma}T\TI{^\rho_{\rho\lambda}} + g^{\mu\lambda}\sRc\TI{^\rho_\sigma}T\TI{^\nu_{\rho\lambda}} \right] = 0.
        \end{aligned}
    \end{equation}
\noindent$\bullet$ $\Lagr=\sR_{\mu\nu}\sRc^{\mu\nu}$
    \begin{equation}
        \sR_{\mu(\alpha|\nu|\beta)}\sR^{\mu\nu} + \half \sR_{\mu\alpha}\sRc\TI{^\mu_\beta} + \half\sR_{\mu\beta}\sRc\TI{^\mu_\alpha} - \half g_{\alpha\beta}\sR_{\mu\nu}\sRc^{\mu\nu} = 0.
    \end{equation}
    \begin{equation}
        \begin{aligned}
            &\cd_\lambda(\sg \sRc^{\mu\lambda})\delta^\nu_\sigma - \cd_\sigma(\sg \sRc^{\mu\nu})\\ 
            &+\cd_\lambda(\sg g^{\mu\lambda} \sR\TI{_\sigma^\nu}) - \cd_\lambda(\sg g^{\mu\nu}\sR\TI{_\sigma^\lambda})\\
            &+2\sg\left[\sRc^{\mu\nu}T\TI{^\lambda_{\lambda\sigma}} - \sRc^{\nu\lambda}T\TI{^\rho_{\rho\lambda}}\delta^\mu_\sigma + \sRc^{\nu\lambda}T\TI{^\mu_{\sigma\lambda}}\right]\\ 
            &+ 2\sg\left[g^{\mu\nu}\sR\TI{_\sigma^\lambda}T\TI{^\rho_{\rho\lambda}} - g^{\mu\lambda}\sR\TI{_\sigma^\nu}T\TI{^\rho_{\rho\lambda}} + g^{\mu\rho}\sR\TI{_\sigma^\lambda}T\TI{^\nu_{\lambda\rho}}\right] = 0.
        \end{aligned}
    \end{equation}
\noindent$\bullet$ $\Lagr=\sR_{\mu\nu}\sRc^{\nu\mu}$
    \begin{equation}
        \sR_{\mu(\alpha|\nu|\beta)}\sR^{\nu\mu} + \half \sR_{\beta\nu}\sRc\TI{^\nu_\alpha} + \half\sR_{\alpha\nu}\sRc\TI{^\nu_\beta} - \half g_{\alpha\beta}\sR_{\mu\nu}\sRc^{\nu\mu} = 0.
    \end{equation}
    \begin{equation}
        \begin{aligned}
            &\cd_\lambda(\sg \sRc^{\lambda\mu})\delta^\nu_\sigma - \cd_\sigma(\sg \sRc^{\nu\mu})\\ 
            &+\cd_\lambda(\sg g^{\mu\lambda} \sR\TI{^\nu_\sigma}) - \cd_\lambda(\sg g^{\mu\nu}\sR\TI{^\lambda_\sigma})\\
            &+2\sg\left[\sRc^{\nu\mu}T\TI{^\lambda_{\lambda\sigma}} - \sRc^{\lambda\nu}T\TI{^\rho_{\rho\lambda}}\delta^\mu_\sigma + \sRc^{\lambda\nu}T\TI{^\mu_{\sigma\lambda}}\right]\\ 
            &+ 2\sg\left[g^{\mu\nu}\sR\TI{^\lambda_\sigma}T\TI{^\rho_{\rho\lambda}} - g^{\mu\lambda}\sR\TI{^\nu_\sigma}T\TI{^\rho_{\rho\lambda}} + g^{\mu\rho}\sR\TI{^\lambda_\sigma}T\TI{^\nu_{\lambda\rho}}\right] = 0.
        \end{aligned}
    \end{equation}
\noindent$\bullet$ $\Lagr=\sR^\prime_{\mu\nu}\sR^{\prime\mu\nu}$
    \begin{equation}
        2\sR^\prime_{\alpha\mu}\sR\TI{_\beta^\mu} - \half g_{\alpha\beta}\sR^\prime_{\mu\nu}\sR^{\prime\mu\nu}=0.
    \end{equation}
    \begin{equation}
        \begin{aligned}
            \left[4\cd_\lambda(\sg \sR^{\prime\nu\lambda}) + 4\sg\sR^{\prime\lambda\nu}T\TI{^\rho_{\rho\lambda}} + 2\sg\sR^{\prime\lambda\rho}T\TI{^\nu_{\lambda\rho}}\right]\delta^\mu_\sigma = 0.
        \end{aligned}
    \end{equation}
\noindent$\bullet$ $\Lagr=\sR_{[\mu\nu]}\sR^{\prime\mu\nu}$
    \begin{equation}
        \sR_{[\beta\nu]}\sR\TI{^\prime_\alpha^\nu} + \sR_{[\alpha\nu]}\sR\TI{^\prime_\beta^\nu}  -\half g_{\alpha\beta}\sR_{[\mu\nu]}\sR^{\prime\mu\nu} = 0.
    \end{equation}
    \begin{equation}
        \begin{aligned}
            &\left[4\cd_\lambda(\sg\sR^{[\nu\lambda]})+4\sg T\TI{^\rho_{\rho\lambda}}\sR^{[\lambda\nu]}+2\sg\sR^{[\lambda\rho]}T\TI{^\nu_{\lambda\rho}}\right]\delta^\mu_\sigma \\
            &+\half \left[ \cd_\sigma(\sg \sR^{\prime\nu\mu})-\cd_\lambda(\sg \sR^{\prime\mu\lambda})\delta^\nu_\sigma + \sg\sR^{\prime\mu\lambda}T\TI{^\rho_{\rho\lambda}}\delta^\nu_\sigma \right.\\
            &\left. - \sg \sR^{\prime\nu\mu} T\TI{^\rho_{\rho\sigma}} + \sg T\TI{^\nu_{\sigma\lambda}}\sR^{\prime\mu\lambda} \right] = 0.
        \end{aligned}
    \end{equation}
\noindent$\bullet$ $\Lagr=\sRc_{[\mu\nu]}\sR^{\prime\mu\nu}$
    \begin{equation}
        \sR_{\mu(\alpha|\nu|\beta)}\sR^{\prime\mu\nu} + \sRc_{[\mu\beta]}\sR\TI{^\prime^\mu_\alpha} -\half g_{\alpha\beta}\sRc_{[\mu\nu]}\sR^{\prime\mu\nu} = 0.
    \end{equation}
    \begin{equation}
        \begin{aligned}
            &\left[ 4\cd_\lambda(\sg \sRc^{[\nu\lambda]}) + 4\sg\sRc^{[\lambda\nu]}T\TI{^\rho_{\rho\lambda}} + \sg \sRc^{[\lambda\rho]}T\TI{^\nu_{\lambda\rho}} \right]\delta^\mu_\sigma\\
            &\cd_\lambda(\sg g^{\mu\lambda}\sR\TI{^\prime_\sigma^\nu}) - \cd_\lambda(\sg g^{\mu\nu}\sR\TI{^\prime_\sigma^\nu}) + \sg \left[ g^{\mu\nu}\sR\TI{^\prime_\sigma^\lambda}T\TI{^\rho_{\rho\lambda}}\right.\\
            &\left. -g^{\mu\lambda}\sR\TI{^\prime_\sigma^\nu}T\TI{^\rho_{\rho\lambda}} + g^{\mu\lambda}\sR\TI{^\prime_\sigma^\rho}T\TI{^\nu_{\rho\lambda}} \right]=0.
        \end{aligned}
    \end{equation}
\\
\newpage
\noindent$\bullet$ $\Lagr=\sR_{\mu\nu\rho\lambda}\sR^{\mu\nu\rho\lambda}$
    \begin{equation}
       \sR_{\mu\alpha\rho\lambda}\sR\TI{^\mu_\beta^{\rho\lambda}} + 2\sR_{\mu\nu\rho\alpha}\sR\TI{^{\mu\nu\rho}_\beta} - \sR_{\alpha\nu\rho\lambda}\sR\TI{_\beta^{\nu\rho\lambda}} - \half g_{\alpha\beta}\sR_{\mu\nu\rho\lambda}\sR^{\mu\nu\rho\lambda}=0.
    \end{equation}
    \begin{equation}
        \begin{aligned}
            4\cd_\lambda(\sg \sR\TI{_\sigma^{\mu\nu\lambda}})+2\sg\left[ 2\sR\TI{_\sigma^{\mu\lambda\nu}}T\TI{^\rho_{\rho\lambda}} + \sR\TI{_\sigma^{\mu\lambda\rho}}T\TI{^\nu_{\lambda\rho}} \right] = 0.
        \end{aligned}
    \end{equation}
\noindent$\bullet$ $\Lagr=\sR_{\mu\nu\rho\lambda}\sR^{\nu\mu\rho\lambda}$
    \begin{equation}
        2\sR_{\mu\nu\alpha\lambda}\sR\TI{^{\nu\mu}_\beta^\lambda} - \half g_{\alpha\beta}\sR_{\mu\nu\rho\lambda}\sR^{\nu\mu\rho\lambda}=0.
    \end{equation}
    \begin{equation}
        \begin{aligned}
            4\cd_\lambda(\sg \sR\TI{^\mu_\sigma^{\nu\lambda}})+2\sg\left[ 2\sR\TI{^\mu_\sigma^{\lambda\nu}}T\TI{^\rho_{\rho\lambda}} + \sR\TI{^\mu_\sigma^{\lambda\rho}}T\TI{^\nu_{\lambda\rho}} \right] = 0.
        \end{aligned}
    \end{equation}
    \noindent$\bullet$ $\Lagr=\sR_{\mu\nu\rho\lambda}\sR^{\rho\lambda\mu\nu}$
    \begin{equation}
        \sR\TI{^{\mu\nu}_{\rho\alpha}}\sR\TI{^\rho_{\beta\mu\nu}} + \sR\TI{^{\mu\nu}_{\rho\beta}}\sR\TI{^\rho_{\alpha\mu\nu}} + - \half g_{\alpha\beta}\sR_{\mu\nu\rho\lambda}\sR^{\rho\lambda\mu\nu}=0.
    \end{equation}
    \begin{equation}
        \begin{aligned}
            &2\cd_\lambda(\sg \sR\TI{^\nu^\lambda_\sigma^\mu}) - 2\cd_\lambda(\sg\sR\TI{^\lambda^\nu_\sigma^\mu})\\
            &+2\sg\left[ \sR\TI{^\lambda^\nu_\sigma^\mu}T\TI{^\rho_{\rho\lambda}} -\sR\TI{^\nu^\lambda_\sigma^\mu}T\TI{^\rho_{\rho\lambda}} + \sR\TI{^\lambda^\rho_\sigma^\mu}T\TI{^\nu_{\lambda\rho}} \right] = 0.
        \end{aligned}
    \end{equation}
\noindent$\bullet$ $\Lagr=\sR_{\mu\nu\rho\lambda}\sR^{\rho\nu\mu\lambda}$
    \begin{equation}
        \sR\TI{^\mu_\beta_\nu^\lambda}\sR\TI{^\nu_{\alpha\mu\lambda}} + \sR\TI{^{\mu\nu}_{\rho\beta}}\sR\TI{^\rho_{\nu\mu\alpha}} + -\half g_{\alpha\beta}\sR_{\mu\nu\rho\lambda}\sR^{\rho\nu\mu\lambda}=0.
    \end{equation}
    \begin{equation}
        \begin{aligned}
            &2\cd_\lambda(\sg\sR\TI{^\nu^\mu_\sigma^\lambda})-2\cd_\lambda(\sg\sR\TI{^\lambda^\mu_\sigma^\nu})\\
            &+2\sg\left[ \sR\TI{^\lambda^\mu_\sigma^\nu}T\TI{^\rho_{\rho\lambda}} - \sR\TI{^\nu^\mu_\sigma^\lambda}T\TI{^\rho_{\rho\lambda}} + \sR\TI{^\rho^\mu_\sigma^\lambda}T\TI{^\nu_{\lambda\rho}} \right]=0.
        \end{aligned}
    \end{equation}
\noindent$\bullet$ $\Lagr=\sR_{\mu\nu\rho\lambda}\sR^{\nu\rho\mu\lambda}$
    \begin{equation}
        \sR\TI{^\mu_{\nu\beta}^\lambda}\sR\TI{^\nu_{\alpha\mu\lambda}}+ \sR\TI{^\mu_\nu^\lambda_\beta}\sR\TI{^\nu_{\lambda\mu\alpha}} -\half g_{\alpha\beta}\sR_{\mu\nu\rho\lambda}\sR^{\nu\rho\mu\lambda}=0.
    \end{equation}
    \begin{equation}
        \begin{aligned}
            &\cd_\lambda(\sg \sR\TI{^\nu_\sigma^\mu^\lambda})-\cd_\lambda(\sg \sR\TI{^\lambda_\sigma^{\mu\nu}})\\
            &+\cd_\lambda(\sg\sR\TI{^{\nu\mu}_\sigma^\lambda}) - \cd_\lambda(\sg \sR\TI{^{\mu\lambda}_\sigma^\nu})\\
            &+\sg\left[ \sR\TI{^\lambda_\sigma^{\mu\nu}}T\TI{^\rho_{\rho\lambda}} - \sR\TI{^\nu_\sigma^{\mu\lambda}}T\TI{^\rho_{\rho\lambda}}+\sR\TI{^\lambda_\sigma^{\mu\rho}}T\TI{^\nu_{\lambda\rho}} \right]\\
            &+\sg\left[ \sR\TI{^{\mu\lambda}_\sigma^\nu}T\TI{^\rho_{\rho\lambda}} - \sR\TI{^\nu^\mu_\sigma^\lambda}T\TI{^\rho_{\rho\lambda}}+\sR\TI{^\lambda^\mu_\sigma^{\rho}}T\TI{^\nu_{\lambda\rho}} \right] = 0.
        \end{aligned}
    \end{equation}
\noindent$\bullet$ $\Lagr=\sR_{\mu\nu\rho\lambda}\sR^{\mu\rho\nu\lambda}$
    \begin{equation}
        2\sR_{\mu\nu(\beta|\lambda}\sR\TI{^\mu_{\alpha)}^{\nu\lambda}} + \sR_{\lambda\mu\nu\beta}\sR\TI{^{\lambda\nu\mu}_\alpha} - \sR_{\beta\mu\nu\lambda}\sR\TI{_\alpha^{\nu\mu\lambda}} -\half g_{\alpha\beta}\sR_{\mu\nu\rho\lambda}\sR^{\mu\rho\nu\lambda}=0.
    \end{equation}
    \begin{equation}
        \begin{aligned}
            &2\cd_\lambda(\sg \sR\TI{_\sigma^{\nu\mu\lambda}})-2\cd_\lambda(\sg \sR\TI{_\sigma^{\lambda\mu\nu}})\\
            &+2\sg\left[ \sR\TI{_\sigma^{\lambda\mu\nu}}T\TI{^\rho_{\rho\lambda}} - \sR\TI{_\sigma^{\nu\mu\lambda}}T\TI{^\rho_{\rho\lambda}} + \sR\TI{_\sigma^{\lambda\mu\rho}}T\TI{^\nu_{\lambda\rho}} \right] = 0.
        \end{aligned}
    \end{equation}

\chapter{Details of the series solution}
\section{Variation of transformed action}\label{app:SigmaVariation}
        Here the variation of the Legendre transformed action \eqref{eq:legendre2}, where the auxillary field $\Sigma_{\mu\nu}$ is written in terms of $q^{\mu\nu}$ and $g^{\mu\nu}$ is performed with respect to the metric $g_{\mu\nu}$.

        The part of the Lagrangian density depending on $g_{\mu\nu}$ is
        \begin{equation}
            \sg\Lagr_g\equiv\mathfrak{L}_g = -\sg\left[ \half\frac{\pd F}{\pd \Sigma_{\mu\nu}}\Sigma_{\mu\nu}(q,g)-\half F(q,g)-\Lagr_m(g,\chi) \right].
        \end{equation}
        Written out in terms of $\Sigma_{\mu\nu}$
        \begin{equation}
            \mathfrak{L}_g =-\sg\left[\half\beta \Sigma^2 +\half\alpha \Sigma_{\lambda\sigma}\Sigma^{\lambda\sigma} + U(\chi) + \half g^{\lambda\sigma}\pd_\lambda \chi \pd_\sigma \chi \right],
        \end{equation}
        where $\Sigma \equiv g^{\mu\nu}\Sigma_{\mu\nu}$.
        The variation is 
        \begin{equation}\label{eq:lagSigmavar}
            \begin{aligned}
                \delta\mathfrak{L}_g =&-\sg\left[ \beta \Sigma \delta\Sigma + \alpha g^{\lambda\gamma}\Sigma_{\lambda\sigma}\delta(g^{\rho\sigma}\Sigma_{\gamma\rho}) + \half\pd_\lambda \chi \pd_\sigma \chi \delta g^{\lambda\sigma} \right]\\
                &+\half\sg g^{\mu\nu} \Lagr_g \delta g_{\mu\nu}
            \end{aligned}
        \end{equation}
        First looking at $\delta\Sigma$ separately
        \begin{equation}
            \delta\Sigma = \delta \left( \frac{-4}{2\alpha+8\beta}+\frac{1}{2\alpha+8\beta}\frac{\sq}{\sg}q^{\sigma\lambda}g_{\sigma\lambda} \right),
        \end{equation}
        with the identity $\delta \sg = 1/2 \sg g^{\mu\nu}\delta g_{\mu\nu}$ we have
        \begin{equation}
            \delta\Sigma = \frac{1}{2\alpha+8\beta}\frac{\sq}{\sg}\left( q^{\mu\nu} -\half q^{\sigma\lambda}g_{\sigma\lambda}g^{\mu\nu}\right)\delta g_{\mu\nu}.    
        \end{equation}
        Now separately looking at $\delta\Sigma_{\mu\nu}$ 
        \begin{equation}
            \delta\Sigma_{\mu\nu} =  \delta\left[\frac{1}{2\alpha}\frac{\sq}{\sq}q^{\sigma\lambda}g_{\sigma\mu}g_{\lambda\nu} - \frac{1}{2\alpha+8\beta}\left(1+\frac{\beta}{\alpha}\frac{\sq}{\sq}q^{\sigma\lambda}g_{\sigma\lambda}\right)g_{\mu\nu}  \right],
        \end{equation}
        which becomes
        \begin{equation}
            \begin{aligned}
                \delta\Sigma_{\mu\nu} =& -\frac{\delta g_{\mu\nu}}{2\alpha+8\beta}+\frac{\sq}{\sg}\left[ -\frac{1}{4\alpha}q^{\sigma\lambda}g_{\sigma\mu}g_{\lambda\nu}g^{\rho\gamma} \delta g_{\rho\gamma}+\frac{1}{\alpha}q^{\sigma\lambda}g_{\sigma\mu}\delta g_{\lambda\nu}\right.\\
                &\left.-\frac{\beta}{\alpha(2\alpha+8\beta)}\left( q^{\lambda\sigma}g_{\lambda\sigma}\delta g_{\mu\nu}-\half q^{\sigma\lambda}g_{\sigma\lambda}g_{\mu\nu}g^{\rho\gamma}\delta g_{\rho\gamma} + q^{\sigma\lambda}g_{\mu\nu}\delta g_{\sigma\lambda}\right) \right].
            \end{aligned}
        \end{equation}
        With these out of the way it is just a question of plugging these in to \eqref{eq:lagSigmavar} and simplifying the expression. In the end we get
        \begin{equation}
        \begin{aligned}
            \delta\mathfrak{L}_g &=\sg\left\{\vphantom{\half} \right. - \frac{1}{4(\alpha+4\beta)}\frac{\sq}{\sg}q^{\sigma\lambda}g_{\sigma\mu}g_{\lambda\nu} \\
            &+ \frac{1}{4\alpha}\frac{q}{g} \left[ q^{\sigma\lambda}q^{\rho\delta}g_{\lambda\delta}g_{\rho\nu}g_{\sigma\mu}-\frac{\beta}{(\alpha+4\beta)}q^{\delta\rho}g_{\delta\rho}q^{\sigma\lambda}g_{\sigma\mu}g_{\lambda\nu}\right]\\
            &+\frac{1}{2}g_{\mu\nu}\left[\frac{1}{\alpha+4\beta}\left(\frac{\beta}{8\alpha}\frac{q}{g}q^{\lambda\sigma}g_{\lambda\sigma}q^{\rho\delta}g_{\rho\delta} + \half \right) -\frac{q}{g}\frac{1}{8\alpha} q^{\lambda\sigma}q^{\delta\rho}g_{\lambda\delta}g_{\sigma\rho}\right]\\
            &-\half \pd_\mu \chi \pd_\nu \chi + \half g_{\mu\nu}\left[U(\chi) + \half g^{\lambda\sigma}\pd_\lambda \chi \pd_\sigma \chi\right]\left. \vphantom{\half} \right\}\delta g^{\mu\nu}.
        \end{aligned}
    \end{equation}

    \section{Series solution of the modified scalar sector}\label{app:coeffs}
    Here a few coefficients for the approximate solution \eqref{eq:seriesAnz} of the disformal transformation \eqref{eq:metricAnsatz} parameters and the scalar sector with a few more terms are presented.
    \begin{equation}
        \begin{aligned}
            &a_0 = \frac{1}{1 + (2\alpha + 8 \beta) U(\chi)}\\
            &b_0 = \frac{2\alpha }{\left[1 + (2\alpha  + 8 \beta) U(\chi)\right] \left[1 + (4 \alpha + 
            8 \beta) U(\chi)\right]}\\
            &a_1=-\frac{\alpha + 2 \beta}{1 + (4 \alpha + 8 \beta) U}\\
            &b_1 = \frac{4\alpha \left[-2 (\alpha + \beta) - (3 \alpha + 
            4 \beta) (3 \alpha + 8 \beta) U - 
         8 (\alpha + 2 \beta)^2 (\alpha + 4 \beta) U^2\right]}{
          (1 + 2\alpha U + 8 \beta U) (1 + 4 \alpha U + 
            8 \beta U)^3}\\
            &a_2= \frac{\alpha \left[7 \alpha + 10 \beta + 4 (3 \alpha + 5 \beta) (3 \alpha + 8 \beta) U + 40 (\alpha + 2 \beta)^2 (\alpha + 4 \beta) U^2\right] }{ 2 (1 + 2\alpha U + 8 \beta U) (1 + 
            4 \alpha U + 8 \beta U)^3}\\
            &b_2=\frac{2\alpha}{ (1 + 2\alpha U + 8 \beta U) (1 + 4 \alpha U + 8 \beta U)^5}\\ 
            &\times\left[\vphantom{\half}\right.  64 \alpha^6 U^4 + 4 \beta^2 (1 + 8 \beta U)^4 + 128 \alpha^5 U^3 (3 + 8 \beta U) \\
            &+ 5 \alpha \beta (1 + 8 \beta U)^3 (5 + 16 \beta U) + 2\alpha^3 U (1 + 8 \beta U) (85 + 128 \beta U (7 + 11 \beta U)) \\
            &+ 4\alpha^4 U^2 (111 + 
               8 \beta U (131 + 208 \beta U)) + (\alpha + 
               8 \alpha \beta U)^2 (21 + 2\beta U (173 + 
                  328 \beta U))  \left.\vphantom{\half}\right]\\
        \end{aligned}
    \end{equation}
    The scalar sector with few more terms
    \begin{equation}
        \begin{aligned}
            \Lagr_m =& -\frac{U}{
                1 + (2\alpha + 8 \beta) U}
                -\frac{\half X_q}{1 + (2\alpha + 8 \beta) U} 
                \\
                &+ \frac{\alpha^2 U + 
                \beta (1 + 8 \beta U) + \alpha (1 + 6 \beta U)}{2 \left[1 + (2\alpha + 8 \beta) U\right] \left[1 + (4 \alpha + 8 \beta) U\right]}X_q^2
                        \\
                        &- \frac{
                             \alpha \left[12 \alpha^3 U^2 + 
               3 \beta (1 + 8 \beta U)^2 + 2\alpha^2 U (7 + 
               48 \beta U) + 
               3 \alpha (1 + 18 \beta U + 80 \beta^2 U^2)\right]}{
                   2 (1 + 2\alpha U + 8 \beta U) (1 + 4 \alpha U + 
                   8 \beta U)^3 }X_q^3
                   \\
                   &+ \frac{3\alpha}{8 (1 + 2\alpha U + 
                   8 \beta U) (1 + 4 \alpha U + 8 \beta U)^5}X_q^4\left[\vphantom{\half}\right. 64 \alpha^6 U^4 \dots 
                \end{aligned}
                \end{equation}

\end{appendices}

\newpage
\bibliographystyle{JHEP}
\bibliography{masters}

\providecommand{\href}[2]{#2}\begingroup\raggedright\begin{thebibliography}{10}

\bibitem{Carroll04}
S.~M. Carroll, \emph{{Spacetime and geometry: An introduction to general
  relativity}}.
\newblock 2004.

\bibitem{Martin13}
J.~Martin, C.~Ringeval and V.~Vennin, \emph{{Encyclopædia Inflationaris}},
  \href{http://dx.doi.org/10.1016/j.dark.2014.01.003}{\emph{Phys. Dark Univ.}
  {\bfseries 5-6} (2014) 75--235},
  [\href{https://arxiv.org/abs/1303.3787}{{\ttfamily 1303.3787}}].

\bibitem{Ferraris82}
M.~{Ferraris}, M.~{Francaviglia} and C.~{Reina}, \emph{{Variational formulation
  of general relativity from 1915 to 1925 ``Palatini's method'' discovered by
  Einstein in 1925}}, \href{http://dx.doi.org/10.1007/BF00756060}{\emph{General
  Relativity and Gravitation} {\bfseries 14} (Mar., 1982) 243--254}.

\bibitem{HawGib77}
G.~W. Gibbons and S.~W. Hawking, \emph{Action integrals and partition functions
  in quantum gravity},
  \href{http://dx.doi.org/10.1103/PhysRevD.15.2752}{\emph{Phys. Rev. D}
  {\bfseries 15} (May, 1977) 2752--2756}.

\bibitem{York72}
J.~W. York, \emph{Role of conformal three-geometry in the dynamics of
  gravitation},
  \href{http://dx.doi.org/10.1103/PhysRevLett.28.1082}{\emph{Phys. Rev. Lett.}
  {\bfseries 28} (Apr, 1972) 1082--1085}.

\bibitem{Hawking73}
S.~W. Hawking and G.~F.~R. Ellis, \emph{{The Large Scale Structure of
  Space-Time}}.
\newblock Cambridge Monographs on Mathematical Physics. Cambridge University
  Press, 2011,
  \href{http://dx.doi.org/10.1017/CBO9780511524646}{10.1017/CBO9780511524646}.

\bibitem{GCT10}
A.~Guarnizo, L.~Castaneda and J.~M. Tejeiro, \emph{{Boundary Term in Metric
  f(R) Gravity: Field Equations in the Metric Formalism}},
  \href{http://dx.doi.org/10.1007/s10714-010-1012-6}{\emph{Gen. Rel. Grav.}
  {\bfseries 42} (2010) 2713--2728},
  [\href{https://arxiv.org/abs/1002.0617}{{\ttfamily 1002.0617}}].

\bibitem{WaldGR}
R.~M. Wald, \emph{{General Relativity}}.
\newblock Chicago Univ. Pr., Chicago, USA, 1984,
  \href{http://dx.doi.org/10.7208/chicago/9780226870373.001.0001}{10.7208/chicago/9780226870373.001.0001}.

\bibitem{Dyer08}
E.~Dyer and K.~Hinterbichler, \emph{{Boundary Terms, Variational Principles and
  Higher Derivative Modified Gravity}},
  \href{http://dx.doi.org/10.1103/PhysRevD.79.024028}{\emph{Phys. Rev.}
  {\bfseries D79} (2009) 024028},
  [\href{https://arxiv.org/abs/0809.4033}{{\ttfamily 0809.4033}}].

\bibitem{Burton97}
H.~Burton and R.~B. Mann, \emph{{Palatini variational principle for an extended
  Einstein-Hilbert action}},
  \href{http://dx.doi.org/10.1103/PhysRevD.57.4754}{\emph{Phys. Rev.}
  {\bfseries D57} (1998) 4754--4759},
  [\href{https://arxiv.org/abs/gr-qc/9711003}{{\ttfamily gr-qc/9711003}}].

\bibitem{Vitagliano10}
V.~Vitagliano, T.~P. Sotiriou and S.~Liberati, \emph{{The dynamics of
  metric-affine gravity}}, \href{http://dx.doi.org/10.1016/j.aop.2011.02.008,
  10.1016/j.aop.2012.11.002}{\emph{Annals Phys.} {\bfseries 326} (2011)
  1259--1273}, [\href{https://arxiv.org/abs/1008.0171}{{\ttfamily 1008.0171}}].

\bibitem{Schouten54}
J.~A. Schouten, \emph{Ricci-Calculus}.
\newblock Springer-Verlag Berlin Heidelberg, 1954,
  \href{http://dx.doi.org/10.1007/978-3-662-12927-2}{10.1007/978-3-662-12927-2}.

\bibitem{Hehl76}
F.~W. Hehl, P.~von~der Heyde, G.~D. Kerlick and J.~M. Nester, \emph{General
  relativity with spin and torsion: Foundations and prospects},
  \href{http://dx.doi.org/10.1103/RevModPhys.48.393}{\emph{Rev. Mod. Phys.}
  {\bfseries 48} (Jul, 1976) 393--416}.

\bibitem{NJ10}
N.~Dadhich and J.~M. Pons, \emph{{On the equivalence of the Einstein-Hilbert
  and the Einstein-Palatini formulations of general relativity for an arbitrary
  connection}}, \href{http://dx.doi.org/10.1007/s10714-012-1393-9}{\emph{Gen.
  Rel. Grav.} {\bfseries 44} (2012) 2337--2352},
  [\href{https://arxiv.org/abs/1010.0869}{{\ttfamily 1010.0869}}].

\bibitem{hehl78}
F.~W. {Hehl} and G.~D. {Kerlick}, \emph{{Metric-affine variational principles
  in general relativity. I - Riemannian space-time}},
  \href{http://dx.doi.org/10.1007/BF00760141}{\emph{General Relativity and
  Gravitation} {\bfseries 9} (Aug., 1978) 691--710}.

\bibitem{Bernal16}
A.~N. Bernal, B.~Janssen, A.~Jimenez-Cano, J.~A. Orejuela, M.~Sanchez and
  P.~Sanchez-Moreno, \emph{{On the (non-)uniqueness of the Levi-Civita solution
  in the Einstein–Hilbert–Palatini formalism}},
  \href{http://dx.doi.org/10.1016/j.physletb.2017.03.001}{\emph{Phys. Lett.}
  {\bfseries B768} (2017) 280--287},
  [\href{https://arxiv.org/abs/1606.08756}{{\ttfamily 1606.08756}}].

\bibitem{SL07}
T.~P. Sotiriou and S.~Liberati, \emph{{Metric-affine f(R) theories of
  gravity}}, \href{http://dx.doi.org/10.1016/j.aop.2006.06.002}{\emph{Annals
  Phys.} {\bfseries 322} (2007) 935--966},
  [\href{https://arxiv.org/abs/gr-qc/0604006}{{\ttfamily gr-qc/0604006}}].

\bibitem{schrodinger85}
E.~Schrödinger, \emph{Space-Time Structure}.
\newblock Cambridge Science Classics. Cambridge University Press, 1985,
  \href{http://dx.doi.org/10.1017/CBO9780511586446}{10.1017/CBO9780511586446}.

\bibitem{SF10}
T.~P. Sotiriou and V.~Faraoni, \emph{{f(R) Theories Of Gravity}},
  \href{http://dx.doi.org/10.1103/RevModPhys.82.451}{\emph{Rev. Mod. Phys.}
  {\bfseries 82} (2010) 451--497},
  [\href{https://arxiv.org/abs/0805.1726}{{\ttfamily 0805.1726}}].

\bibitem{Lyth09}
D.~H. Lyth and A.~R. Liddle, \emph{{The primordial density perturbation:
  Cosmology, inflation and the origin of structure}}.
\newblock 2009.

\bibitem{Baumann09}
D.~Baumann, \emph{{Inflation}},  in \emph{{Physics of the large and the small,
  TASI 09, proceedings of the Theoretical Advanced Study Institute in
  Elementary Particle Physics, Boulder, Colorado, USA, 1-26 June 2009}},
  pp.~523--686, 2011, \href{https://arxiv.org/abs/0907.5424}{{\ttfamily
  0907.5424}}, \href{http://dx.doi.org/10.1142/9789814327183_0010}{DOI}.

\bibitem{Weinberg08}
S.~Weinberg, \emph{{Cosmology}}.
\newblock 2008.

\bibitem{BirrellDavies}
N.~D. Birrell and P.~C.~W. Davies, \emph{{Quantum Fields in Curved Space}}.
\newblock Cambridge Monographs on Mathematical Physics. Cambridge Univ. Press,
  Cambridge, UK, 1984,
  \href{http://dx.doi.org/10.1017/CBO9780511622632}{10.1017/CBO9780511622632}.

\bibitem{Lyth04}
D.~H. Lyth, K.~A. Malik and M.~Sasaki, \emph{{A General proof of the
  conservation of the curvature perturbation}},
  \href{http://dx.doi.org/10.1088/1475-7516/2005/05/004}{\emph{JCAP} {\bfseries
  0505} (2005) 004}, [\href{https://arxiv.org/abs/astro-ph/0411220}{{\ttfamily
  astro-ph/0411220}}].

\bibitem{Albrecht92}
A.~Albrecht, P.~Ferreira, M.~Joyce and T.~Prokopec, \emph{{Inflation and
  squeezed quantum states}},
  \href{http://dx.doi.org/10.1103/PhysRevD.50.4807}{\emph{Phys. Rev.}
  {\bfseries D50} (1994) 4807--4820},
  [\href{https://arxiv.org/abs/astro-ph/9303001}{{\ttfamily
  astro-ph/9303001}}].

\bibitem{Mukhanov88}
V.~F. Mukhanov, \emph{{Quantum Theory of Gauge Invariant Cosmological
  Perturbations}}, {\emph{Sov. Phys. JETP} {\bfseries 67} (1988) 1297--1302}.

\bibitem{Isidori07}
G.~Isidori, V.~S. Rychkov, A.~Strumia and N.~Tetradis, \emph{{Gravitational
  corrections to standard model vacuum decay}},
  \href{http://dx.doi.org/10.1103/PhysRevD.77.025034}{\emph{Phys. Rev.}
  {\bfseries D77} (2008) 025034},
  [\href{https://arxiv.org/abs/0712.0242}{{\ttfamily 0712.0242}}].

\bibitem{Hamada13}
Y.~Hamada, H.~Kawai and K.-y. Oda, \emph{{Minimal Higgs inflation}},
  \href{http://dx.doi.org/10.1093/ptep/ptt116}{\emph{PTEP} {\bfseries 2014}
  (2014) 023B02}, [\href{https://arxiv.org/abs/1308.6651}{{\ttfamily
  1308.6651}}].

\bibitem{Fairbairn14}
M.~Fairbairn, P.~Grothaus and R.~Hogan, \emph{{The Problem with False Vacuum
  Higgs Inflation}},
  \href{http://dx.doi.org/10.1088/1475-7516/2014/06/039}{\emph{JCAP} {\bfseries
  1406} (2014) 039}, [\href{https://arxiv.org/abs/1403.7483}{{\ttfamily
  1403.7483}}].

\bibitem{Flanagan04}
E.~E. Flanagan, \emph{{The Conformal frame freedom in theories of
  gravitation}}, \href{http://dx.doi.org/10.10880264-93812115N02}{\emph{Class.
  Quant. Grav.} {\bfseries 21} (2004) 3817},
  [\href{https://arxiv.org/abs/gr-qc0403063}{{\ttfamily gr-qc0403063}}].

\bibitem{Tsujikawa04}
S.~Tsujikawa and B.~Gumjudpai, \emph{{Density perturbations in generalized
  Einstein scenarios and constraints on nonminimal couplings from the Cosmic
  Microwave Background}},
  \href{http://dx.doi.org/10.1103/PhysRevD.69.123523}{\emph{Phys. Rev.}
  {\bfseries D69} (2004) 123523},
  [\href{https://arxiv.org/abs/astro-ph/0402185}{{\ttfamily
  astro-ph/0402185}}].

\bibitem{Postma14}
M.~Postma and M.~Volponi, \emph{{Equivalence of the Einstein and Jordan
  frames}}, \href{http://dx.doi.org/10.1103/PhysRevD.90.103516}{\emph{Phys.
  Rev.} {\bfseries D90} (2014) 103516},
  [\href{https://arxiv.org/abs/1407.6874}{{\ttfamily 1407.6874}}].

\bibitem{Rubio18}
J.~Rubio, \emph{{Higgs inflation}},
  \href{http://dx.doi.org/10.3389/fspas.2018.00050}{\emph{Front. Astron. Space
  Sci.} {\bfseries 5} (2019) 50},
  [\href{https://arxiv.org/abs/1807.02376}{{\ttfamily 1807.02376}}].

\bibitem{Bezrukov07}
F.~L. Bezrukov and M.~Shaposhnikov, \emph{{The Standard Model Higgs boson as
  the inflaton}},
  \href{http://dx.doi.org/10.1016/j.physletb.2007.11.072}{\emph{Phys. Lett.}
  {\bfseries B659} (2008) 703--706},
  [\href{https://arxiv.org/abs/0710.3755}{{\ttfamily 0710.3755}}].

\bibitem{Bezrukov10}
F.~Bezrukov, A.~Magnin, M.~Shaposhnikov and S.~Sibiryakov, \emph{{Higgs
  inflation: consistency and generalisations}},
  \href{http://dx.doi.org/10.1007/JHEP01(2011)016}{\emph{JHEP} {\bfseries 01}
  (2011) 016}, [\href{https://arxiv.org/abs/1008.5157}{{\ttfamily 1008.5157}}].

\bibitem{GarciaBellido08}
J.~Garcia-Bellido, D.~G. Figueroa and J.~Rubio, \emph{{Preheating in the
  Standard Model with the Higgs-Inflaton coupled to gravity}},
  \href{http://dx.doi.org/10.1103/PhysRevD.79.063531}{\emph{Phys. Rev.}
  {\bfseries D79} (2009) 063531},
  [\href{https://arxiv.org/abs/0812.4624}{{\ttfamily 0812.4624}}].

\bibitem{Repond16}
J.~Repond and J.~Rubio, \emph{{Combined Preheating on the lattice with
  applications to Higgs inflation}},
  \href{http://dx.doi.org/10.1088/1475-7516/2016/07/043}{\emph{JCAP} {\bfseries
  1607} (2016) 043}, [\href{https://arxiv.org/abs/1604.08238}{{\ttfamily
  1604.08238}}].

\bibitem{Bezrukov08}
F.~Bezrukov, D.~Gorbunov and M.~Shaposhnikov, \emph{{On initial conditions for
  the Hot Big Bang}},
  \href{http://dx.doi.org/10.1088/1475-7516/2009/06/029}{\emph{JCAP} {\bfseries
  0906} (2009) 029}, [\href{https://arxiv.org/abs/0812.3622}{{\ttfamily
  0812.3622}}].

\bibitem{Akrami18}
{\scshape Planck} collaboration, Y.~Akrami et~al., \emph{{Planck 2018 results.
  X. Constraints on inflation}},
  \href{https://arxiv.org/abs/1807.06211}{{\ttfamily 1807.06211}}.

\bibitem{Atkins12}
M.~Atkins and X.~Calmet, \emph{{Bounds on the Nonminimal Coupling of the Higgs
  Boson to Gravity}},
  \href{http://dx.doi.org/10.1103/PhysRevLett.110.051301}{\emph{Phys. Rev.
  Lett.} {\bfseries 110} (2013) 051301},
  [\href{https://arxiv.org/abs/1211.0281}{{\ttfamily 1211.0281}}].

\bibitem{Schwartz14}
M.~D. Schwartz, \emph{{Quantum Field Theory and the Standard Model}}.
\newblock Cambridge University Press, 2014.

\bibitem{Barbon09}
J.~L.~F. Barbon and J.~R. Espinosa, \emph{{On the Naturalness of Higgs
  Inflation}}, \href{http://dx.doi.org/10.1103/PhysRevD.79.081302}{\emph{Phys.
  Rev.} {\bfseries D79} (2009) 081302},
  [\href{https://arxiv.org/abs/0903.0355}{{\ttfamily 0903.0355}}].

\bibitem{Bauer08}
F.~Bauer and D.~A. Demir, \emph{{Inflation with Non-Minimal Coupling: Metric
  versus Palatini Formulations}},
  \href{http://dx.doi.org/10.1016/j.physletb.2008.06.014}{\emph{Phys. Lett.}
  {\bfseries B665} (2008) 222--226},
  [\href{https://arxiv.org/abs/0803.2664}{{\ttfamily 0803.2664}}].

\bibitem{RubioEemeli19}
J.~Rubio and E.~S. Tomberg, \emph{{Preheating in Palatini Higgs inflation}},
  \href{http://dx.doi.org/10.1088/1475-7516/2019/04/021}{\emph{JCAP} {\bfseries
  1904} (2019) 021}, [\href{https://arxiv.org/abs/1902.10148}{{\ttfamily
  1902.10148}}].

\bibitem{Bauer11}
F.~Bauer and D.~A. Demir, \emph{{Higgs-Palatini Inflation and Unitarity}},
  \href{http://dx.doi.org/10.1016/j.physletb.2011.03.042}{\emph{Phys. Lett.}
  {\bfseries B698} (2011) 425--429},
  [\href{https://arxiv.org/abs/1012.2900}{{\ttfamily 1012.2900}}].

\bibitem{Clifton11}
T.~Clifton, P.~G. Ferreira, A.~Padilla and C.~Skordis, \emph{{Modified Gravity
  and Cosmology}},
  \href{http://dx.doi.org/10.1016/j.physrep.2012.01.001}{\emph{Phys. Rept.}
  {\bfseries 513} (2012) 1--189},
  [\href{https://arxiv.org/abs/1106.2476}{{\ttfamily 1106.2476}}].

\bibitem{DeWitt62}
R.~Utiyama and B.~S. DeWitt, \emph{Renormalization of a classical gravitational
  field interacting with quantized matter fields},
  \href{http://dx.doi.org/10.1063/1.1724264}{\emph{Journal of Mathematical
  Physics} {\bfseries 3} (1962) 608--618},
  [\href{https://arxiv.org/abs/https://doi.org/10.1063/1.1724264}{{\ttfamily
  https://doi.org/10.1063/1.1724264}}].

\bibitem{PnS}
M.~E. Peskin and D.~V. Schroeder, \emph{{An Introduction to quantum field
  theory}}.
\newblock Addison-Wesley, Reading, USA, 1995.

\bibitem{Stelle76}
K.~S. Stelle, \emph{{Renormalization of Higher Derivative Quantum Gravity}},
  \href{http://dx.doi.org/10.1103/PhysRevD.16.953}{\emph{Phys. Rev.} {\bfseries
  D16} (1977) 953--969}.

\bibitem{Boulware85}
D.~G. Boulware and S.~Deser, \emph{{String Generated Gravity Models}},
  \href{http://dx.doi.org/10.1103/PhysRevLett.55.2656}{\emph{Phys. Rev. Lett.}
  {\bfseries 55} (1985) 2656}.

\bibitem{Deser87}
S.~Deser, \emph{Gravity from strings},
  \href{http://dx.doi.org/10.1088/0031-8949/1987/t15/018}{\emph{Physica
  Scripta} {\bfseries T15} (jan, 1987) 138--142}.

\bibitem{Tong09}
D.~Tong, \emph{{String Theory}},
  \href{https://arxiv.org/abs/0908.0333}{{\ttfamily 0908.0333}}.

\bibitem{Woodard09}
R.~P. Woodard, \emph{{How Far Are We from the Quantum Theory of Gravity?}},
  \href{http://dx.doi.org/10.1088/0034-4885/72/12/126002}{\emph{Rept. Prog.
  Phys.} {\bfseries 72} (2009) 126002},
  [\href{https://arxiv.org/abs/0907.4238}{{\ttfamily 0907.4238}}].

\bibitem{Parker93}
L.~Parker and J.~Z. Simon, \emph{{Einstein equation with quantum corrections
  reduced to second order}},
  \href{http://dx.doi.org/10.1103/PhysRevD.47.1339}{\emph{Phys. Rev.}
  {\bfseries D47} (1993) 1339--1355},
  [\href{https://arxiv.org/abs/gr-qc/9211002}{{\ttfamily gr-qc/9211002}}].

\bibitem{BJ19}
J.~Beltrán~Jiménez and A.~Delhom, \emph{{Ghosts in metric-affine higher order
  curvature gravity}},  \href{https://arxiv.org/abs/1901.08988}{{\ttfamily
  1901.08988}}.

\bibitem{Sotiriou06}
T.~P. Sotiriou, \emph{{f(R) gravity and scalar-tensor theory}},
  \href{http://dx.doi.org/10.1088/0264-9381/23/17/003}{\emph{Class. Quant.
  Grav.} {\bfseries 23} (2006) 5117--5128},
  [\href{https://arxiv.org/abs/gr-qc/0604028}{{\ttfamily gr-qc/0604028}}].

\bibitem{Lanczos38}
C.~Lanczos, \emph{A remarkable property of the riemann-christoffel tensor in
  four dimensions}, {\emph{Annals of Mathematics} {\bfseries 39} (1938)
  842--850}.

\bibitem{Alvarez16}
E.~Alvarez and S.~Gonzalez-Martin, \emph{{Weyl Gravity Revisited}},
  \href{http://dx.doi.org/10.1088/1475-7516/2017/02/011}{\emph{JCAP} {\bfseries
  1702} (2017) 011}, [\href{https://arxiv.org/abs/1610.03539}{{\ttfamily
  1610.03539}}].

\bibitem{Salvio18}
A.~Salvio, \emph{{Quadratic Gravity}},
  \href{http://dx.doi.org/10.3389/fphy.2018.00077}{\emph{Front.in Phys.}
  {\bfseries 6} (2018) 77}, [\href{https://arxiv.org/abs/1804.09944}{{\ttfamily
  1804.09944}}].

\bibitem{Stelle78}
K.~S. Stelle, \emph{Classical gravity with higher derivatives},
  \href{http://dx.doi.org/10.1007/BF00760427}{\emph{General Relativity and
  Gravitation} {\bfseries 9} (Apr, 1978) 353--371}.

\bibitem{Lovelock69}
D.~Lovelock, \emph{The uniqueness of the einstein field equations in a
  four-dimensional space},
  \href{http://dx.doi.org/10.1007/BF00248156}{\emph{Archive for Rational
  Mechanics and Analysis} {\bfseries 33} (Jan, 1969) 54--70}.

\bibitem{lovelock71}
D.~Lovelock, \emph{The einstein tensor and its generalizations},
  \href{http://dx.doi.org/10.1063/1.1665613}{\emph{Journal of Mathematical
  Physics} {\bfseries 12} (1971) 498--501},
  [\href{https://arxiv.org/abs/https://doi.org/10.1063/1.1665613}{{\ttfamily
  https://doi.org/10.1063/1.1665613}}].

\bibitem{BJB08}
M.~Borunda, B.~Janssen and M.~Bastero-Gil, \emph{{Palatini versus metric
  formulation in higher curvature gravity}},
  \href{http://dx.doi.org/10.1088/1475-7516/2008/11/008}{\emph{JCAP} {\bfseries
  0811} (2008) 008}, [\href{https://arxiv.org/abs/0804.4440}{{\ttfamily
  0804.4440}}].

\bibitem{Alvarez18}
E.~Alvarez, J.~Anero, S.~Gonzalez-Martin and R.~Santos-Garcia, \emph{{Physical
  content of Quadratic Gravity}},
  \href{http://dx.doi.org/10.1140/epjc/s10052-018-6250-x}{\emph{Eur. Phys. J.}
  {\bfseries C78} (2018) 794},
  [\href{https://arxiv.org/abs/1802.05922}{{\ttfamily 1802.05922}}].

\bibitem{Alvarez-Gaume15}
L.~Alvarez-Gaume, A.~Kehagias, C.~Kounnas, D.~Lüst and A.~Riotto,
  \emph{{Aspects of Quadratic Gravity}},
  \href{http://dx.doi.org/10.1002/prop.201500100}{\emph{Fortsch. Phys.}
  {\bfseries 64} (2016) 176--189},
  [\href{https://arxiv.org/abs/1505.07657}{{\ttfamily 1505.07657}}].

\bibitem{Afonso17}
V.~I. Afonso, C.~Bejarano, J.~Beltran~Jimenez, G.~J. Olmo and E.~Orazi,
  \emph{{The trivial role of torsion in projective invariant theories of
  gravity with non-minimally coupled matter fields}},
  \href{http://dx.doi.org/10.1088/1361-6382/aa9151}{\emph{Class. Quant. Grav.}
  {\bfseries 34} (2017) 235003},
  [\href{https://arxiv.org/abs/1705.03806}{{\ttfamily 1705.03806}}].

\bibitem{Moffat}
J.~W. Moffat, \emph{New theory of gravitation},
  \href{http://dx.doi.org/10.1103/PhysRevD.19.3554}{\emph{Phys. Rev. D}
  {\bfseries 19} (Jun, 1979) 3554--3558}.

\bibitem{Damour92}
T.~Damour, S.~Deser and J.~G. McCarthy, \emph{{Nonsymmetric gravity theories:
  Inconsistencies and a cure}},
  \href{http://dx.doi.org/10.1103/PhysRevD.47.1541}{\emph{Phys. Rev.}
  {\bfseries D47} (1993) 1541--1556},
  [\href{https://arxiv.org/abs/gr-qc/9207003}{{\ttfamily gr-qc/9207003}}].

\bibitem{Domenech15}
G.~Domènech, A.~Naruko and M.~Sasaki, \emph{{Cosmological disformal
  invariance}},
  \href{http://dx.doi.org/10.1088/1475-7516/2015/10/067}{\emph{JCAP} {\bfseries
  1510} (2015) 067}, [\href{https://arxiv.org/abs/1505.00174}{{\ttfamily
  1505.00174}}].

\bibitem{Takahashi17}
K.~Takahashi, H.~Motohashi, T.~Suyama and T.~Kobayashi, \emph{{General
  invertible transformation and physical degrees of freedom}},
  \href{http://dx.doi.org/10.1103/PhysRevD.95.084053}{\emph{Phys. Rev.}
  {\bfseries D95} (2017) 084053},
  [\href{https://arxiv.org/abs/1702.01849}{{\ttfamily 1702.01849}}].

\bibitem{Zumalacarregui13}
M.~Zumalacárregui and J.~García-Bellido, \emph{{Transforming gravity: from
  derivative couplings to matter to second-order scalar-tensor theories beyond
  the Horndeski Lagrangian}},
  \href{http://dx.doi.org/10.1103/PhysRevD.89.064046}{\emph{Phys. Rev.}
  {\bfseries D89} (2014) 064046},
  [\href{https://arxiv.org/abs/1308.4685}{{\ttfamily 1308.4685}}].

\bibitem{Motohashi15}
H.~Motohashi and J.~White, \emph{{Disformal invariance of curvature
  perturbation}},
  \href{http://dx.doi.org/10.1088/1475-7516/2016/02/065}{\emph{JCAP} {\bfseries
  1602} (2016) 065}, [\href{https://arxiv.org/abs/1504.00846}{{\ttfamily
  1504.00846}}].

\bibitem{Watanabe15}
Y.~Watanabe, A.~Naruko and M.~Sasaki, \emph{{Multi-disformal invariance of
  non-linear primordial perturbations}},
  \href{http://dx.doi.org/10.1209/0295-5075/111/39002}{\emph{EPL} {\bfseries
  111} (2015) 39002}, [\href{https://arxiv.org/abs/1504.00672}{{\ttfamily
  1504.00672}}].

\bibitem{Tsujikawa14}
S.~Tsujikawa, \emph{{Disformal invariance of cosmological perturbations in a
  generalized class of Horndeski theories}},
  \href{http://dx.doi.org/10.1088/1475-7516/2015/04/043}{\emph{JCAP} {\bfseries
  1504} (2015) 043}, [\href{https://arxiv.org/abs/1412.6210}{{\ttfamily
  1412.6210}}].

\bibitem{Minamitsuji14}
M.~Minamitsuji, \emph{{Disformal transformation of cosmological
  perturbations}},
  \href{http://dx.doi.org/10.1016/j.physletb.2014.08.037}{\emph{Phys. Lett.}
  {\bfseries B737} (2014) 139--150},
  [\href{https://arxiv.org/abs/1409.1566}{{\ttfamily 1409.1566}}].

\bibitem{EERW18}
V.-M. Enckell, K.~Enqvist, S.~Rasanen and L.-P. Wahlman, \emph{{Inflation with
  $R^2$ term in the Palatini formalism}},
  \href{https://arxiv.org/abs/1810.05536}{{\ttfamily 1810.05536}}.

\bibitem{Enckell18}
V.-M. Enckell, K.~Enqvist, S.~Rasanen and L.-P. Wahlman, \emph{{Higgs-$R^2$
  inflation - full slow-roll study at tree-level}},
  \href{https://arxiv.org/abs/1812.08754}{{\ttfamily 1812.08754}}.

\bibitem{LiteBIRD20}
H.~Sugai, P.~A.~R. Ade, Y.~Akiba, D.~Alonso, K.~Arnold, J.~Aumont et~al.,
  \emph{Updated design of the cmb polarization experiment satellite litebird},
  2020.

\end{thebibliography}\endgroup

\end{document}